\documentclass[longauth]{aa}

\usepackage{graphicx}
\usepackage{array,multirow,makecell}
\usepackage{txfonts}
\usepackage{natbib}
\usepackage{amsmath}
\usepackage{ulem} 
\usepackage{booktabs}
\usepackage{arydshln}
\usepackage{textcomp}
\usepackage{gensymb}
\usepackage{color}
\usepackage{afterpage}
\usepackage[caption = false]{subfig}
\usepackage{float}
\bibpunct{(}{)}{;}{a}{}{,} 

\newcommand{\Msol}{\ensuremath{M_{\odot}}}

\newcommand{\hii}{H\mbox{\sc ii} }

\newcommand*{\rom}[1]{\expandafter\@slowromancap\romannumeral #1@}

\usepackage[colorlinks=True, filecolor=blue, linkcolor=blue]{hyperref}
\usepackage[capitalise]{cleveref}
\hypersetup{colorlinks,linkcolor={blue},citecolor={blue},urlcolor={red}}

\usepackage[caption=false]{subfig}

\usepackage[normal]{threeparttable}
\usepackage{longtable}
\usepackage{rotating} 

\usepackage{amsmath}
\DeclareFontEncoding{LS1}{}{}
\DeclareFontSubstitution{LS1}{stix}{m}{n}
\DeclareMathAlphabet\mathscr{LS1}{stixscr}{m}{n}

\usepackage{enumitem} 
\usepackage{stfloats} 

\newcommand{\bsens}{\texttt{bsens}\xspace}
\newcommand{\cleanest}{\texttt{cleanest}\xspace}
\newcommand{\denoised}{\texttt{denoised}\xspace}

\newcommand{\NHtwo}{N_{\rm H_2}}
\newcommand{\NHtwoUnits}{$\times 10^{21}$ cm$^{-2}$}
\newcommand{\st}{$^{\star}$}

\makeatletter
\renewcommand*\aa@pageof{, page \thepage{} of \pageref*{LastPage}}
\makeatother

\begin{document}

\title{ALMA-IMF. VI. Investigating the origin of stellar masses: Core mass function evolution in the W43-MM2\&MM3 mini-starburst}
\titlerunning{ALMA-IMF VI: CMF evolution in the W43-MM2\&MM3 mini-starburst}

    \author{Y. Pouteau\inst{1}
        \and F. Motte\inst{1}
        \and T. Nony\inst{2}
        \and M. Gonz\'alez\inst{1, 3}
        \and I. Joncour\inst{1}
        \and J.-F. Robitaille\inst{1}
        \and G. Busquet\inst{4,5,6}
        \and R. Galv\'an-Madrid\inst{2}
        \and A. Gusdorf\inst{7,8}
        \and P. Hennebelle\inst{9}
        \and A. Ginsburg\inst{10}
        \and T. Csengeri\inst{11}
        \and P. Sanhueza\inst{12,13}
        \and P. Dell'Ova\inst{7,8}
        \and A. M.\ Stutz\inst{14,15}
        \and A. P.\ M.\ Towner\inst{10}
        \and N. Cunningham\inst{1}
        \and F. Louvet\inst{1,16}
        \and A. Men'shchikov\inst{9}
        \and M. Fern\'andez-L\'opez\inst{17}
        \and N. Schneider\inst{18}
        \and M. Armante\inst{7,8}
        \and J. Bally\inst{19}
        \and T. Baug\inst{20}
        \and M. Bonfand\inst{11}
        \and S. Bontemps\inst{11}
        \and L. Bronfman\inst{21}
        \and N. Brouillet\inst{11}
        \and D. D\'iaz-Gonz\'alez\inst{2}
        \and F. Herpin\inst{11}
        \and B. Lefloch\inst{1}
        \and H.-L. Liu\inst{14,22}
        \and X. Lu\inst{23}
        \and F. Nakamura\inst{12,13,24}
        \and Q.~Nguy$\tilde{\hat{\rm e}}$n~Lu{\hskip-0.65mm\small'{}\hskip-0.5mm}o{\hskip-0.65mm\small'{}\hskip-0.5mm}ng\inst{25}
        \and F. Olguin\inst{26}
        \and K. Tatematsu\inst{27,13}
        \and M. Valeille-Manet\inst{11}
        }
        
    \institute{Univ. Grenoble Alpes, CNRS, IPAG, 38000 Grenoble, France
            \and Instituto de Radioastronom\'ia y Astrof\'isica, Universidad Nacional Aut\'onoma de M\'exico, Morelia, Michoac\'an 58089, M\'exico 
            \and Universidad Internacional de Valencia (VIU), C/Pintor Sorolla 21, E-46002 Valencia, Spain 
            \and Departament de F\'isica Qu\`antica i Astrof\'isica (FQA), Universitat de Barcelona (UB),  c. Mart\'i i Franqu\`es, 1, 08028 Barcelona, Spain 
            \and Institut de Ci\`encies del Cosmos (ICCUB), Universitat de Barcelona (UB), c. Mart\'i i Franqu\`es, 1, 08028 Barcelona, Spain 
            \and Institut d'Estudis Espacials de Catalunya (IEEC), c. Gran Capit\`a, 2-4, 08034 Barcelona, Spain 
            \and Laboratoire de Physique de l'{\'E}cole Normale Sup{\'e}rieure, ENS, Universit{\'e} PSL, CNRS, Sorbonne Universit{\'e}, Universit{\'e} de Paris, 75005, Paris, France 
            \and Observatoire de Paris, PSL University, Sorbonne Universit{\'e}, LERMA, 75014, Paris, France 
            \and Universit{\'e} Paris-Saclay, Universit{\'e} Paris Cit{\'e}, CEA, CNRS, AIM, 91191, Gif-sur-Yvette, France 
            \and Department of Astronomy, University of Florida, PO Box 112055, USA 
            \and Laboratoire d'astrophysique de Bordeaux, Univ. Bordeaux, CNRS, B18N, all\'ee Geoffroy Saint-Hilaire, 33615 Pessac, France  
            \and National Astronomical Observatory of Japan, National Institutes of Natural Sciences, 2-21-1 Osawa, Mitaka, Tokyo 181-8588, Japan 
            \and Department of Astronomical Science, SOKENDAI (The Graduate University for Advanced Studies), 2-21-1 Osawa, Mitaka, Tokyo 181-8588, Japan   
            \and Departamento de Astronom\'{i}a, Universidad de Concepci\'{o}n, Casilla 160-C, 4030000 Concepci\'{o}n, Chile 
            \and Max-Planck-Institute for Astronomy, K\"{o}nigstuhl 17, 69117 Heidelberg, Germany   
            \and DAS, Universidad de Chile, 1515 camino el observatorio, Las Condes, Santiago, Chile 
            \and Instituto Argentino de Radioastronom\'\i a (CCT-La Plata, CONICET; CICPBA), C.C. No. 5, 1894, Villa Elisa, Buenos Aires, Argentina   
            \and I. Physikalisches Institut, Universit\"{a}t zu K\"{o}ln, Z\"{u}lpicher Str. 77, 50937 K\"{o}ln, Germany 
            \and Department of Astrophysical and Planetary Sciences, University of Colorado, Boulder, Colorado 80389, USA    
            \and S. N. Bose National Centre for Basic Sciences, Block JD, Sector III, Salt Lake, Kolkata 700106, India    
            \and Departamento de Astronom\'{i}a, Universidad de Chile, Casilla 36-D, Santiago, Chile   
            \and Department of Astronomy, Yunnan University, Kunming, 650091, PR China    
            \and Shanghai Astronomical Observatory, Chinese Academy of Sciences, 80 Nandan Road, Shanghai 200030, People’s Republic of China  
            \and The Graduate University for Advanced Studies (SOKENDAI), 2-21-1 Osawa, Mitaka, Tokyo 181-0015, Japan 
            \and CSMES, The American University of Paris, 2bis, Passage Landrieu 75007 Paris, France 
            \and Institute of Astronomy, National Tsing Hua University, Hsinchu 30013, Taiwan   
            \and Nobeyama Radio Observatory, National Astronomical Observatory of Japan, National Institutes of Natural Sciences, Nobeyama, Minamimaki, Minamisaku, Nagano 384-1305, Japan 
                        }

    \date{}

    \abstract
    {Among the most central open questions regarding the initial mass function (IMF) of stars is the impact of environment on the shape of the core mass function (CMF) and thus potentially on the IMF. }
%
    {The ALMA-IMF Large Program aims to investigate the variations in the core distributions (CMF and mass segregation) with cloud characteristics, such as the density and kinematic of the gas, as diagnostic observables of the formation process and evolution of clouds. The present study focuses on the W43-MM2\&MM3 mini-starburst, whose CMF has recently been found to be top-heavy with respect to the Salpeter slope of the canonical IMF.}
%
    {W43-MM2\&MM3 is a useful test case for environmental studies because it harbors a rich cluster that contains a statistically significant number of cores (specifically, 205 cores), which was previously characterized in Paper~III. We applied a multi-scale decomposition technique to the ALMA 1.3~mm and 3~mm continuum images of W43-MM2\&MM3 to define six subregions, each 0.5-1~pc in size. For each subregion we characterized the high column density probability distribution function, $\eta$-PDF, and the shape of the cloud gas using the 1.3~mm image. Using the core catalog, we investigate correlations between the CMF and cloud and core properties, such as the $\eta$-PDF and the core mass segregation.} 
%
    {We classify the W43-MM2\&MM3 subregions into different stages of evolution, from quiescent to burst to post-burst, based on the surface number density of cores, number of outflows, and ultra-compact  \hii presence. The high-mass end ($>$1~$\Msol$) of the subregion CMFs varies from close to the Salpeter slope (quiescent) to top-heavy (burst and post-burst). Moreover, the second tail of the $\eta$-PDF varies from steep (quiescent) to flat (burst and post-burst), as observed for high-mass star-forming clouds. We find that subregions with flat second $\eta$-PDF tails display top-heavy CMFs.}
%
    {In dynamical environments such as W43-MM2\&MM3, the high-mass end of the CMF appears to be rooted in the cloud structure, which is at high column density and surrounds cores. This connection stems from the fact that cores and their immediate surroundings are both determined and shaped by the cloud formation process, the current evolutionary state of the cloud, and, more broadly, the star formation history. The CMF may evolve from Salpeter to top-heavy throughout the star formation process from the quiescent to the burst phase. This scenario raises the question of if the CMF might revert again to Salpeter as the cloud approaches the end of its star formation stage, a hypothesis that remains to be tested.}
  
    \keywords{stars: formation – stars: IMF – ISM: clouds}

    \maketitle


\section{Introduction} \label{sect:intro}

Star formation takes place in molecular clouds, which are partly supported by thermal pressure, turbulence, and magnetic fields but are, above all, formed and shaped by gravity, stellar feedback, and Galactic motions \citep[see reviews by, e.g.,][]{krumholz2015, ballesteros-paredes2020}. The gas reservoir available to form a star is difficult to define and depends strongly on whether the processes of cloud and star formation are quasi-static or dynamic. In the quasi-static scenario, which has long been the most generally accepted scenario, the cores are mass reservoirs for the collapse of protostars that will form a single star or, at most, a small stellar system \citep[e.g.,][]{shu1987,chabrier2003,mckeeOstriker2007}. Observationally, we characterize cores as dense, $n_{\rm H_2}=10^4-10^8$~cm$^{-3}$, gravitationally bound cloud fragments that are referred to either as pre-stellar cores, when they are on the verge of collapse, or as protostellar cores, when a stellar embryo exists at the center of the collapsing core.

However, in dynamical scenarios of cloud and star formation, cores may not be the only mass reservoirs available for the collapse of protostars. In scenarios such as those described in competitive accretion, global hierarchical  collapse, or inertial inflow models, the environment plays a crucial role in the assembly of core mass and the protostellar accretion of gas onto the newborn star \citep[e.g.,][]{smith2009,wang2010,vazquez2019, pelkonen2021}. Clouds are observed to globally collapse toward a few, very specific sites, named filament hubs and ridges; they are cloud structures a few parsecs in size  with mean volume densities similar to, or greater than, the minimum value observed for isolated starless cores, $n_{\rm H_2}=10^4$~cm$^{-3}$ \citep[e.g.,][]{schneider2010, peretto2013, busquet2013, galvanmadrid2013}. Cores at these locations are fed, during both their pre-stellar and protostellar phases, by gas inflows originating from the global infall of ridges and hubs \citep[e.g.,][]{smith2009, motte2018a}.
In the present study, we define clouds and cores as molecular cloud structures of a few parsecs and $\sim$0.01~pc in size, respectively. Neither clouds nor cores, however, can be fully considered in isolation from their respective environment: they are dynamically evolving \citep[e.g.,][]{csengeri2011, olguin2021, sanhueza2021}. 

For decades, the stellar initial mass function (IMF), which characterizes the mass distribution of stars above 0.01~\Msol, has been found to display a universal form that can be schematized by a lognormal function, peaking near $0.3~\Msol$, and a power law of the form $\frac{{\rm d}N}{{\rm d}\log M} \propto M^{-1.35}$, or in its cumulative forms  $N(>{\rm log}(M))\propto M^{-1.35}$, above $\sim$1$~\Msol$ \citep{chabrier2005,bastian2010,hopkins2012, kroupa2013}. 
The analogous function for cores, the core mass function (CMF), has also been long found to have a similar shape, or at least the same slope at the high-mass end as those of the canonical IMF \citep[e.g.,][]{motte1998, testi-sargent1998,enoch2008, konyves2015, konyves2020, takemura2021}. These results led to the simple interpretation of a direct mapping between the CMF and the IMF. In the meantime, numerical simulations computed CMFs that also showed agreement with the IMF shape  \citep[e.g.,][]{klessen2000, padoan2011} in some cases (but see \citealt{smith2009}). 
All this taken together suggests that cores on the verge of collapse are the direct progenitors of stars and that their mass could be directly accreted, with a given efficiency, by the nascent star \citep[e.g.,][]{chabrier2005,andre2014}. 

More recently, however, top-heavy IMFs have been revealed in young massive clusters of the Milky Way \citep{kim2006, lu2013, maia2016, hosek2019}, in nearby galaxies \citep{schneider2018}, and in high-redshift galaxies \citep{smith2014, zhang2018}. According to \cite{marks2012}, the slope of the high-mass-end IMF could be directly related to cloud volume density for the precursors of extreme star clusters. When focusing on young, very dense, and massive Galactic clouds, \cite{motte2018b}, \cite{kong2019}, \cite{lu2020}, and \cite{pouteau2022} all observed atypical CMFs. The CMFs of the W43-MM1 and W43-MM2\&MM3 mini-starburst ridges, that of the G28.37 hub, and those of the massive clouds in the Central Molecular Zone are indeed top-heavy when compared to the canonical IMF. In detail, their CMF high-mass ends, which range from $\sim$1~$\Msol$ to $\sim$100~$\Msol$, can be fitted by a power law of the form $N(> \log M) \propto M^\alpha$ with $\alpha \in [-0.95;-0.85]$, compared to the canonical $\alpha_{\rm IMF} = -1.35$ Salpeter IMF slope \citep{salpeter1955}.
Recent numerical simulations used density-threshold-defined sink particles, which are interpreted as the numerical simulation equivalent of a forming star, and found variations in the sink mass function. In particular, the power-law index of the high-mass CMF is observed to depend on whether the initial cloud support is thermally dominated or turbulence-dominated \citep{ballesteros-paredes2015, leeHennebelle2018a,hennebelle2022}.

In this framework, the ALMA-IMF\footnote{
    ALMA project \#2017.1.01355.L; see \url{http://www.almaimf.com} (PIs: Motte, Ginsburg, Louvet, Sanhueza).}
Large Program \citep{motte2022} was set up to investigate the variations in the CMF with cloud characteristics, especially the density structure, which is likely related to the cloud formation process and evolutionary stage. ALMA-IMF imaged 15 massive protoclusters at different evolutionary stages and with cloud characteristics that cover a wide range in terms of mass \citep[$2.5-33\times 10^3~\Msol$ in $1-8$~pc$^2$;][]{motte2022} and density. 
The W43 complex hosts two such clouds, W43-MM1 and W43-MM2\&MM3, whose column densities  are greater than the column density threshold taken to define ridges ($>$10$^{23}$~cm$^{-2}$; \citealt{hill2011, hennemann2012, nguyen2013}). These high-density, parsec-size filamentary structures have been qualified as mini-starbursts because they harbor protoclusters that efficiently form high-mass stars \citep{motte2003, louvet2014}, as in other ridges or hubs \citep[e.g.,][]{nguyen2011b, galvanmadrid2013, nony2021} and in cloud complexes \citep{nguyen2016}. Although they are located within 10~pc of an association of OB and Wolf-Rayet stars \citep{blum1999,bik2005}, these two ridges are mainly composed of cold gas (21--28~K; see Fig.~2 of \citealt{nguyen2013}). The ALMA-IMF Large Program imaged them with $\sim$0.5$\arcsec$ resolution \citep[see companion papers, Paper~I and Paper~II,][]{motte2022, ginsburg2022}, which corresponds to $\sim$2500~au at the $5.5$~kpc distance of W43 \citep{zhang2014}. Its 1.3~mm images cover the massive clouds W43-MM1 and W43-MM2, $\sim$$1.3\times10^4~\Msol$ and $\sim$$1.2\times10^4~\Msol$ over 7~pc$^2$ and 6~pc$^2$, respectively, and their less massive neighbor W43-MM3, $\sim$$0.5\times10^4~\Msol$ over 6~pc$^2$ \citep{motte2022}. Paper~III of this series \citep{pouteau2022} found that the W43-MM2\&MM3 mini-starburst has an atypical CMF, with a top-heavy high-mass end with respect to the Salpeter slope of the canonical IMF. 

The present study aims to investigate the CMF variations in W43-MM2\&MM3 and establish the link between the CMF shape and local conditions. The variation in these conditions is quantified by estimating the median properties, density structure, and evolutionary state of the cloud, as well as the spatial distribution and mass segregation of cores.
Among the statistical tools used to characterize molecular clouds, probability distribution functions (PDFs) of density and column density are one of the most widely used in both theoretical and numerical studies \citep[e.g.,][]{klessen2000, girichidis2014, leeHennebelle2018a, jaupartChabrier2020} and in observational studies \citep[e.g.,][]{kainulainen2009, schneider2012, schneider2016, schneider2022, stutzKainulainen2015}. In simulations, PDFs are usually obtained from cloud volume density, while in observations they are obtained from column density ($\eta$-PDF; see the definition in Sect.~\ref{sect:PDFs}), which can be related to radial density distributions in clouds but only in idealized cases \citep{federrathKlessen2013, myers2015}. In agreement with observations, PDFs are expected to be lognormal if isothermal supersonic turbulence is dominant and to develop a power-law tail at high column densities when gravity is no longer negligible \citep{klessen2000, kainulainen2009, ballesteros-paredes2011, schneider2013, pokhrel2016}. The $\eta$-PDF sometimes harbors a second, flatter tail at column densities higher than $5-10\times 10^{22}$~cm$^{-2}$ \citep{schneider2015, lin2016, schneider2022}, whose nature is still a matter of debate \citep[e.g.,][]{sadavoy2014, tremblin2014, motte2018a, khullar2021, donkov2021}.
Because the high-column-density parts of a cloud host most of the cores, understanding the link between the tails of the $\eta$-PDF and the shape of the CMF may provide insights into the possible universality of the CMF.

As for mass segregation, it measures whether the most massive objects, in particular stars or cores, show a significantly different distribution than their lower-mass counterparts and whether they are more clustered. The mass segregation of young stars and cores is generally studied to determine whether the mass segregation observed in stellar clusters is inherited by a primordial mass segregation of cores in clouds \citep{hetem2019, dib2018segreg} or is rapidly erased through dynamical interactions of the stars and expulsion of the gas \citep{parkerMeyer2012, FujiiPortegies-Zwart2016}. A dozen studies have investigated the mass segregation of 0.002--0.1~pc cloud structures, but only five found clear evidence of segregation \citep{lane2016, parker2018, plunkett2018, dibHenning2019, nony2021}. 
\cite{dibHenning2019} and \cite{nony2021} proposed that the level of core mass segregation depends on how the gas mass is assembled to form the molecular clouds. Environmental factors could therefore have an impact on both core mass segregation and CMF shape.

\begin{figure*}[htbp!]
\centering
\begin{minipage}{1.\textwidth}
    \centering
    \includegraphics[width=1.\textwidth]{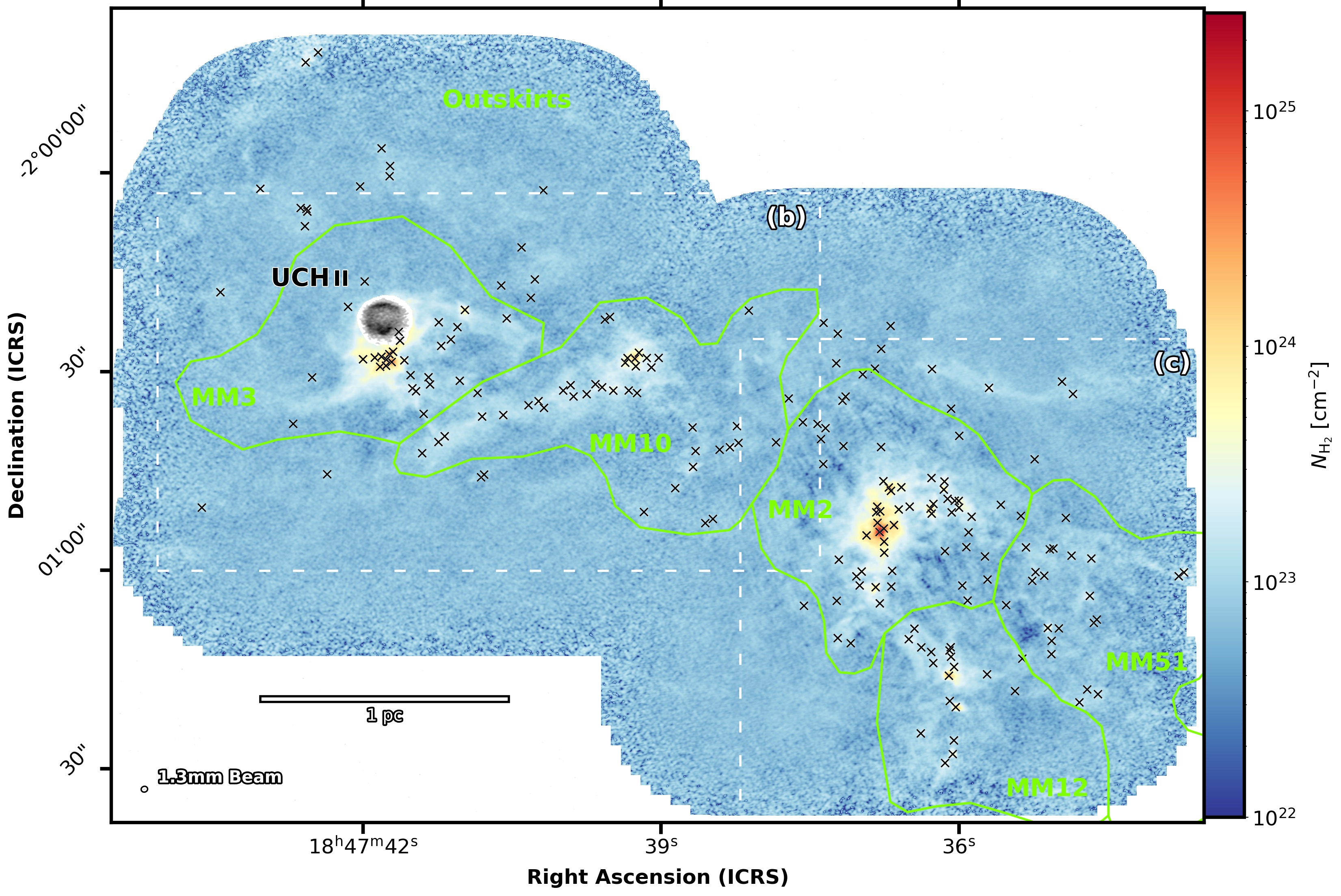}
\end{minipage}
\caption{W43-MM2\&MM3 mini-starburst ridge traced by its column density (color scale), derived from the ALMA 1.3~mm continuum image of \cite{pouteau2022}. It hosts a rich cluster of 205 cores (gray crosses), as reported by \cite{pouteau2022}, and the UC\hii region W43-MM3, which is traced by its H41$\alpha$ emission line (Galv\'an-Madrid et al. in prep.; gray scale) and where the column density is not computed. Green polygons outline the W43-MM2\&MM3 subregions defined in Sect.~\ref{sect:subregions def}. The ellipse in the lower-left corner indicates the $0.46\arcsec$ angular resolution of the column density image, while the scale bar indicates the length scale at a distance of 5.5~kpc. \textit{a)} Dashed white boxes outline the zoomed-in views shown in panels \textit{b--c}.}
\label{fig:nh2 and cores}
\end{figure*}

\setcounter{figure}{0}
\begin{figure*}[htbp!]
\centering
\begin{minipage}{1.\textwidth}
    \centering
    \includegraphics[width=0.7\textwidth]{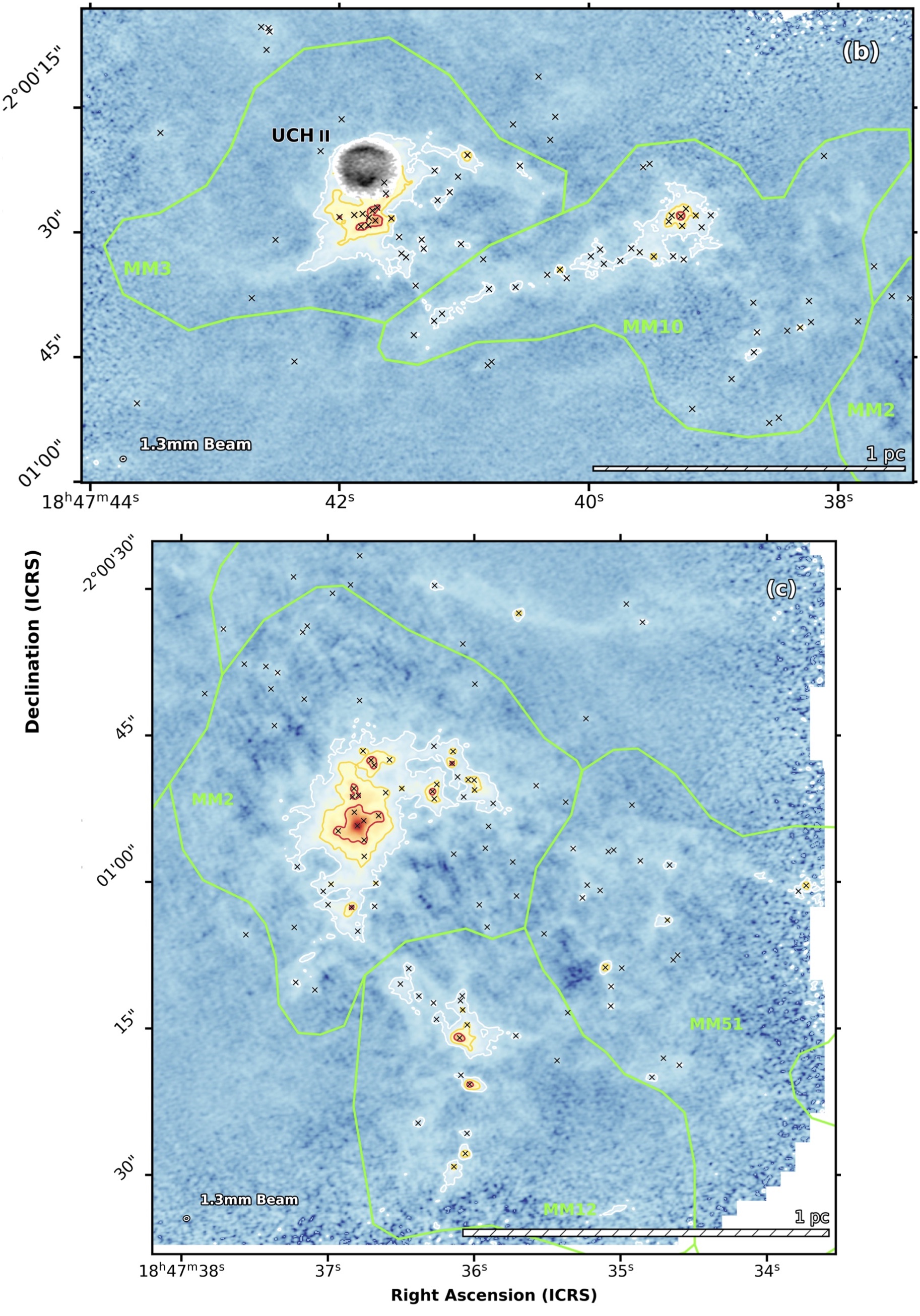}
\end{minipage}
\caption{(Continued) Zoomed-in view of the MM3 and MM10 subregions (\textit{b}) and the MM2, MM12, and MM51 subregions (\textit{c}). White, orange, and red contours correspond to column densities of $1.0$, $3.5$, and $10\times10^{23}$~cm$^{-2}$. Pixels located between the white and orange contours and between the white and red contours constitute the $s_2$ tails of the MM10 and MM51 subregions and the MM2, MM3, and MM12 subregions, respectively.}
\end{figure*}

The paper is organized as follows: In Sect.~\ref{sect:database and basic properties} we present the database used in this work, separate the W43-MM2\&MM3 into six subregions, compute their column density maps, and estimate their characteristics, such as mass, volume density, and column density PDF. 
In Sect.~\ref{sect:analysis} we then search for variations: within these six subregions, in the CMF shape, in the cloud characteristics, and in the core mass segregation. In Sect.~\ref{sect:discussion} we investigate the correlation between the power-law index of the CMF high-mass end and the cloud characteristics, including the $\eta$-PDF, the spatial core distribution, and the cloud and star formation histories. We summarize the paper and present our conclusions in Sect.~\ref{sect:conclusions}.

\begin{figure*}[htbp!]
\centering
\begin{minipage}{1.\textwidth}
  \centering
  \includegraphics[width=.8\textwidth]{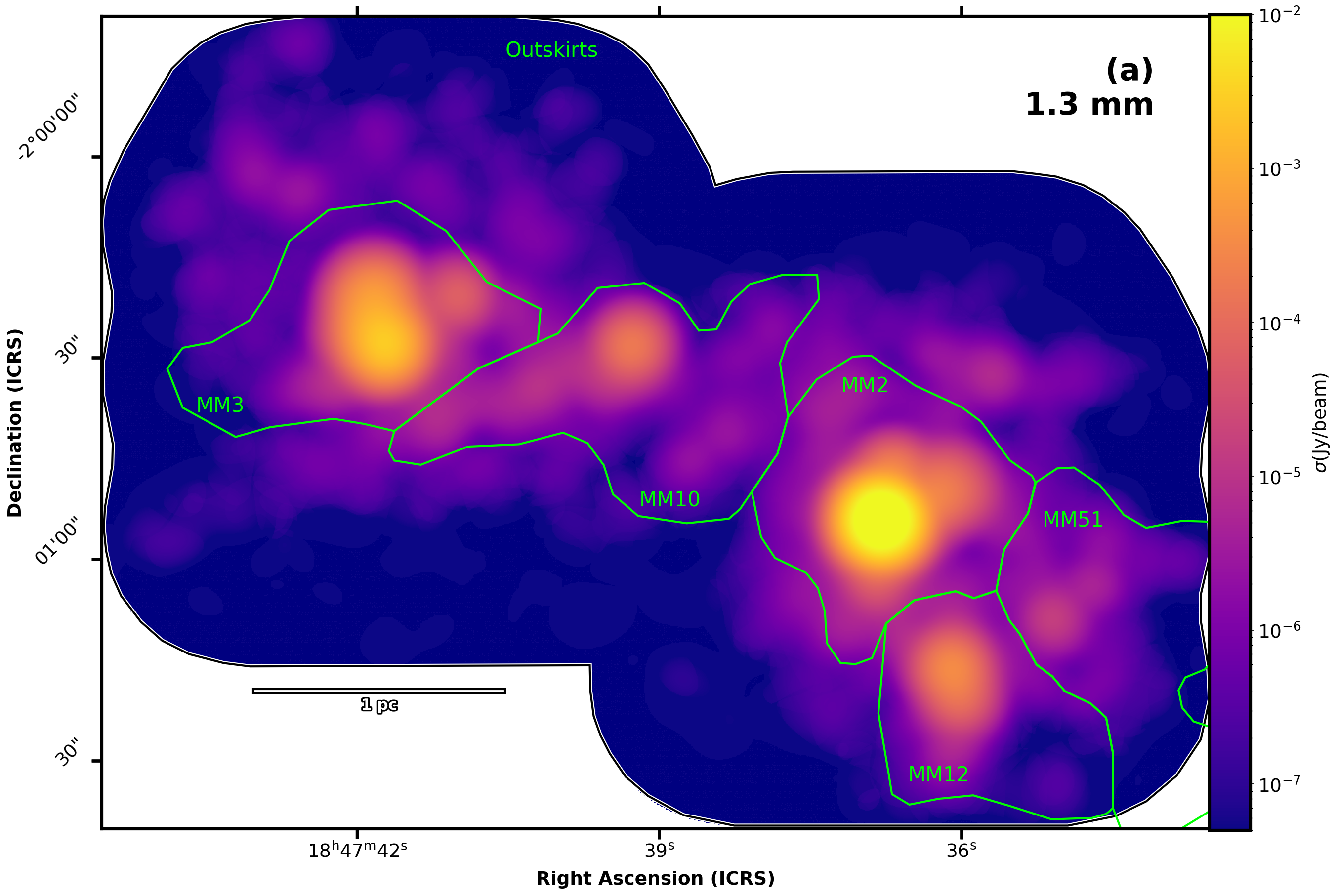}
\end{minipage}
\begin{minipage}{1.\textwidth}
  \centering
  \includegraphics[width=.8\textwidth]{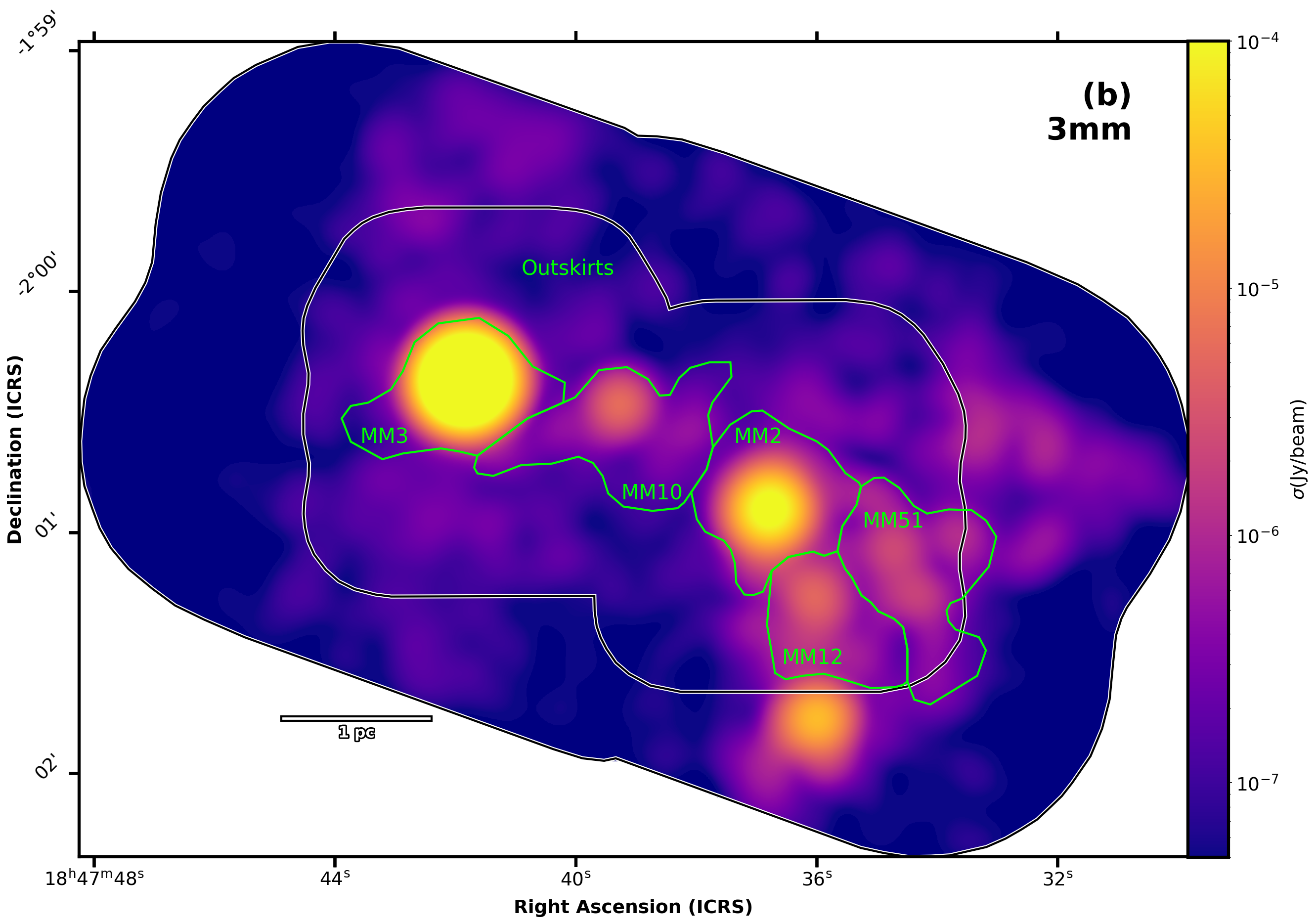}
\end{minipage}
\caption{Subregions of $0.5-1$~pc sizes identified in the W43-MM2\&MM3 ridge from the ALMA 12~m array \denoised \bsens continuum images of \cite{pouteau2022} at 1.3~mm (\textit{panel a}) and 3~mm (\textit{panel b}). A multi-scale decomposition was performed, decomposition scales of $\sim$4$-7\arcsec$ were chosen, and subregion outlines (green contours) were set to follow inflection points in the column density distribution. Subregions are labeled MM2, MM3, MM10, MM12, and MM51 following the names of fragments extracted from $\sim$10$\arcsec$ resolution (sub)millimeter images \citep{motte2003}. The remaining subregion at 1.3~mm is labeled Outskirts. The white contour outlines the 1.3~mm coverage in panel \textit{(b),} and a 1 pc scale bar is shown in the lower-left corner of each image.}
\label{fig:mngseg scales}
\end{figure*}

\section{Database and basic properties}
\label{sect:database and basic properties}

\subsection{Observations and core catalog} \label{sect:obs and catalog}

The W43-MM2\&MM3 mini-starburst ridge is among the targets of the ALMA-IMF Large Program \citep[Paper~I,][]{motte2022}, whose observations were taken between December 2017 and December 2018. The present paper uses the 12~m array continuum images in the ALMA band~6 (1.3~mm, central frequency $\nu_{\rm c}\simeq 228.4$~GHz) and band~3 (3~mm, $\nu_{\rm c}\simeq99.66$~GHz) and the core catalog\footnote{
    Full catalog is available at the CDS via anonymous ftp to cdsarc.u-strasbg.fr (ftp://130.79.128.5) or via \url{http://cdsarc.u-strasbg.fr/viz-bin/cat/J/A+A/664/A26}} 
all reported in Paper~III \citep{pouteau2022}. Images of individual ALMA-IMF fields W43-MM2 and W43-MM3 were first cleaned using the iterative self-calibration pipeline scripts\footnote{
    This pipeline performs several iterations of phase self-calibration using custom masks in order to better define the self-calibration model and clean more and more deeply using the TCLEAN task and refined parameters after each pass. The pipeline is available at \url{https://github.com/ALMA-IMF/reduction} .} 
developed by the ALMA-IMF consortium and fully described in  Paper~II \citep{ginsburg2022}. 
Paper~III \citep{pouteau2022} used the multi-scale option of the TCLEAN task, with  parameters of 0, 3, 9, 27 pixels (up to 81 at 3~mm), to minimize interferometric artifacts associated with missing short spacings. Images at 3~mm were then smoothed to the 1.3~mm synthesized beam of $\Theta_{\rm beam} \simeq 0.46\arcsec$ of both 1.3~mm fields, corresponding to $\sim$2500~au (or 0.013~pc) at the 5.5~kpc distance of W43. According to Paper~II, the maximum recoverable scales are $\sim$5.6~$\arcsec$ at 1.3~mm and $\sim$8.1~$\arcsec$ at 3~mm \citep{ginsburg2022}, corresponding to $\sim$0.15~pc and $\sim$0.2~pc, respectively. 
Because W43-MM2 and W43-MM3 share a common area in both bands, the mosaics were combined using primary-beam shape weights in Paper~III \citep{pouteau2022}. The multi-scale non-Gaussian segmentation technique \citep[{MnGSeg}; see][]{robitaille2019} was also applied to reduce the noise level of the best-sensitivity (\bsens) 1.3~mm and 3~mm images by $\sim$30\% \citep[for more information, see Sect.~2 and Appendix~A of][]{pouteau2022}. The resulting images are qualified as \bsens \denoised, with a continuum sensitivity of $\sim$80~$\mu$Jy$\,$beam$^{-1}$ and $\sim$21~$\mu$Jy$\,$beam$^{-1}$ at 1.3~mm and 3~mm, respectively.

Paper~III identified cores by extracting compact emission peaks from their surrounding background in the ALMA 12~m array continuum images at 1.3~mm, with a $5\sigma$ sensitivity per beam unit of $\sim$0.25~$\Msol$ \citep{pouteau2022}. The \textsl{getsf} algorithm \citep{men2021getsf} was applied on the \denoised, \bsens and \cleanest images at 1.3~mm and 3~mm that were built by using all frequency channels and selecting those without line emission, respectively \citep[see detailed definition in][]{pouteau2022}. The \textsl{getsf} catalog at 1.3~mm was then filtered to exclude free-free emission peaks and continuum fluxes were corrected for line contamination and optically thick thermal dust emission \citep[for details, see][]{pouteau2022}. The resulting core catalog contains 205 cores, with a median size full width at half maximum (FWHM) of 3\,400~au and masses, $M^{\rm core}$, ranging from 0.1 to 70~$\Msol$ (see Tables~E.1--E.2 of \citealt{pouteau2022}). Figure~\ref{fig:nh2 and cores} locates the rich protocluster of cores discovered and characterized by Paper~III \citep{pouteau2022} in the W43-MM2\&MM3 ridge. A companion paper, Paper~V \citep{nony2023} investigates the protostellar or pre-stellar nature of the W43-MM2\&MM3 cores and finds that between 41 and 51 ($46\pm5$) protostars are driving outflows or igniting hot cores.


{\renewcommand{\arraystretch}{1.5}%
\begin{table*}[ht]
\centering
\resizebox{\textwidth}{!}{
\begin{threeparttable}[c]
\caption{Main characteristics of the cloud and core population in the parsec-scale subregions of W43-MM2\&MM3.}
\label{tab:subregions chara}
\begin{tabular}{cccccccccc}
    \hline\hline
    Subregion   &   \multirow{2}{*}{$A$}   &    $T_{\rm dust}$   &   $N_{\rm H_2}$   & \multirow{2}{*}{$M^{\rm subregion}$}   &
    \multirow{2}{*}{$n_{\rm H_2}$}   &  Number   &   Core   &   \multirow{2}{*}{$\sum \limits_{\rm subregion}M^{\rm core}$} & $\sum \limits_{\rm subregion}M^{\rm core}$ \\
    name  &  & range  &  range  &  &  &  of  cores  &  density  &  & $/ M^{\rm subregion}$ \\
     & [pc$^2$] & [K] & [$\times 10^{23}\,$cm$^{-2}$] & [\Msol] & [$\times 10^{3}\,$cm$^{-3}$] & $N_{\rm core}$ & [pc$^{-2}$] & [\Msol] & [\%] \\
    (1) & (2) & (3) & (4) & (5) & (6) & (7) & (8) & (9) & (10) \\
    \hline
    Outskirts &  $\sim$7$\pm2$ & $21-42$ &   $0.1-5.5$ &      $>$120 &       $>$0.07 & $35\pm20$ &   5$\pm2$ &   $26\pm2$ & $<22$ \\
    MM51      &    0.7$\pm0.3$ & $22-28$ &   $0.1-8.7$ &       $>$35 &         $>$25 &  22$\pm1$ & 31$\pm18$ &   $27\pm2$ & $<77$ \\
    MM10      &    0.9$\pm0.3$ & $21-40$ &  $0.3-30.0$ &  $340\pm10$ &    $101\pm45$ &  41$\pm6$ & 46$\pm19$ &   $61\pm4$ & $\sim$18
    \\
    MM12      &    0.6$\pm0.2$ & $21-33$ &  $0.1-22.0$ &  $200\pm10$ &    $188\pm25$ &  19$\pm1$ & 32$\pm14$ &   $48\pm3$ & $\sim$24
    \\
    MM2       &    0.8$\pm0.2$ & $20-65$ & $0.4-260.6$ & $1300\pm20$ &   $490\pm150$ & 59$\pm11$ &  74$\pm5$ & $240\pm11$ & $\sim$18
    \\
    MM3       &    0.8$\pm0.3$ & $22-35$ &  $0.5-80.2$ & $1020\pm50$ &    $375\pm90$ &  29$\pm7$ &  36$\pm6$ &  $140\pm7$ & $\sim$14
    \\ \hline
    Total     &       $\sim$11 & $20-65$ & $0.1-107.3$ &     $>$3000 &        $>$0.5 &       205 &        19 & $540\pm25$ & $<$18
    \\ \hline
\end{tabular}
\begin{tablenotes}
\item (2) Area of the subregions defined in Sect.~\ref{sect:subregions def} (see also Figs.~\ref{fig:nh2 and cores}--\ref{fig:mngseg scales}). Except for the full area, labeled as Total, the uncertainties reflect the inherent difficulty in defining the subregion outlines and are the basis for the uncertainties given in cols.~5--8.
\item (3) Temperature range measured on the dust temperature image of \cref{appendixfig:PPMAP with NH2}
\citep[see also][]{pouteau2022}.
\item (4) Column density range measured on the column density image of \cref{fig:nh2 and cores}.
\item (5) Subregion mass computed from the 1.3~mm and dust temperature images of \ref{appendixfig:PPMAP with NH2} \citep[see also][]{pouteau2022} following \cref{eq:optically thin mass}. These values are lower limits for the MM51 and Outskirts subregions, and for the Total region (see Sect.~\ref{sect:subregions coldens and mass}).
\item (6) Volume densities computed from cols.~2 and 5 following \cref{eq:density}. 
\item (7)--(8) Number and surface number density of cores, detected by \cite{pouteau2022}, which are located over subregion areas of col.~2 (see also \cref{fig:mngseg scales}).
\item (9) Cumulative mass of cores, taken from Table~E.2 of \cite{pouteau2022}. Uncertainties arise from those associated with individual core mass estimates that depend on flux and temperature uncertainties, thus the ignoring uncertainties of col.~7.
\item (10) Concentration of subregion gas mass within cores computed from cols.~5 and 9.

\end{tablenotes}
\end{threeparttable}}
\end{table*}}

\subsection{Splitting the W43-MM2\&MM3 ridge into six subregions} \label{sect:subregions def}

As shown in \cref{fig:nh2 and cores}, the W43-MM2\&MM3 ridge displays an inhomogeneous gas distribution. We separated it into subregions of $\sim$0.5--1~pc scales, which, in the framework of dynamical cloud formation, would correspond to cloud subparts with more homogeneous characteristics and star formation activities than the whole cloud.
To define the boundaries of these subregions, we used the {MnGSeg} technique developed by \cite{robitaille2019}. This algorithm, based on complex wavelet decomposition, has the main objective of separating coherent structures associated with star formation from the turbulent cloud structures, which are incoherent from one scale to another and referred to as Gaussian. We used it here solely for its multi-scale decomposition ability of the \bsens continuum images at 1.3~mm and 3~mm. Decomposed spatial scales range from the beam, $\Theta_{\rm beam}$ corresponding to $\sim$0.013~pc, to the largest scales traced by the present interferometric images, $\sim$0.15~pc at 1.3~mm and $\sim$0.2~pc at 3~mm \citep{ginsburg2022, motte2022}. 

Figure~\ref{fig:mngseg scales} shows, for the W43-MM2\&MM3 ridge, the wavelet filtered spatial scale tracing cloud structures of 0.11--0.19~pc ($4\arcsec-7\arcsec$ at 5.5~kpc), which are thus the largest traced by the ALMA-IMF configurations of the 12~m array at 1.3~mm and 3~mm. Five subregions of $\sim$0.5--1~pc sizes were identified in both Figs.~\ref{fig:mngseg scales}a--b; their boundaries were visually defined by following the minimum inflection points, corresponding to minima or saddle points. Discovered as separate cloud structures in (sub)millimeter continuum images with $\sim$0.3~pc spatial resolutions, we keep their labeling that comes from the rank of their extraction \citep{motte2003}. The W43-MM2\&MM3 subregions identified here are thus called MM2, MM3, MM10, MM12, and MM51. 
Figures~\ref{fig:mngseg scales}a-b are dominated by the two centrally concentrated subregions MM2 and MM3, the latter of which hosts a well-known ultra-compact (UC) \hii region \citep{nguyen2017}. Additionally, Fig.~\ref{fig:mngseg scales}a displays three networks of filaments with lower density: MM10, MM12, and MM51. In \cref{fig:mngseg scales}b, the bright region south of MM12, known as MM13, is mainly associated with free-free emission from an UC\hii region \citep{motte2003, nguyen2017}. Since this subregion was not imaged at 1.3~mm, we ignore it in the following analysis. Once the five subregions mentioned above are subtracted from the area imaged at 1.3~mm, an area remains that we call here the Outskirts subregion.

\cref{tab:subregions chara} lists the characteristics of these six subregions (MM2, MM3, MM10, MM12, MM51, and Outskirts) as derived from measurements in the 1.3~mm \bsens \denoised image obtained with the 12~m array of ALMA (see Sect.~\ref{sect:obs and catalog}). We chose the 1.3~mm image because, unlike the 3~mm image, it is less contaminated by free-free emission. \cref{tab:subregions chara} lists their spatial area, temperature and column density ranges, mass, and density, along with the basic properties of their core populations (number, surface density, and cumulative mass of cores, as well as subregion mass concentration in core mass). These values are computed in the following Sects.~\ref{sect:subregions coldens and mass} and \ref{sect:spatial variations of core pop}. Uncertainties are estimated by slightly varying the integration areas of Figs.~\ref{fig:nh2 and cores}--\ref{fig:mngseg scales} (see \cref{tab:subregions chara}) to take into account uncertainties in the definition of subregion outlines and areas.

\subsection{Subregion column density and mass estimates} 
\label{sect:subregions coldens and mass}

Because the thermal dust emission of clouds is mostly optically thin at 1.3~mm, we computed the column densities and masses of the W43-MM2\&MM3 subregions, under the assumption of optically thin thermal dust emission. The column density image of \cref{fig:nh2 and cores}, $\NHtwo$, is computed from the 1.3~mm fluxes, $S_{\rm 1.3mm}^{\rm peak}$, measured in the 12~m array continuum image with a $0.46\arcsec$-beam. 
Using the correction for optical thickness proposed by Paper~III, would only change its values by 15\% toward the center of four cores \citep{pouteau2022}. We used the following equation, and provide a numerical application for typical values of $S_{\rm 1.3mm}^{\rm peak}$, dust temperature and opacity per unit (gas $+$ dust) mass column density at 1.3~mm, $T_{\rm dust}$ and $\kappa_{\rm 1.3mm}$. After the numerical application using the Planck function, the dependence on each physical variable is given, for simplicity, in the Rayleigh-Jeans approximation:
\begin{equation}
    \begin{split}
    \NHtwo \; [{\rm cm}^{-2}] \: &=  \frac{S_{\rm 1.3mm}^{\rm peak}}{ \Omega_{\rm beam}\;\mu\,m_{\rm H} \: \kappa_{\rm 1.3mm}\: B_{\rm \tiny 1.3mm}(T_{\rm dust})} + \NHtwo^{\rm bckg} \\
    &\simeq\:  1.3\times10^{23}~{\rm cm^{-2}} \times \left(\frac{S_{\rm 1.3mm}^{\rm peak}}{\rm mJy\, beam^{-1}}\right)
    \, \left(\frac {T_{\rm dust}}{\rm 23~K}\right)^{-1}\\
    & \times
     \left(\frac {\kappa_{\rm 1.3\,mm}}{\rm 0.01\,cm^2\, g^{-1}}\right)^{-1}
    + 0.7\times10^{23}~\rm cm^{-2},
    \end{split}
\label{eq:nh2 region} 
\end{equation}
where $\Omega_{\rm beam}=\frac{\pi \; 0.51\arcsec \times 0.40\arcsec}{4\,\ln{2}}$ is the beam solid angle, $\mu = 2.8$ is the mean molecular weight per hydrogen molecule (assuming an helium abundance of 10\%), $m_{\rm H}$ is the mass of atomic hydrogen, $B_{\rm \tiny 1.3mm}(T_{\rm dust})$ is the Planck function for $T_{\rm dust}$ at $\nu_{\rm 1.3mm}$, and $\NHtwo^{\rm bckg}$ is the background column density, filtered by ALMA observations. 

The dust temperature image is taken from Paper~III \citep{pouteau2022}, with values ranging from 20~K to 65~K (see \cref{appendixfig:PPMAP with NH2}). 
It has an angular resolution of $2.5\arcsec$ over the majority of the map but of $0.46\arcsec$ toward the protostellar cores \citep[see Sect.~4.2 of][]{pouteau2022}. The frequency $\nu_{\rm 1.3mm} = 228.9$~GHz is taken from Paper~II \citep{ginsburg2022} assuming a spectral index of $3.5;$ which corresponds to a spectral index of $\beta = 1.5$ for the dust opacity and which is suitable for optically thin dense gas \citep[see][]{andre1993, juvela2015}. Because the W43-MM2\&MM3 ridge is a dense cloud \citep{nguyen2013}, we adopted a dust opacity per unit (gas~$+$~dust) mass at 1.3~mm and adapted for cold cloud structures: $\kappa_{\rm 1.3mm}=0.01\,\rm cm^2\,g^{-1}$ \citep{OssenkopfHenning1994}.
This dust opacity corresponds to that of dust grains that have developed thick ice mantles during a cold phase preceding the strong (external) heating of the W43-MM2\&MM3 ridge by the OB and Wolf-Rayet stellar cluster \citep{blum1999}. This value should be adapted even for the Outskirts region, whose 1.3~mm flux restored by our interferometric image is constituted, for a large fraction, of the sum of core fluxes (see \cref{tab:subregions chara}).

Since, due to interferometer filtering, scales larger than $\sim$5.6$\arcsec$ are missing (see Sect.~\ref{sect:obs and catalog}), we added a $\NHtwo^{\rm bckg}\sim 0.7\times10^{23}~\rm cm^{-2}$ level measured at the periphery of the cloud ridge on the \textit{Herschel} column density image\footnote{
    The \textit{Herschel} column density image used here is taken from the archival data products of the HOBYS key program \citep{motte2010}. See \url{https://www.hobys.org/data.html}.
    }, 
with $25\arcsec$ resolution, by \cite{nguyen2013}.
Figure~\ref{appendixfig:pdf W43 herschel}a compares the $\eta$-PDF (see definition in Sect.~\ref{sect:PDFs}) of the column density images derived from the present ALMA data (see above) and these \textit{Herschel} data. 
It shows that a value of $\NHtwo^{\rm bckg}\sim 0.7\times10^{23}~\rm cm^{-2}$ indeed allows these $\eta$-PDF functions to be consistent with each other. This first-order correction probably overestimates, by up to $\sim$$0.15\times10^{23}~\rm cm^{-2}$, the column density values in low-density subregions such as Outskirts and MM51, and underestimates it, by up to $\sim$$0.3\times10^{23}~\rm cm^{-2}$, in centrally concentrated subregions such as MM2 and MM3. 
We therefore assumed a different background column density for each subregion: $\NHtwo^{\rm bckg} = (0.55\pm 0.10) \times10^{23}~\rm cm^{-2}$ for Outskirts and MM51, $\NHtwo^{\rm bckg} = (0.70\pm 0.15) \times10^{23}~\rm cm^{-2}$ for MM10 and MM12, and $\NHtwo^{\rm bckg} = (1.0\pm 0.2) \times10^{23}~\rm cm^{-2}$ for MM2 and MM3. Future studies will be performed on maps combining present ALMA 12~m array data with ALMA-IMF data of the 7~m array and total power antennas. Their reduction and combination is not simple but the ALMA-IMF consortium is working on it \citep{ginsburg2022,cunningham2023}.
Figure~\ref{fig:nh2 and cores} presents the resulting column density image of the W43-MM2\&MM3 ridge, with $\NHtwo$ values varying from $\sim$1$\times10^{22}~\rm cm^{-2}$, in region where missing large-scale emission creates interferometric artifacts consisting of negative bowls, to $\sim$2.6$\times10^{25}~\rm cm^{-2}$ toward the peak of MM2.
This column density image is consistent with the one of Dell'Ova et al. (in prep.), which was produced by the point process mapping of the temperature (PPMAP) procedure \citep{marsh2015} and achieves a $2.5\arcsec$ resolution. Except for the column density range of each subregion, measurements performed in Tables~\ref{tab:subregions chara}--\ref{tab:cmf and s} do not depend on the exact value chosen for the background column density.

Subregion masses, $M^{\rm subregion}$, are computed from the 1.3~mm peak fluxes and dust temperatures measured for each pixel $i$ within the subregion area $A$, $\left(S^{\rm peak}_{\rm 1.3\,mm} \right)_{\rm i}$ and $(T_{\rm dust})_{i}$ read in images provided by  Paper~III \citep[][see also \cref{appendixfig:PPMAP with NH2}]{pouteau2022}. We used the following equation:
\begin{equation}
    \begin{split}
    M^{\rm subregion} \; [\Msol] \: &= \sum \limits_{\rm pixel\, i}\limits^{A} \frac{ \left(S^{\rm peak}_{\rm 1.3\,mm} \right)_{\rm i} \; d^2}{ \kappa_{\rm 1.3\,mm}\; B_{\rm 1.3\,mm}\left[(T_{\rm dust})_{i}\right]} \times \frac{\Omega_{\rm pixel}}{\Omega_{\rm beam}},
    \end{split}
\label{eq:optically thin mass}
\end{equation}
where $d$ is the $5.5$~kpc distance of W43 \citep{zhang2014} and $\Omega_{\rm pixel}=(0.1\arcsec)^2$ is the pixel area of our images. Subregion masses, listed in \cref{tab:subregions chara}, range from $\sim$$35~\Msol$ to $\sim$$1\,300~\Msol$. We assumed that the ALMA-IMF configurations filter out the emission of the cloud surrounding each subregion. This assumption is correct for the centrally concentrated subregions MM2, MM3, MM10, and MM12, but totally incorrect for the MM51 and Outskirts subregions as well as the total imaged area.

Volume densities of subregions are computed assuming a spherical geometry and a line-of-sight radius equal to the equivalent plane-of-the-sky radius of its defined area, $A$: 
\begin{equation}
    \begin{split}
    n_{\rm H_2} \; [{\rm cm}^{-3}] &= \frac{1}{\mu\;m_{\rm H}} \times \frac{M^{\rm subregion}}{\frac{4}{3}\pi \; \left(\frac{A}{\pi}\right)^{3/2}} \\
    &\simeq 2.1\times 10^5\,{\rm cm}^{-3} \times \left( \frac{M^{\rm subregion}}{10^3~\Msol}\right) \left( \frac{A}{1~{\rm pc}^2} \right)^{-3/2}.
    \end{split}
    \label{eq:density}
\end{equation}
The volume densities for MM10, MM51, and Outskirts are uncertain because the spherical assumption is more questionable for these subregions, and even more important for Outskirts, because its size along the line of sight is poorly approximated by the equivalent radius of its defined area. 
Uncertainties given for $M^{\rm subregion}$ and $n_{\rm H_2}$ take into account uncertainties in the definition of subregion outlines, which constitute the relative uncertainties from one subregion to another. We estimate that the absolute uncertainties in column densities, subregion masses, and volume densities arise primarily from our poor knowledge of the absolute value of the dust opacity per unit mass and are a factor of about two.

In summary, the W43-MM2\&MM3 ridge consists of six subregions with a large diversity of environments. They have maximum column densities, masses, and volume densities covering more than one order of magnitude: $6-110 \times 10^{23}$~cm$^{-2}$, $35-1\,300~\Msol$, and $0.07-500 \times 10^{3}$~cm$^{-3}$ (see \cref{tab:subregions chara}).

\section{Analysis} \label{sect:analysis}

Paper~III \citep{pouteau2022} studied the CMF of the W43-MM2\&MM3 mini-starburst ridge and revealed it is top-heavy with respect to the Salpeter slope of the canonical IMF \citep{salpeter1955, kroupa2002}. We here investigate variations throughout the W43-MM2\&MM3 subregions of the core properties and cloud density structure (see Sects.~\ref{sect:spatial variations of core pop}--\ref{sect:mass segregation}), with the aim of searching for correlations between them in Sect.~\ref{sect:discussion}.
\cref{tab:cmf and s} provides the parameters (fit ranges and power-law indices) characterizing the CMF and $\eta$-PDF of the W43-MM2\&MM3 subregions studied in Sects.~\ref{sect:spatial variations of core pop}--\ref{sect:PDFs}. It also gives measurements of the protostellar fraction of their core population and an estimation of their evolutionary stage, performed in Sects.~\ref{sect:spatial variations of core pop} and \ref{sect:link with sf history}.

\subsection{Spatial variations in the core populations}
\label{sect:spatial variations of core pop}

We have taken advantage of the good statistics allowed by the large core catalog of Paper~III \citep{pouteau2022} to examine whether the core mass range of one subregion is statistically differentiable from another (see Sect.~\ref{sect:core populations}) and whether the CMF is top-heavy everywhere in the W43-MM2\&MM3 ridge (see Sect.~\ref{sect:CMFs}).

\subsubsection{Subregion core populations} \label{sect:core populations}

We separated the 205 cores of W43-MM2\&MM3 into six core populations associated with the subregions defined in Sect.~\ref{sect:subregions def}. 
The numbers of cores hosted in the subregions range from $19\pm1$ for MM12 to $59\pm11$ for MM2, the most massive, the densest, and therefore also the most populated subregion of W43-MM2\&MM3 (see \cref{tab:subregions chara}).
Cores are not evenly spatially distributed, with surface number densities varying from 5~pc$^{-2}$ to 74~pc$^{-2}$. As for the concentration of the subregion mass within cores, it is similar within the subregions, $\sum \limits_{\rm subregion}M^{\rm core}/ M^{\rm subregion}\sim 18\%$, in agreement with the values measured for ALMA-IMF clouds qualified as Young (see \cref{tab:subregions chara} and Table~5 of \citealt{motte2022}).
The completeness levels of each subregion catalog were estimated from the core extraction simulations of Paper~III \citep[see Appendix~C of][]{pouteau2022}. They vary from $\sim$0.45$\pm 0.2~\Msol$ to $\sim$1.1$\pm 0.2~\Msol$ for the Outskirts and MM2 subregions, respectively and are listed in \cref{tab:cmf and s}.

We used the protostellar versus pre-stellar core nature determined by Paper~V \citep{nony2023} to compute the fraction of protostellar cores, $f_{\rm proto}$, in the six W43-MM2\&MM3 subregions:
\begin{equation}
    f_{\rm proto}= N_{\rm proto}/N_{\rm core},
\end{equation} 
where $N_{\rm proto}$ and $N_{\rm core}$ are the number of protostellar cores and the total number of cores including protostellar and pre-stellar cores. On average, the protostellar fraction above the $90\%$ completeness levels of core catalogs is $f_{\rm proto} \simeq 30\%$ (see \cref{tab:cmf and s}). While the protostellar fraction is enhanced, $f_{\rm proto} \simeq 40\%$, in the MM2 and MM12 subregions, it is significantly low, $f_{\rm proto} < 10\%$, in the Outskirts.

\begin{figure}[htbp!]
  \centering
  \includegraphics[width=0.49\textwidth]{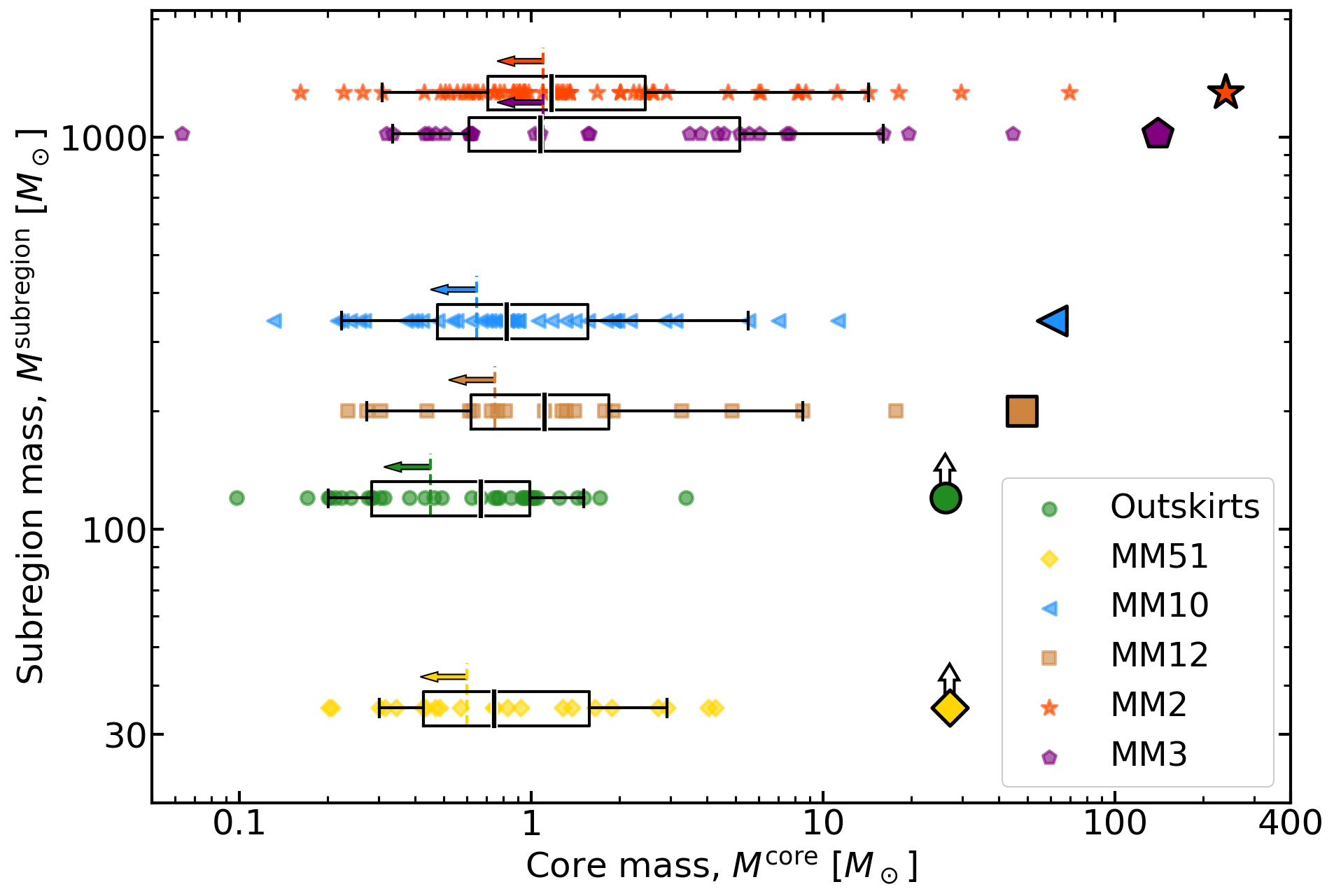}
  \caption{Distribution of the W43-MM2\&MM3 subregion masses as a function of the masses of the cores they host, taken from \cref{tab:cmf and s} and Paper~III \citep{pouteau2022}. Small markers represent the cores associated with each subregion; larger markers correspond to the cumulative mass of their cores. For each subregion, the box plot presents the first and third quartiles and whiskers indicate the 5\% and 95\% limits of the core sample. Colored arrows and vertical dashed lines represent the completeness limit for each subregion.}
  \label{fig:subregion mass vs core mass}
\end{figure}

{\renewcommand{\arraystretch}{1.5}%
\begin{table*}[htbp!]
\centering
\resizebox{\textwidth}{!}{
\begin{threeparttable}[c]
\caption{Characterization of the CMF and $\eta$-PDF, and evaluation of the evolutionary stage of the parsec-scale subregions of W43-MM2\&MM3.}
\label{tab:cmf and s}
\begin{tabular}{c|cc|ccccc|cc}
    \hline\hline
    Subregion & Core mass & \multirow{2}{*}{$\alpha$} & $\NHtwo^{\rm bckg}$ & Fitted & \multirow{2}{*}{$s_2$} & Fitted & \multirow{2}{*}{$s_3$} & Protostellar & Evolutionary \\
    name      &      range  &                         &        offset & range ($s_2$) &               & range ($s_3$) &              &   fraction & stage      \\
              &    [\Msol]  &                         & [\NHtwoUnits] & [\NHtwoUnits] &               & [\NHtwoUnits] &              &       [\%] &            \\
    (1)       &     (2)     &           (3)           &       (4)     &     ( 5)      &      (6)      &      (7)      &      (8)     &     (9)    &    (10)    \\ \hline
    \multicolumn{4}{l}{{\bf CMF high-mass ends close to the Salpeter slope or steeper}} &             &               &              &            &            \\
    Outskirts &  $0.45-3.4$ & $-1.54_{-0.42}^{+0.32}$ &     $55\pm10$ &      100--550 &         $>-5$ &            -- &           -- &  $7\%\pm1$ & Quiescent  \\
    MM51      &   $0.6-4.3$ & $-1.17_{-0.35}^{+0.18}$ &     $55\pm10$ &      100--250 &         $>-5$ &      250--900 & $-0.8\pm0.2$ & $25\%\pm5$ & Quiescent  \\ 
    MM10      & $0.65-11.2$ & $-1.16_{-0.30}^{+0.16}$ &     $70\pm15$ &      100--300 &  $-2.4\pm0.2$ &     300--3000 & $-0.9\pm0.3$ & $21\%\pm2$ & Pre-burst? \\ \hline
    \multicolumn{2}{l}{{\bf Top-heavy CMFs}} &        &               &               &               &               &              &            &            \\
    MM12      & $0.75-17.8$ & $-0.95_{-0.36}^{+0.15}$ &     $70\pm15$ &      100--500 &  $-2.1\pm0.3$ &     500--2200 & $-0.1\pm0.2$ & $41\%\pm5$ & Main-burst?\\
    MM2       &  $1.1-69.9$ & $-0.93_{-0.21}^{+0.11}$ &    $100\pm20$ &     150--1000 &  $-0.6\pm0.2$ & 1000--26\,000 & $-0.2\pm0.3$ & $37\%\pm2$ & Main burst \\
    MM3       &  $1.1-44.6$ & $-0.59_{-0.12}^{+0.07}$ &    $100\pm20$ &     150--1200 &  $-0.8\pm0.2$ &    1200--8000 & $-0.7\pm0.3$ & $22\%\pm3$ & Post-burst \\ \hline
    Total     &  $0.8-69.9$ & $-0.93_{-0.10}^{+0.07}$ &     $70\pm15$ &     100--1000 &  $-0.9\pm0.3$ & 1000--26\,000 & $-0.3\pm0.2$ & $27\%\pm3$ & --         \\ \hline
    \end{tabular}
\begin{tablenotes}
%
\item (2) Mass range used to fit a power law to the subregion CMF. The lower limit of this mass range is the 90\% completeness limit of each subregion, with an error of $\pm 0.2~\Msol$ \citep[see][]{pouteau2022}. Its upper limit corresponds to the maximum mass of cores detected in the subregion.
\item (3) Power-law index of the high-mass end of the subregion CMF in its cumulative form, $N(> \log M) \propto M^\alpha$, measured over the mass range of col.~2. Uncertainties are estimated by taking into account the fit uncertainties associated with the small-sample statistics, the completeness limit uncertainty of col.~2, and by varying core masses according to flux, dust temperature, and emissivity uncertainties (see Sect.~\ref{sect:CMFs}).
\item (4) Column density offset assumed to correspond to the background of each subregion and used in \cref{eq:nh2 region} to compute their PDF (see Sect.~\ref{sect:subregions coldens and mass}).
\item (5) and (7) Ranges of column density used to fit two consecutive power laws to the subregion $\eta$-PDF tail. The limits of these ranges are defined in Sect.~\ref{sect:PDFs}.
\item (6) and (8) Power-law indices of second an third tails of the subregion $\eta$-PDF, $p_\eta \propto (\NHtwo)^s$ (see Sect.~\ref{sect:PDFs}), measured over the column density ranges of cols.~5 and 7, respectively. Uncertainties are estimated by varying the number of bins, the limits of the fitted ranges, and the column density offset of col.~4.
\item (9) Protostellar fraction of the core catalog above $0.8~\Msol$, as derived from the pre-stellar versus protostellar nature of cores determined by \cite{nony2023}.
\item (10) Subregion evolutionary stage defined in Sect.~\ref{sect:link with sf history}, based on the ratio of pre-stellar versus protostellar cores (col.~9), the surface number density of cores (col.~8 of \cref{tab:subregions chara}), and the presence of \hii regions.
\end{tablenotes}
\end{threeparttable}
}
\end{table*}}

Figure~\ref{fig:subregion mass vs core mass} displays, for the W43-MM2\&MM3 subregions, their mass as a function of the individual and integrated masses of the cores they host. A correlation trend appears between the total mass of the subregions and the integrated mass of the cores within them, in agreement with the almost constant concentration of the subregion mass into cores discussed above.
Moreover, the most massive regions, MM2 and MM3, host the most massive cores (see \cref{fig:subregion mass vs core mass}). A similar correlation was observed by \cite{motte2022} for cores in the ALMA-IMF clouds, and also by \cite{lin2019} for structures mostly on larger scales than cores. In that respect, the MM10 subregion is an exception among the four subregions more massive than $200~\Msol$ because its most massive core is only $\sim$11.5~$\Msol$ and its median mass is 1.5 times lower (see \cref{fig:subregion mass vs core mass} and \cref{tab:cmf and s}). At the opposite side of the core mass range, the population of low-mass cores with respect to intermediate-mass cores appears to be under-numerous in the MM3 subregion compared with other W43-MM2\&MM3 subregions. We ran a two-sample Kolmogorov-Smirnov (KS) test to assess the likelihood that two distributions are drawn from the same parent sample (null hypothesis). In the case of the MM3 subregion, the KS test provides significant evidence that its core sample is drawn from a different population than the complementary core catalog consisting of all cores of the MM2, MM10, MM12, MM51, and Outskirts subregions (with a KS statistic of 0.3 and a p-value of $p = 0.014$).

We have therefore observed correlation trends of the mass of the subregion with the integrated mass of the cores they host as well as with the mass of their most massive core. However, the small number of W43-MM2\&MM3 subregions and the large uncertainties in the masses prevent us from making a reliable estimate of the significance and strength of these correlations. Similar analysis performed for the 15 clouds observed by the ALMA-IMF Large Program should allow us to analyze these potential relationships and their implications.

\begin{figure*}[htbp!]
\centering
\begin{minipage}{0.48\textwidth}
  \centering
  \includegraphics[width=1.\textwidth]{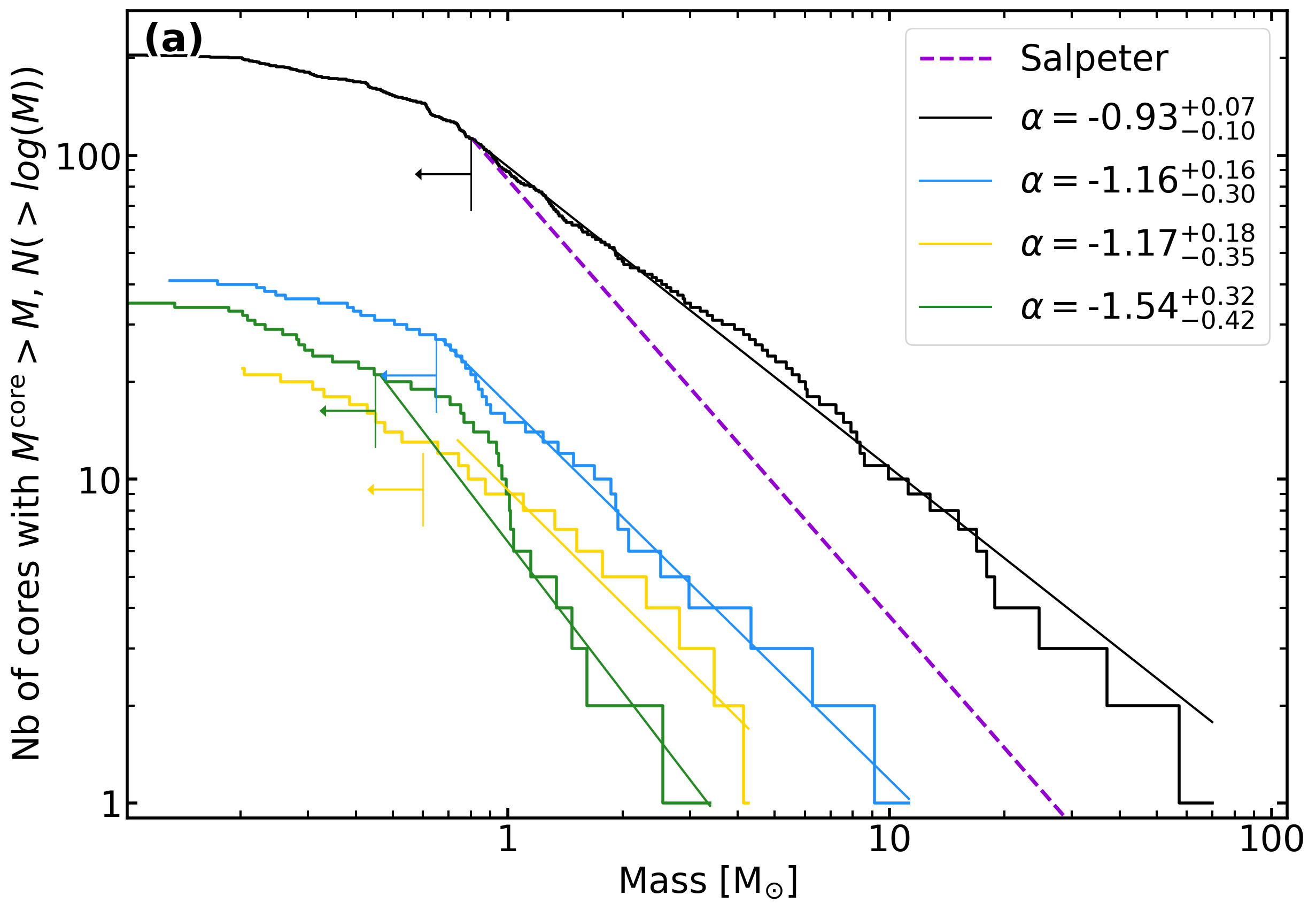}
\end{minipage}%
\hskip 0.0199\textwidth
\begin{minipage}{0.48\textwidth}
  \centering
  \includegraphics[width=1.\textwidth]{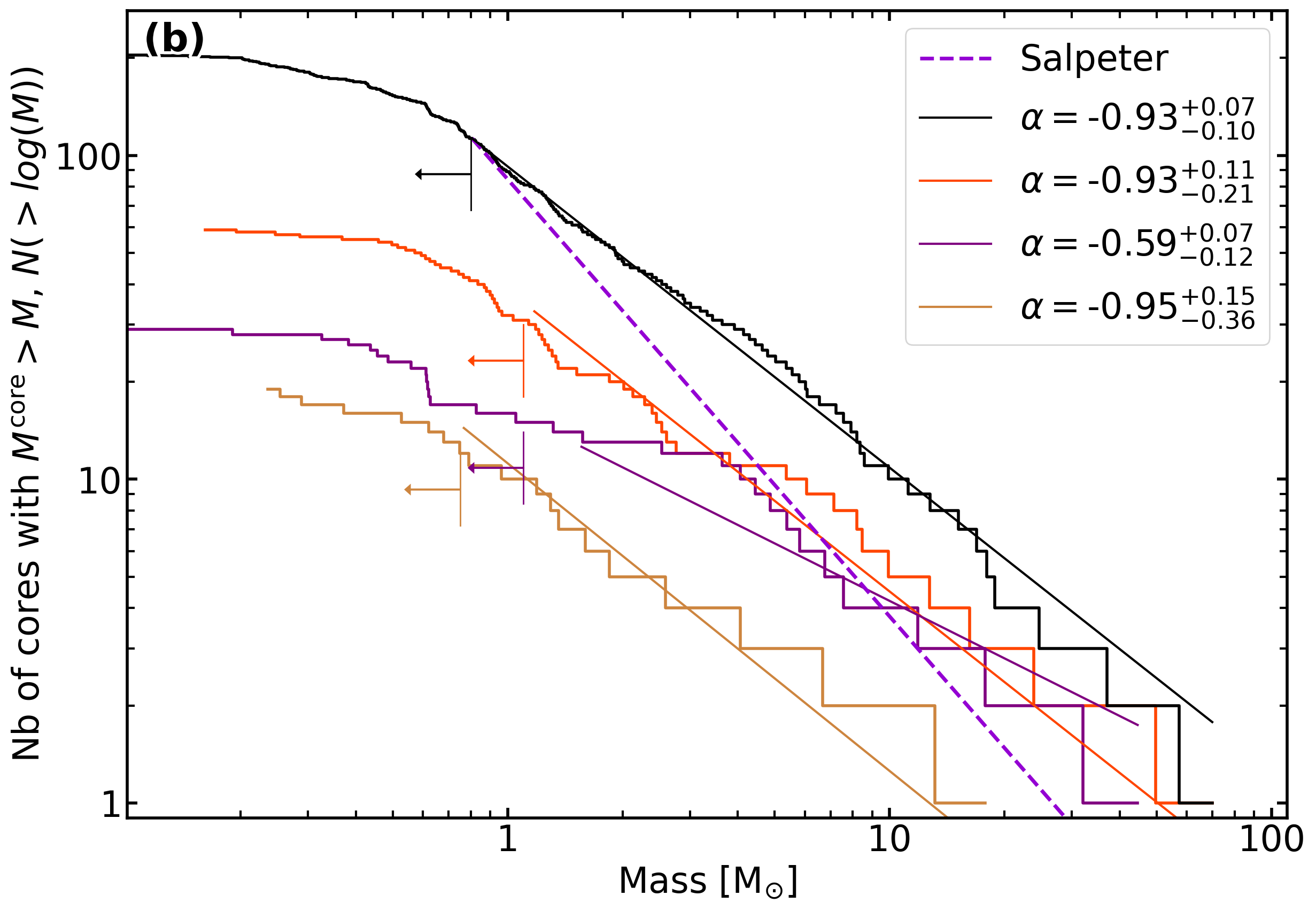}
\end{minipage}
\vskip 0.1cm
\caption{CMFs of the W43-MM2\&MM3 subregions, plotted in their cumulative form, with shapes at the high-mass end varying from steeper than or close to the Salpeter slope of the canonical IMF (MM10, MM51, and Outskirts, \textit{panel a}) to top-heavy (MM2, MM3, and MM12, \textit{panel b}). CMFs (colored histograms) are fitted by single power laws of the form $N(>{\rm log}(M))\propto M^{\alpha}$ (lines), above the associated 90\% completeness limits (left arrows and vertical lines; see \cref{tab:cmf and s}) and using a bootstrapping method that uses a MLE method (see Sect.~\ref{sect:CMFs}). The high-mass end of the canonical IMF, which has a power-law index of $\alpha_{\rm IMF} = -1.35$ \citep[dashed magenta lines;][]{salpeter1955}, is shown for comparison. The top-heavy CMF of the W43-MM2\&MM3 ridge (black histogram) mostly originates from that of the MM2 subregion. The MM3 CMF is complex, being both top-heavy and bottom-light.}
\label{fig:various cmfs}
\end{figure*}

\begin{figure*}[htbp!]
    \centering
    \includegraphics[width=1.\textwidth]{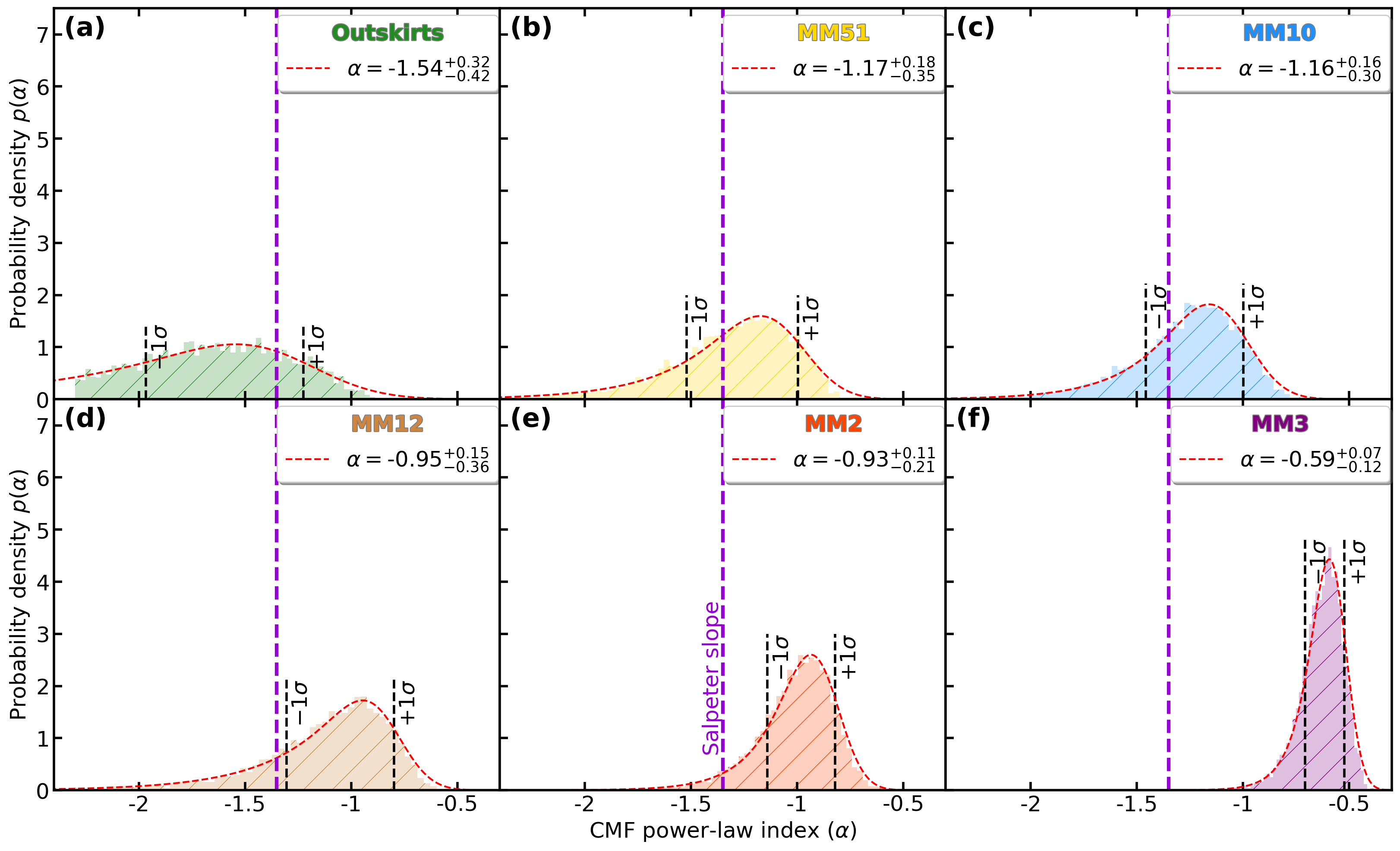}
    \caption{Determination of the index and uncertainty of the power law fitted to the high-mass end of the cumulative CMF observed for the six W43-MM2\&MM3 subregions (\textit{panels~a--f}). Each bootstrapping probability density function (colored and hatched histogram) is built from 5000 slopes fitted by a MLE method for data sets generated from a random draw with discount of the subregion core sample. This bootstrap procedure quantifies the effect of small-number statistics on $\alpha$ and its uncertainty and also includes uncertainties on the core mass and catalog completeness level (see Sect.~\ref{sect:core populations}). The dashed red curves represent the EMGs with negative skewness, which are fitted to the histograms. Power-law indices, $\alpha$, are taken to be values at the EMG peaks; their asymmetric uncertainties are estimated from the $-1\sigma$ and $+1\sigma$ standard deviations of the EMG (dashed vertical segments). The high-mass end of the canonical IMF, which has a power-law index of $\alpha_{\rm IMF}= -1.35$ \citep[dashed magenta lines;][]{salpeter1955}, is shown for comparison.}
    \label{fig:BS subreg}
\end{figure*}

\subsubsection{Varying subregion CMFs}\label{sect:CMFs}

Figure~\ref{fig:various cmfs} presents the CMFs of the six W43-MM2\&MM3 subregions, in comparison to that of the total imaged area covering the main part of the ridge, where cores are expected to form. The latter was published by Paper~III \citep{pouteau2022} and referred to as the CMF of the W43-MM2\&MM3 ridge. As shown in \cref{fig:various cmfs}b, the ridge CMF is primarily from the MM2 subregion, which accounts for $\sim$45\% of the total subregion mass and of the total mass into cores (see \cref{tab:subregions chara}). The second largest contributor to the W43-MM2\&MM3 CMF is the MM3 subregion, with 35\% of the total subregion mass and 25\% of the mass into cores (see \cref{tab:subregions chara}). 

The CMFs of all six subregions are presented in their complementary cumulative distribution form (cumulative form, for short) and fitted by single power laws of the form $N(> \log M) \propto M^\alpha$, from their 90\% completeness limits to the mass of their most massive core (see \cref{tab:cmf and s}). To do so, we used the maximum likelihood estimation (MLE) method of \cite{clauset2009} and \cite{alstott2014}, which is based on the KS metric and dedicated to probability laws fitted by power laws. 
Given the small-number statistics for the core sample of each W43-MM2\&MM3 subregion, we did not directly apply the \cite{alstott2014} method to their CMF but used it in a bootstrap procedure that simultaneously estimates the most likely CMF power-law index and its uncertainty. In detail, for each subregion, we computed the bootstrapping probability density function of 5000 slopes of CMF high-mass ends, which are fitted by power laws using the \cite{alstott2014} method. These synthetic CMFs are built from core samples generated from a random draw with discount of the observed core sample of each W43-MM2\&MM3 subregion.
To account for uncertainties associated with the core mass estimates, the mass of each randomly drawn core is allowed to vary in a $2\sigma$ Gaussian range of $[M_{\rm min}-M_{\rm max}]$. Here, $M_{\rm max}$ and $M_{\rm min}$ are the maximum and minimum masses, respectively, of each core as computed from its measured flux, estimated temperature, and dust opacity, plus or minus their associated $1\,\sigma$ uncertainties (see Tables~E.1--E.2 and Sect.~5.1 of \citealt{pouteau2022}). 
To account for the uncertainty associated with the sample incompleteness, we used the ability of the \cite{alstott2014} method to fit the initial point of the power-law fit. We allowed the fit of the initial point of the power-law fit to uniformly vary from the 90\% completeness level, plus or minus its uncertainty, $\pm 0.2~\Msol$ (see \cref{tab:cmf and s}). Given that most cores are intermediate-mass and pre-stellar (with $T_{\textrm{dust}}\simeq 23\pm3$~K; see \cref{appendixfig:PPMAP with NH2}) taking a constant dust temperature for all cores, including the most massive ones associated with hot cores, would change the power-law indices fitted to the subregion CMF by at most $18\%$.

Figure~\ref{fig:BS subreg} shows the bootstrapping  probability density functions of the six W43-MM2\&MM3 subregions. The peak of these functions locate the most likely power-law index of their CMF high-mass ends, which vary from $\alpha=-1.54$ for the Outskirts to $\alpha=-0.59$ for the MM3 subregion (see \cref{tab:cmf and s}).
To account for the asymmetry of these bootstrapping functions, their histograms are fitted by exponentially modified Gaussians (EMGs) with a negative skewness. 
For each subregion, uncertainties on the power-law index, $-\sigma$ and $+\sigma$, are estimated as the ranges of indices that are below and above, respectively, the peak of the bootstrapping probability density function and contain 68.2\% of the fitted indices. In W43-MM2\&MM3 subregions, uncertainties range from $\sigma=0.07$ to $\sigma=0.42$, with lower limits about 1.8 times larger than higher limits. The power-law indices and uncertainties derived by this bootstrapping procedure are reported in \cref{fig:various cmfs} and \cref{tab:cmf and s}. 
To test the validity of our approach, we compared the power-law index fitted by the bootstrapping procedure described above with those directly fitted by the method of \cite{alstott2014} and a linear regression on its cumulative form. 
The W43-MM2\&MM3 ridge, whose cumulative CMF was studied by Paper~III \citep{pouteau2022}, has a CMF high-mass end with a sufficiently high number statistics for this purpose. The power-law index we measured with our bootstrapping procedure, $\alpha=-0.93_{-0.10}^{+0.07}$, is consistent with those fitted by the method of \cite{alstott2014} and by linear regression, $\alpha_{\rm Alstott}=-0.96\pm0.11$ and $\alpha_{\rm LinReg}=-0.95\pm0.04$, respectively. 

According to \cref{tab:cmf and s} and \cref{fig:BS subreg}, there is a continuous flattening of the power-law indices measured for the CMF high-mass end of W43-MM2\&MM3 subregions. 
First, the Outskirts have a power-law index steeper than, but still consistent with, the slope of the Salpeter's IMF: $\alpha=-1.54_{-0.42}^{+0.32}$ compared to $\alpha_{\rm IMF}=-1.35$. Second, the MM10 and MM51 subregions have CMF high-mass ends close to the Salpeter slope, $\alpha \simeq -1.16_{-0.32}^{+0.17}$. Third, the MM2 and MM12 subregions present top-heavy CMFs with similar power-law indices, $\alpha \simeq -0.94_{-0.28}^{+0.13}$. 
And fourth, the MM3 subregion displays the flattest CMF with an irregular shape, $\alpha = -0.59_{-0.12}^{+0.07}$. 
These subregions are separated in \cref{tab:cmf and s} but they do not constitute distinct groups because of the index ranges allowed by the power-law uncertainties as defined by the bootstrap procedure overlap (see Figs.~\ref{fig:BS subreg} and \ref{appendixfig:EMG subreg}).
Since the core catalogs of the Outskirts, MM51, and MM12 subregions have small-number statistics, their CMF high-mass end is barely constrained. As for the subregions with larger statistics, MM2, MM3, and MM10, they present more robust CMFs that can be considered different, although not completely separate, in terms of the power-law index of their high-mass end (see \cref{appendixfig:EMG subreg}).

\begin{figure*}[hbtp!]
    \centering
    \includegraphics[width=1.\textwidth]{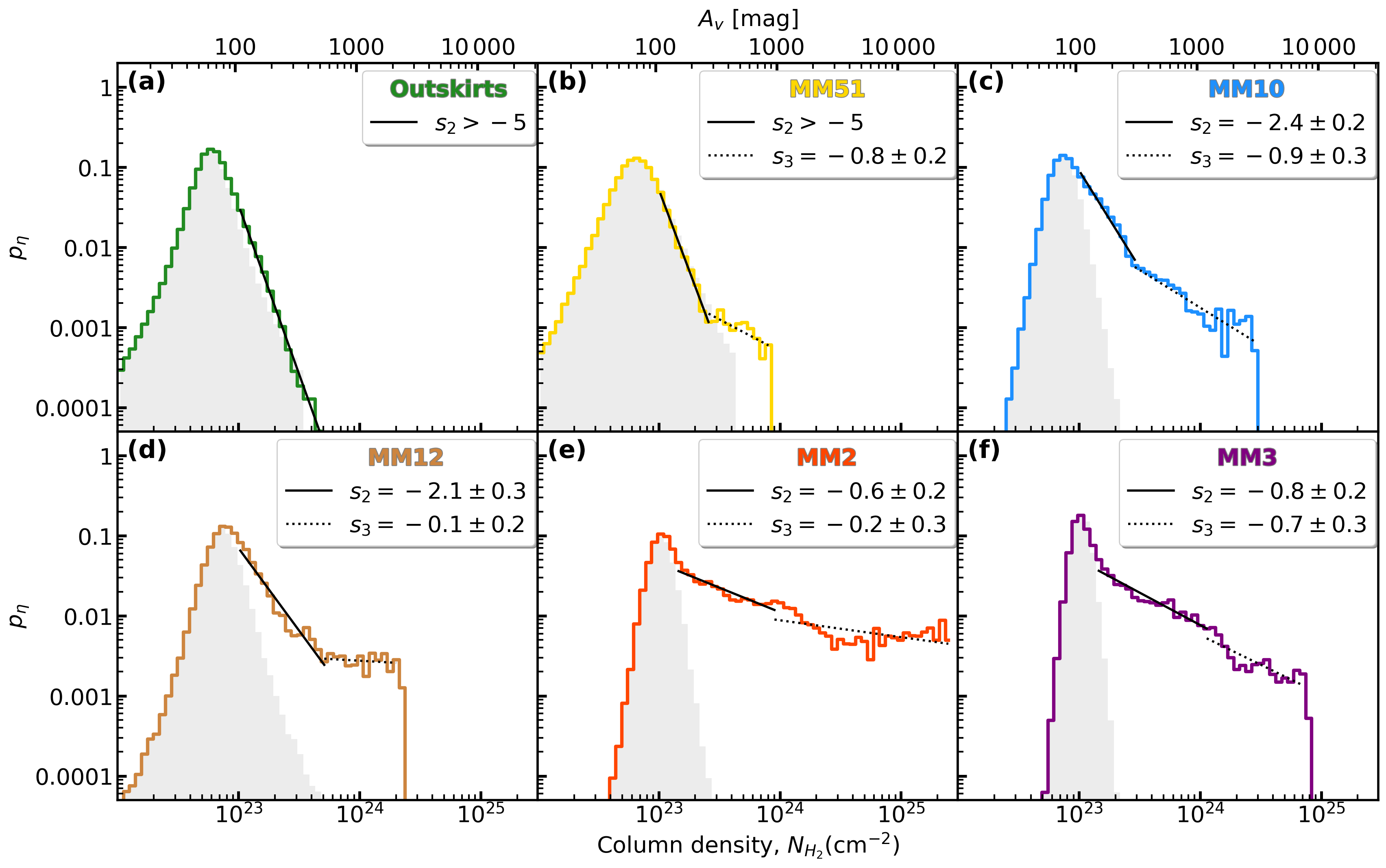}
    \caption{ALMA $\eta$-PDFs of the W43-MM2\&MM3 subregions (colored histograms), whose high-column-density tail is fitted by the two power laws $s_2$ and $s_3$ (solid and dotted black segments, respectively) above the ascending part of their histogram, which we have symmetrized with respect to the peak (gray-shaded histogram; see Sect.~\ref{sect:PDFs}). While the third tail is dominated by pixels at the cores' position, the second tail corresponds to the immediate surroundings of the cores (see \cref{appendixfig:PPMAP with NH2}). The MM2 and MM3 subregions have flatter second tails, a clearer continuity between their second and third tails, and reach much higher column densities in their third tails than the other subregions (\textit{panels~e--f}). Uncertainties on the power-law indices account for uncertainties on the binning, fit ranges, and column density offset (see Sect.~\ref{sect:PDFs}).}
    \label{fig:various eta-PDF}
\end{figure*}

\subsection{Varying column density \texorpdfstring{$\eta$}{n}-PDFs} \label{sect:PDFs}

To trace the density structure of the W43-MM2\&MM3 molecular cloud and its variations within subregions, we used PDFs applied to column density images. Column densities are typically estimated from maps of the dust continuum emission \citep[e.g.,][]{schneider2012, stutzKainulainen2015} but some studies also used dust extinction \citep[e.g.,][]{kainulainen2009,alves2014} and even molecular line emission \citep{carlhoff2013, schneider2016}. In the present study, we used three
column density images. The first column density image is built from the W43-MM2\&MM3 mosaic observed with the ALMA 12~m array (see Sect.~\ref{sect:subregions coldens and mass}), and the second one from the background image of its cores as defined by \textsl{getsf} \citep{men2021getsf} and which corresponds to this same image minus the emission of the cores (see column density image and core locations in \cref{fig:nh2 and cores}). 
The third column density image, provided by the HOBYS\footnote{https://www.hobys.org/} consortium, is computed from \textit{Herschel} images of the W43-Main cloud \citep[see][]{nguyen2013}. While the two column density images derived from ALMA data cannot constrain the low and intermediate column density emission of the W43-MM2\&MM3 mini-starburst, the \textit{Herschel} column density image does not have sufficient angular resolution to trace very high column densities.
For each subregion, we analyzed the PDF of the ALMA image and used the PDF of the background image of its cores to define the column density range where cores dominate and estimate the variation in the power-law exponents of the PDF tail with and without cores.
We also used the PDF of the \textit{Herschel} image to confirm that the first tail of the W43-MM2\&MM3 PDF is not observable by the present ALMA data.

The PDFs of star-forming regions are generally described as the sum of a lognormal function at low column densities and a power-law function at higher column densities (see, e.g., \citealt{froebrichRowles2010,schneider2013, schneider2022}, and the automatic method proposed by \citealt{veltchev2019}). 
The power-law tail is characterized from the point of differentiation of the PDF from a lognormal function, observed to vary from $\sim$$1\times 10^{21}$~cm$^{-2}$ to $\sim$$40\times 10^{21}$~cm$^{-2}$ \citep[see, e.g.,][]{schneider2022}. Our ability to determine the exact value of this departure point depends strongly on whether the lognormal is correctly described in the data \citep[e.g.,][]{Ossenkopf-Okada2016, alves2017}. 
When PDFs are contaminated by low noise levels and/or background or foreground emissions from diffuse clouds, they have their lognormal function and their departure point modified but their power-law tail almost unchanged. In contrast, column density maps computed from interferometric images do not trace the low intensity, large-scale emission associated with the cloud, and as a consequence the measured departure point of the PDF tail and the absolute value of its power-law exponent are uncertain \citep[][see their Fig.~20]{Ossenkopf-Okada2016}.
In order to correct for the interferometric filtering effects at first order, we added to our ALMA 12~m array image different column density offsets simulating the varying background of the W43-MM2\&MM3 subregions (see Sect.~\ref{sect:subregions coldens and mass}). We also used the uncertainty of these offsets to investigate the variation of power-law exponents describing the PDF tail. All these elements taken together provide a good estimate of the uncertainty on the characterization of the PDF tail when using present interferometric images of W43-MM2\&MM3.

The PDF of the column density, N-PDF, is a histogram of discretized probabilities of finding gas in the column density ranges [$b_1$;$b_2$]:
\begin{equation}
    p_{\NHtwo} (\NHtwo \in [b_1;b_2]) = \int^{b_2}_{b_1} p_{\NHtwo} ~{\rm d}\NHtwo.
\end{equation}
To make the sum of all probabilities equal to one, the numbers of map pixels in the $[b_1;b_2]$ range are normalized by the total number of map pixels.
Observations generally analyze the PDF of the natural logarithm of the normalized column density, $\eta \equiv \ln{(\NHtwo / \overline{\NHtwo})}$ with $\overline{\NHtwo}$ the mean column density, and fit the PDF with power laws of exponent $s$:
\begin{equation}
    p_\eta = \NHtwo\times p_{\NHtwo} \propto (\NHtwo)^s .
\end{equation}
Figure~\ref{fig:various eta-PDF} displays the $\eta$-PDF derived from the ALMA images of the W43-MM2\&MM3 subregions. These $\eta$-PDFs exhibit a peaked distribution around the selected value of $\NHtwo^{\rm bckg}$, from $0.55\times 10^{23}$~cm$^{-2}$ to $1.0 \times 10^{23}$~cm$^{-2}$ (see \cref{tab:cmf and s}). 
At much higher column densities than the PDF peak typically derived from \textit{Herschel} images of clouds, $0.1-15\times 10^{21}$~cm$^{-2}$ \citep[see, e.g.,][]{schneider2022}, the $\eta$-PDF peaks derived from our ALMA data (see Figs.~\ref{fig:various eta-PDF}a--f) 
cannot be interpreted as the lognormal function associated with turbulent gas. Rather, they correspond to the remaining information associated with the W43-MM2\&MM3 background, which is largely filtered out in our interferometric image and lies mostly outside our small imaged area \citep[see similar PDFs in][]{lin2016}. 
This is confirmed in \cref{appendixfig:pdf W43 herschel}a by comparison with the $\eta$-PDF obtained from the \textit{Herschel} column density image of W43-Main \citep[50~pc$\times$50~pc cloud; see Fig.~2 of][]{nguyen2013}.

We used the procedure of \cite{schneider2022} and fit the \textit{Herschel} $\eta$-PDF by the sum of a lognormal function at column densities lower than $1.5\times10^{23}~\rm cm^{-2}$ and two consecutive tails (see \cref{appendixfig:pdf W43 herschel}b). The W43-Main $\eta$-PDF has a lognormal distribution that peaks at $\sim$25$\times 10^{21}$~cm$^{-2}$, consistent with but slightly higher than the column density peaks observed in other less dense clouds \citep[e.g.,][]{froebrichRowles2010,schneider2022}. 
As for the $\eta$-PDF tail of W43-Main, it is fitted by two power laws, the second of which is measured on column densities reasonably close to those traced by ALMA, $0.6-4\times10^{23}$~cm$^{-2}$ versus $1-10\times10^{23}$~cm$^{-2}$ (see \cref{appendixfig:pdf W43 herschel}b and \cref{tab:eta-pdf characteristics}).

At column densities higher than their peak, the ALMA $\eta$-PDFs of most W43-MM2\&MM3 subregions exhibit a tail with a continuous flattening that we describe by two consecutive tails and fit by power laws (see \cref{fig:various eta-PDF}). 
These two ALMA $\eta$-PDF tails are called second and third tails ($s_2$ and $s_3$) because the first $\eta$-PDF tail ($s_1$) observed with \textit{Herschel} data is filtered by the ALMA interferometer. 
We explain below the method used to define the column density ranges of these tails, whose boundaries are called departure or end points and whose associated pixels of the column density map are located between the contours in Figs.~\ref{fig:nh2 and cores}b--c and \ref{appendixfig:PPMAP with NH2}. 
We first symmetrized the ascending distribution of each ALMA $\eta$-PDF, badly fitted by a lognormal, and define the first departure point of the $\eta$-PDF tail as the first bin higher by 50\% with respect to the symmetrized function (and $100\times 10^{21}$~cm$^{-2}$ otherwise). Then, the second departure point is on the one hand considered the first inflection point observed on the $\eta$-PDF tail of the subregions MM51, MM10, and MM12 (see Figs.~\ref{fig:various eta-PDF}b--d). It actually corresponds to the maximum column density of the background image of cores in these three subregions. For MM2 and MM3 on the other hand, it is difficult to define a clear inflection point on their $\eta$-PDF tail. We therefore took as second departure point the boundary between ranges of column densities dominated by their cores and by their background image (see Figs.~\ref{fig:various eta-PDF}e--f). 
The range between the first and second departure points is the column density range used to fit a power law on what we call the second tail because it is close to that observed for the \textit{Herschel} $\eta$-PDF (see Tables \ref{tab:cmf and s} and \ref{tab:eta-pdf characteristics}). The ALMA $\eta$-PDF tail of W43-MM2\&MM3 has also a power-law exponent, which is consistent to that observed for the \textit{Herschel} $\eta$-PDF: $s_2^{\rm ALMA}=-0.9\pm0.3$ versus  $s_2^{Herschel}=-1.3\pm0.3$ (see \cref{appendixfig:pdf W43 herschel}b).
Finally, the range between the second departure point and, as an end point, the maximum column density detected in subregions (see \cref{tab:subregions chara}) is used to fit a power law on what we call the third tail. \cref{tab:cmf and s} lists, for each subregion, the ranges of these two $\eta$-PDF tails and their fitted power-law exponents, $s_2$ and $s_3$. Uncertainties on the power-law exponents are estimated by varying the number of bins of the $\eta$-PDFs, the limits of the fit ranges, and by taking into account the uncertainty on the column density offset (see \cref{tab:cmf and s}).
The assumption on the dust temperature used to compute the column density image has a negligible effect on the $\eta$-PDFs since, with a constant temperature on all pixels, the power-law exponents of their $s_2$ and $s_3$ tails are not changed by more than $10\%$.
While the third $\eta$-PDF tail is, by definition, related to cores, the second tail is mainly composed of pixels surrounding the cores, thus excluding their contribution to the $\eta$-PDF tail (see Figs.~\ref{fig:nh2 and cores} and \ref{appendixfig:PPMAP with NH2}).

As shown in Figs.~\ref{fig:various eta-PDF}c--f, the MM10, MM12, MM2, and MM3 subregions all have second and third tails of their $\eta$-PDF that are clearly defined. As for the other two regions, MM51 has only a third tail (see \cref{fig:various eta-PDF}b) and Outskirts no tail that can be revealed with current data (see \cref{fig:various eta-PDF}a). The power-law exponent of their potential second tail must be greater than $s_2 \sim -5$ and could be the continuation of the first tail observed in the \textit{Herschel} $\eta$-PDF, $s_1\sim -2.9$ (see \cref{appendixfig:pdf W43 herschel} and \ref{tab:eta-pdf characteristics}). 
Subregions with top-heavy CMFs, MM12, MM2, and MM3, have flatter second tails and reach much higher column densities in their third tail than the MM10 subregion, whose CMF high-mass end is close to Salpeter (see \cref{fig:various eta-PDF} and \cref{tab:cmf and s}).
Moreover, there is a clearer continuity between the second and third tails of the MM2 and MM3 subregions, than between the tails of MM10, MM12, and even worse MM51.

\begin{figure*}[htbp!]
    \centering
    \includegraphics[width=0.48\textwidth]{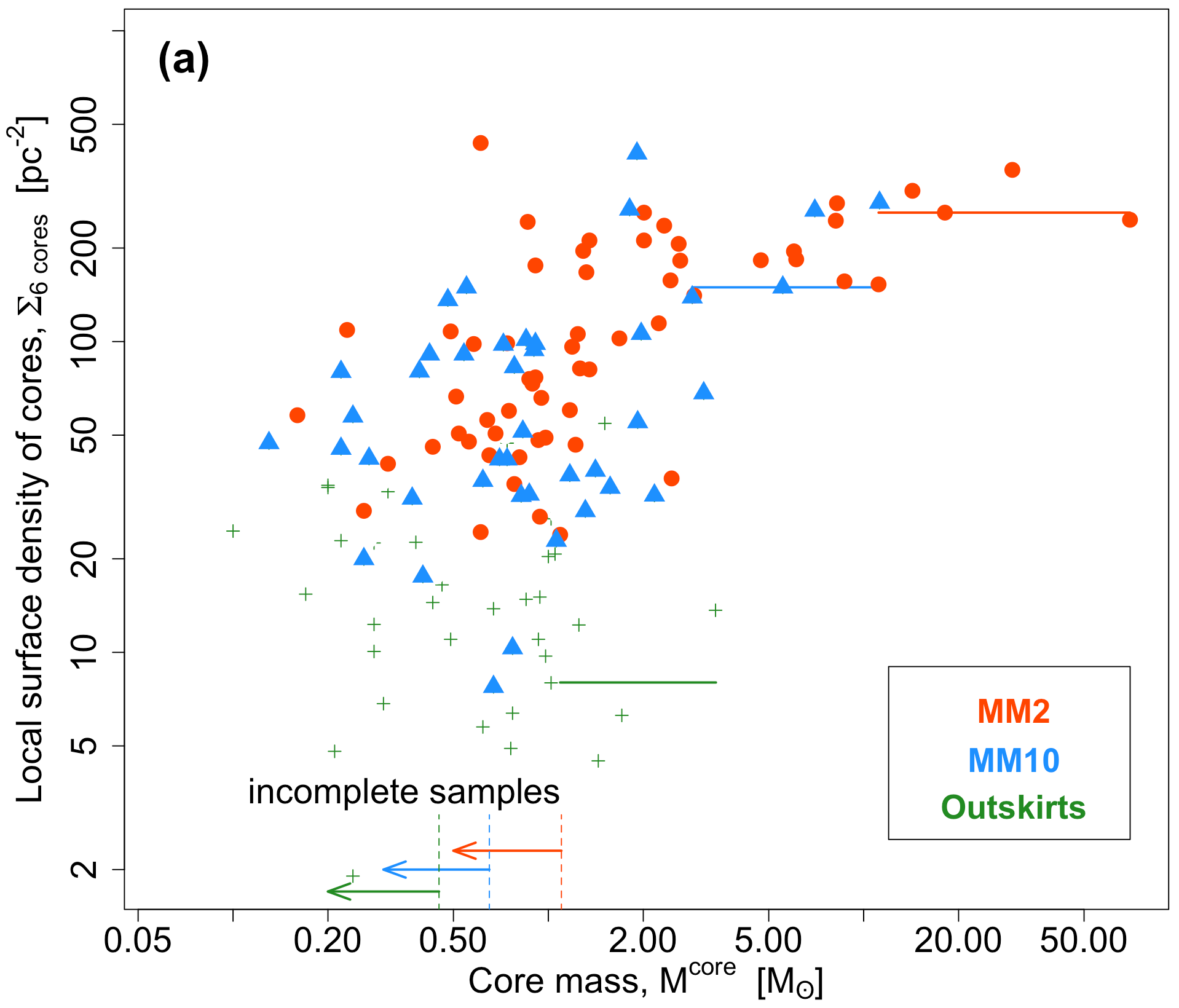} \hskip 0.5cm
    \includegraphics[width=0.48\textwidth]{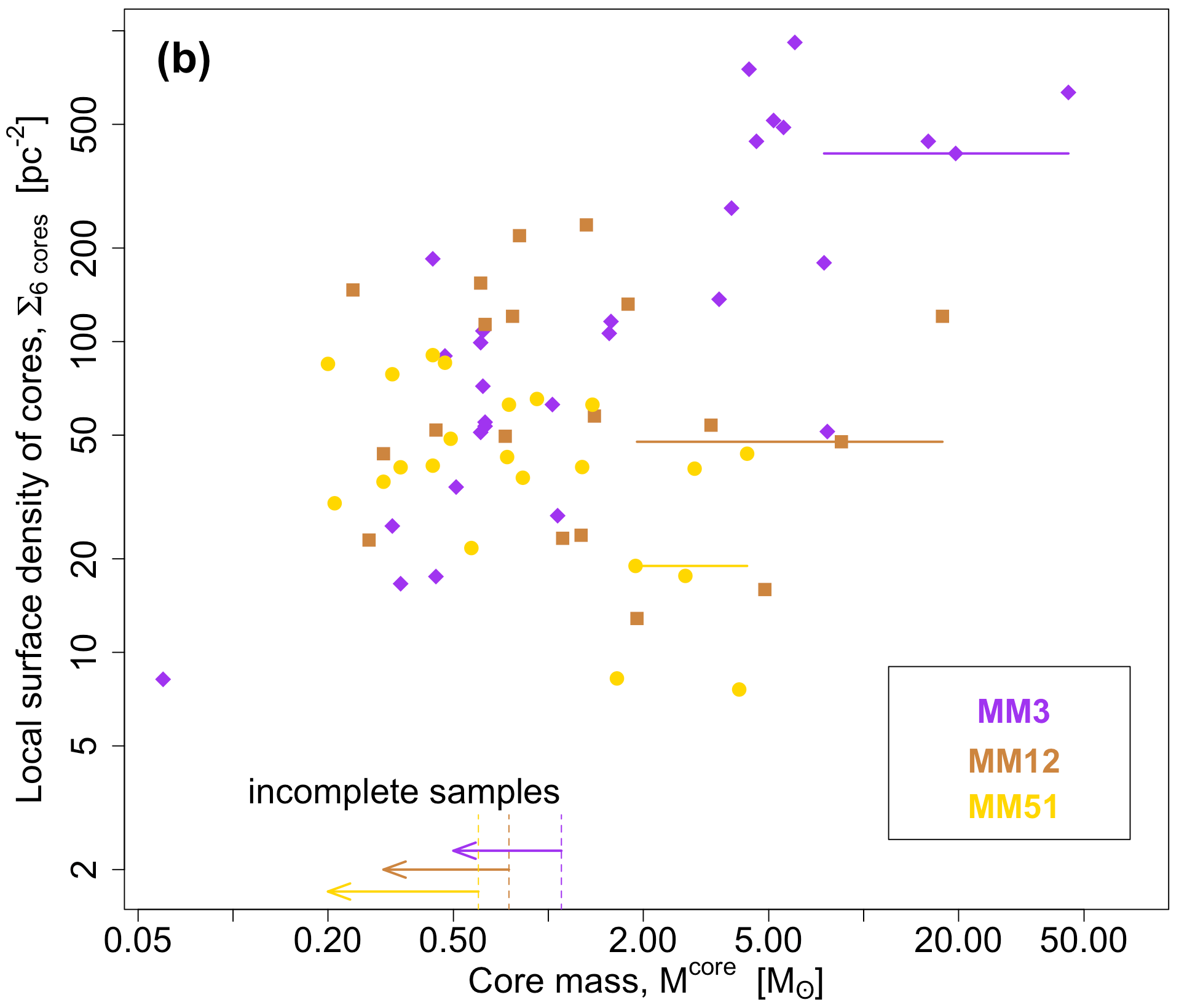}
    \caption{Local surface density of cores as a function of their core mass, for the MM2, MM10, and Outskirts subregions (\textit{panel a}) and for the MM3, MM12, and MM51 subregions (\textit{panel b}). Horizontal segments indicate the local surface densities of the five most massive cores, $\left(\Sigma_{\rm 6\,cores}\right)_{\rm massive}$, in the subregion. Left arrows and vertical dashed segments locate the $90\%$ completeness level of each core catalog (see \cref{tab:cmf and s}). The five most massive cores of MM2, MM3, and MM10 have higher core local surface densities of cores than the median densities measured for their entire core sample (see \cref{tab:mass segregation}).}
    \label{fig:sigma6}
\end{figure*}

\begin{figure*}[htbp!]
   \centering
   \begin{minipage}{0.32\textwidth}
     \centering
     \includegraphics[width=1\textwidth]{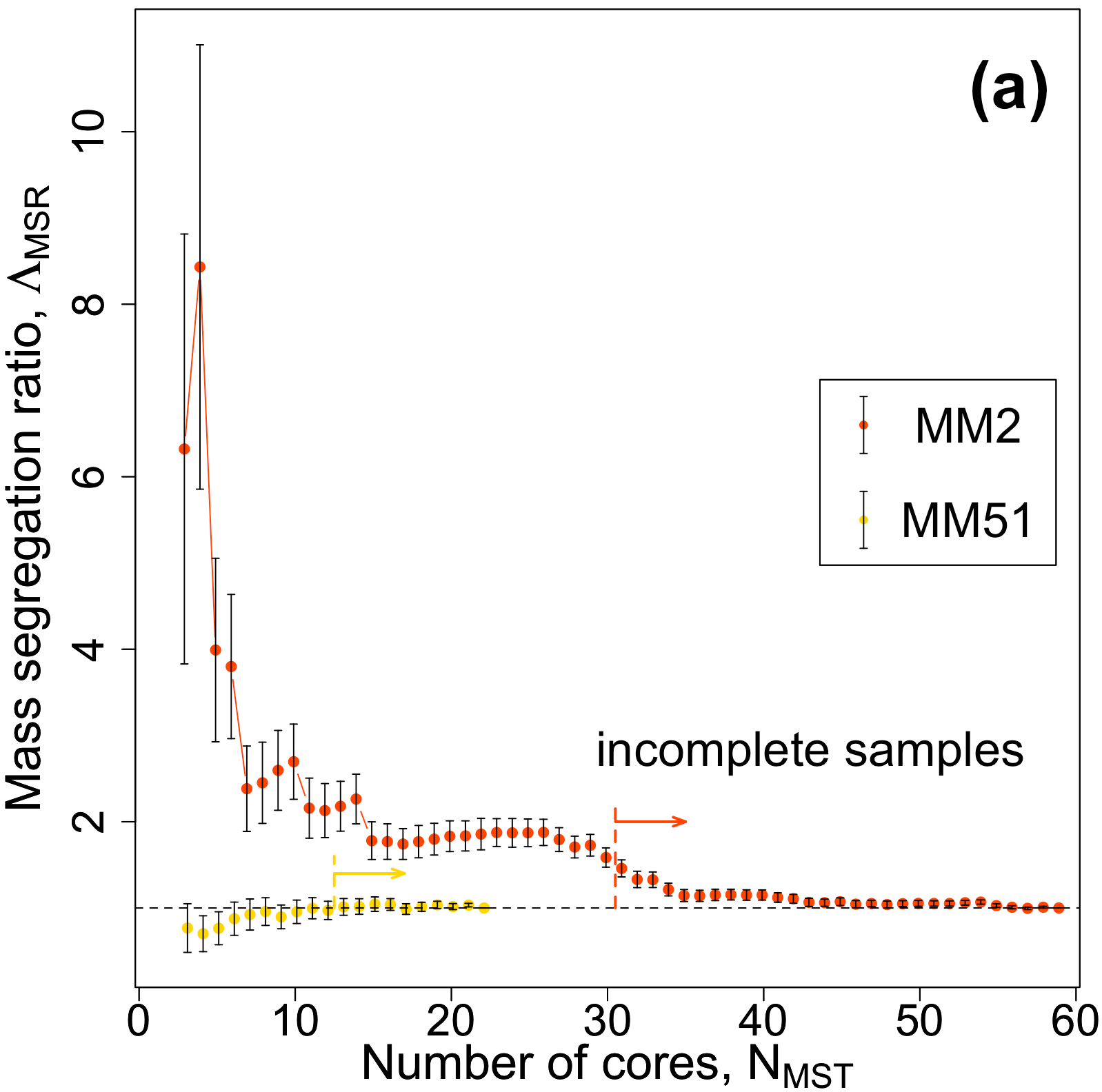}
   \end{minipage}
   \begin{minipage}{0.32\textwidth}
     \centering
     \includegraphics[width=1.\textwidth]{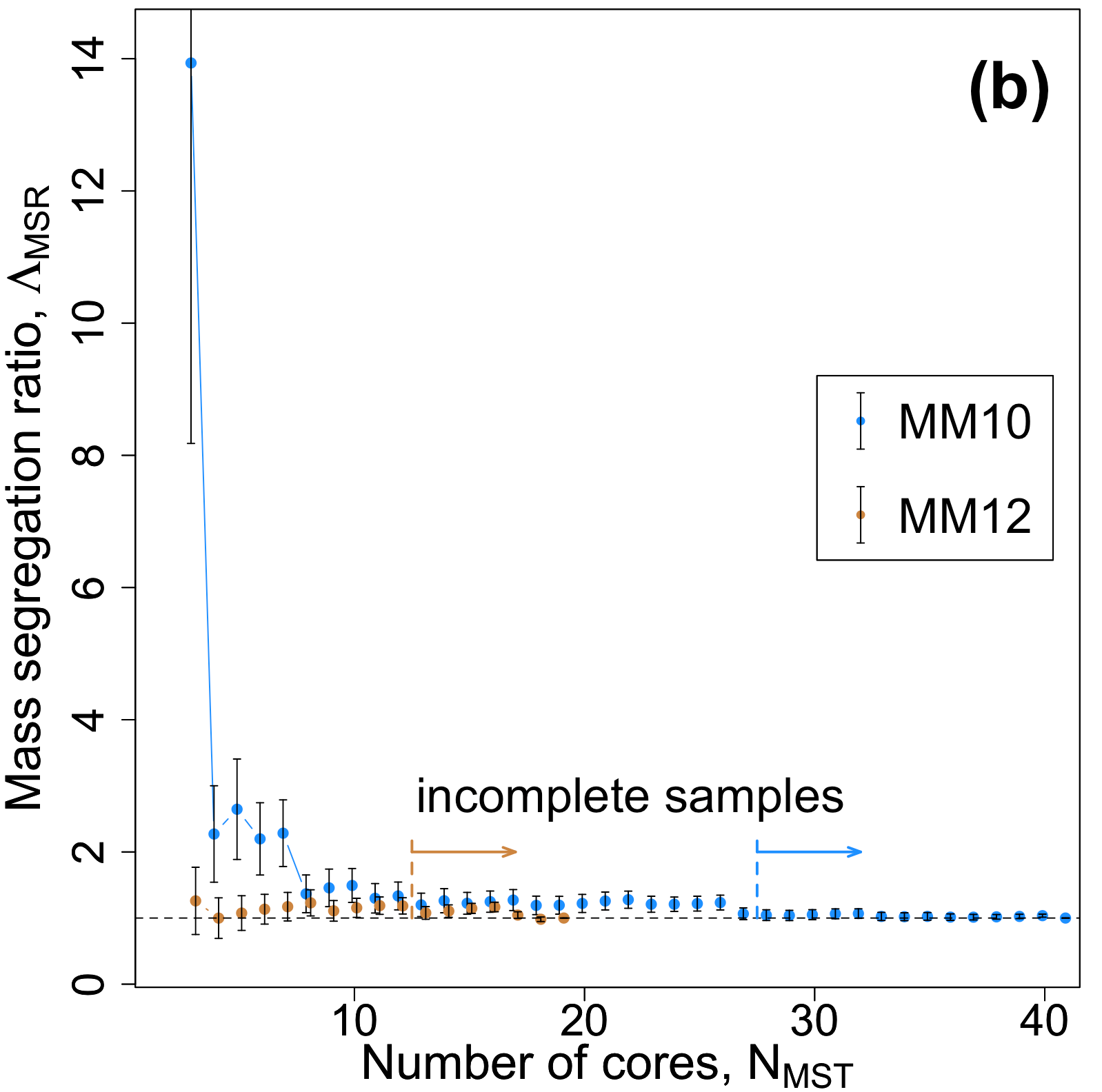}
   \end{minipage}
   \begin{minipage}{0.32\textwidth}
     \centering
     \includegraphics[width=1.\textwidth]{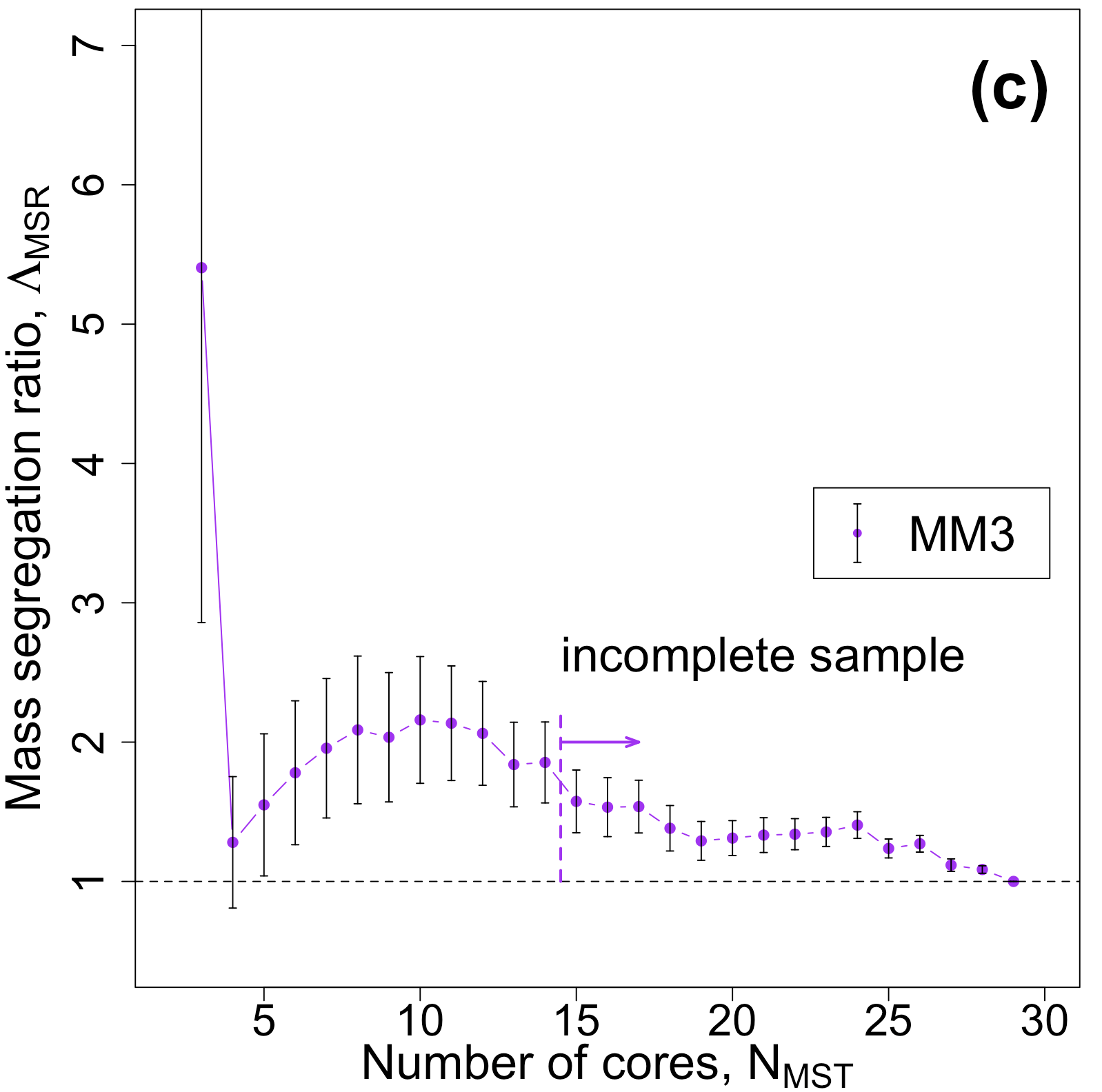}
   \end{minipage}
   \caption{Mass segregation ratio of cores in the W43-MM2\&MM3 subregions. Computed from \cref{eq:lambda msr}, they are measured in the MM2 and MM51 subregions (\textit{panel a}), in MM10 and MM12 (\textit{panel b}), and in MM3 (\textit{panel c}). Error bars are the $\pm1\sigma$ uncertainties. Right arrows and vertical dashed lines locate the $90\%$ completeness levels of the subregion catalogs of cores (see \cref{tab:cmf and s}).  Values well above $\Lambda_{\rm MSR}\simeq1$ (dashed black horizontal line) strongly suggest mass segregation in MM2, MM3, and MM10 (see \cref{tab:mass segregation}).}
\label{fig:lambda msr}
\end{figure*}

\subsection{Varying core mass segregation} \label{sect:mass segregation}

We here quantify the mass segregation of cores and its variations within the W43-MM2\&MM3 subregions, with the aim of revealing any correlation with their CMF shape in Sect.~\ref{sect:BadCorrelations}. There are several ways to measure if a set of objects is more clustered than others (see comparison of methods in \citealt{parkerGoodwin2015}).
We followed the recommendations of \cite{parkerGoodwin2015} and used two different indicators in tandem. The first one measures, for each core using its six closest neighbors, the local core surface density of a subregion, $\Sigma_{\rm 6\,cores}$ \citep{maschbergerClarke2011}. A larger $\Sigma_{\rm 6\,cores}$ value for the most massive cores of a subregion indicates a spatial concentration of cores around the most massive ones. 
In order to assess this spatial concentration around massive cores in the small-statistic samples of cores of the W43-MM2\&MM3 subregions, we measured the $\Sigma_{\rm 6\,cores}$ for the five most massive cores, $\left(\Sigma_{\rm 6\,cores}\right)_{\rm massive}$.
Comparing this value to the general core surface density of each subregion, $\overline{\Sigma_{\rm 6\,cores}}$, can give insights on the level of core mass segregation, which is considered significant for $\left(\Sigma_{\rm6\,cores}\right)_{\rm massive}/\overline{\Sigma_{\rm6\,cores}}>2$.

The second indicator we chose, the mass segregation ratio of cores, $\Lambda_{\rm MSR}$ \citep{allison2009b}, has been used for cores in several high-mass star-forming studies \citep[e.g.,][]{dib2018segreg, sanhueza2019, busquet2019, nony2021}. This indicator is based on the minimum spanning tree (MST) method and is computed with the following equation:
\begin{equation} \label{eq:lambda msr}
    \Lambda_{\rm MSR} (N_{\rm MST}) = \frac{\overline{l_{\rm random}}}{l_{\rm massive}} \pm \frac{\sigma_{\rm random}}{l_{\rm massive}},
\end{equation}
where $l_{\rm massive}$ is the MST length of the $N_{\rm MST}$ most massive cores of a subregion and $\overline{l_{\rm random}}$ is the average MST length of sets of $N_{\rm MST}$ random cores. When $\Lambda_{\rm MSR}$ is plotted against $N_{\rm MST}$, $l_{\rm massive}$ includes an increasing number of cores, which are ordered by decreasing mass. To estimate $\overline{l_{\rm random}}$ and the associated $\sigma_{\rm random}$ of sets of random cores, we used a total of 500 sets of $N_{\rm MST}$ cores uniformly randomly distributed in space.
When $\Lambda_{\rm MSR}\simeq1$, massive cores of a subregion do not show any particular spatial distribution.
In contrast, $\Lambda_{\rm MSR}>1$ indicates a spatial concentration of the massive cores and a core population that then qualifies as segregated in mass. \cref{tab:mass segregation} lists these two indicators of core mass segregation, $\left(\Sigma_{\rm6\,cores}\right)_{\rm massive}/ \overline{\Sigma_{\rm6\,cores}}$ and $\left[ \Lambda_{\rm MSR};~ N_{\rm MST} \right]$, and qualifies the segregation level of each W43-MM2\&MM3 subregion accordingly.

{\renewcommand{\arraystretch}{1.7}%
\begin{table}[ht]
\centering
\resizebox{\linewidth}{!}{
\begin{threeparttable}[c]
\caption{Core mass segregation in  subregions of W43-MM2\&MM3.}
\label{tab:mass segregation}
\begin{tabular}{ccccc}
    \hline\hline
    Subregion & $\left(\Sigma_{\rm 6\,cores}\right)_{\rm massive}$ &  \multirow{2}{*}{$\Lambda_{\rm MSR}$} & $N_{\rm MST}$ & Mass segr. \\
    name & / $\overline{\Sigma_{\rm6\,cores}}$ & & range & level \\
    (1) & (2) & (3) & (4) & (5)  \\
    \hline
    Outskirts             &     $<1$ &           - &                 - &      none \\ 
    MM51                  &     $<1$ &           - &                 - &      none \\ 
    MM10                  &      3.1 & $2.3\pm0.7$ &   $3\rightarrow7$ &      high \\ \hline
    MM12                  &     $<1$ &           - &                 - &      none \\ 
                          &          &             &                   &           \\ 
    \multirow{3}{*}{MM2}  &          &     $5\pm2$ &   $3\rightarrow6$ & very high \\
                          &      2.6 & $2.3\pm0.4$ &  $7\rightarrow14$ &      high \\
                          &          & $1.8\pm0.2$ & $15\rightarrow31$ &  moderate \\
                          &          &             &                   &           \\ 
    MM3                   &      3.6 & $1.8\pm0.4$ &  $3\rightarrow15$ &  moderate \\ \hline
    Total                 &      6.6 &          -- &                -- &        -- \\
    \hline
\end{tabular}
\begin{tablenotes}
\item (2) Ratio of the local core surface density measured for the five most massive cores to that averaged for all cores in each W43-MM2\&MM3 subregion.
\item (3) and (4) Core mass segregation ratio computed from \cref{eq:lambda msr} over the $N_{\rm MST}$ range.
\item (5) Core mass segregation level (see Sect.~\ref{sect:mass segregation}); three regimes for the MM2 subregion corresponding to the plateaus observed in \cref{fig:lambda msr}a.
\end{tablenotes}
\end{threeparttable}}
\end{table}}

Figure~\ref{fig:sigma6} displays the local surface density of cores in the W43-MM2\&MM3 mini-starburst, $\Sigma_{\rm 6\,cores}$. For the MM2, MM3, and MM10 subregions, we observe a correlation trend between the mass of a core and its local core surface density when this core is above the completeness limit of their catalogs (see Figs.~\ref{fig:sigma6}a--b). Such a behavior has already been observed in several studies \citep[e.g.,][]{lane2016, parker2018, dibHenning2019,nony2021} and can be related to the need of high-density gas to form cores, and especially massive cores. While this correlation trend is clear for the MM2 and MM3 subregions, it is only tentative for MM10 and absent for the other subregions.
In addition, in the MM2, MM3, and MM10 subregions, the five most massive cores have local core surface densities $2.6-3.6$ times greater than the core surface densities averaged over the entire subregions (see Figs.~\ref{fig:sigma6}a--b and \cref{tab:mass segregation}). 
This result indicates that in the MM2, MM3, and MM10 subregions, cores spatially concentrate around the most massive ones. It argues in favor of a large clustering of massive cores in these subregions but could also correspond to the peculiar situation where these five massive cores are surrounded by low-mass cores at five different locations. We verified in \cref{fig:nh2 and cores} that this is not the case of any of the MM2, MM3, and MM10 subregions. As for the other three subregions of W43-MM2\&MM3, in each subregion the core surface density of their five most massive cores is lower than that averaged over their entire core sample (see \cref{tab:mass segregation}).

Therefore, according to the local core surface density indicator, $\Sigma_{\rm 6\,cores}$, the MM2, MM3, and MM10 subregions show some evidence of mass segregation and not the other three subregions. We seek here to confirm this result by using the $\Lambda_{\rm MSR}$ indicator.
Figure~\ref{fig:lambda msr} displays the mass segregation ratio of the W43-MM2\&MM3 subregions, excluding the Outskirts, which is not a coherent subregion with convex boundaries. As classically done, we started computing for each subregion, the $\Lambda_{\rm MSR}$ at $N_{\rm MST}=3$ to avoid divergence on the first MST measurements. If massive cores are spatially concentrated, the mass segregation ratio decreases when it is measured over a larger group of cores. This trend is clearly observed for the MM2 and MM10 core catalogs, well above their completeness level (see Figs.~\ref{fig:lambda msr}a--b). The MM2 and MM3 subregions therefore qualify as mass-segregated, in agreement with the local core surface density measurements (see Figs.~\ref{fig:sigma6}a--b and \cref{tab:mass segregation}). 
The $\Lambda_{\rm MSR}$ function observed for the core catalog of MM3 is more complex to interpret. The most massive cores of the MM3 subregion do not appear mass-segregated but taking a larger group of cores reveals some evidence of moderate mass segregation (see \cref{fig:lambda msr}c and \cref{tab:mass segregation}). The unusual shape of the $\Lambda_{\rm MSR}$ function in the MM3 subregion suggests that the structure of its core cluster is not just mass-segregated. This complexity could be a consequence of the MM3 subregion being at an evolved stage, as it already developed an UC\hii region.
According to the $\Lambda_{\rm MSR}$ indicator, the MM12 and MM51 subregions do not show evidence of mass segregation, while the completeness level of their core catalogs would allow for it (see Figs.~\ref{fig:lambda msr}a--b).

In \cref{tab:mass segregation}, we qualified the evidence of core mass segregation using the $\Lambda_{\rm MSR}$ values: moderate for $1.5 <\Lambda_{\rm MSR}< 2$, high for $2 \le \Lambda_{\rm MSR} < 3$, and very high for $\Lambda_{\rm MSR} \ge 3$. Using this classification, the level of mass segregation of MM2, MM10, and MM3 is very high, high, and moderate, respectively (see \cref{tab:mass segregation}). 
The number of points involved in the mass-segregated part of the $\Lambda_{\rm MSR}$ function can sometimes be used to define several regimes. This is the case of the MM2 subregion, whose core mass segregation is so high that $\Lambda_{\rm MSR}$ displays three plateaus. It has a very high level of mass segregation for $N_{\rm MST}=3\rightarrow 6$, high for $N_{\rm MST}=7\rightarrow 14$, and moderate for $N_{\rm MST}=15\rightarrow 31$ out of 58 cores. 
Among the dozen or so studies that have investigated the mass segregation of cloud structures, only two found similarly high segregation levels in the NGC~2264, W43-MM1, and Corona Australis star-forming regions \citep{dibHenning2019, nony2021}. Others notably studying early stages of intermediate- to high-mass star formation found no evidence \citep[e.g.,][]{sanhueza2019, busquet2019}. 
The variety of core mass segregation levels we measured in the W43-MM2\&MM3 subregions is reminiscent of that found by \cite{dibHenning2019} and \cite{nony2021} in five and three (sub)regions, respectively. 

In conclusion, our analysis of the core mass segregation suggests that three out of the six W43-MM2\&MM3 subregions could be mass-segregated, the MM2, MM3, and MM10 subregions (see \cref{tab:mass segregation}). Interestingly, the cores that are mass segregated in the MM10 subregion have low to intermediate, $2.0-11.2~\Msol$, masses, while those in MM2, and MM3 span the $11-70~\Msol$ and $7-45~\Msol$ range, respectively.

\section{Discussion} \label{sect:discussion}

In the subregions of the W43-MM2\&MM3 ridge, the high-mass end of the CMF is observed to vary from steeper or close to the Salpeter slope of the canonical IMF, to top-heavy (see Sect.~\ref{sect:CMFs}). 
In Sect.~\ref{sect:BadCorrelations}, we examine, the link between the CMF shape and the cloud properties and core mass segregation characterized in Sects.~\ref{sect:spatial variations of core pop} and \ref{sect:mass segregation}, respectively. 
In Sect.~\ref{sect:link CMF-PDF}, we then study the link between the power-law index of the CMF high-mass end and the slope of the $\eta$-PDF tail measured in Sect.~\ref{sect:PDFs} for each subregion.
In Sect.~\ref{sect:link with sf history}, we finally determine the likely evolutionary stage of subregions and interpret their CMF variations in the framework of a scenario of the cloud and star formation histories across the W43-MM2\&MM3 ridge.

\subsection{Correlation trends between the CMF and the basic cloud and core properties} \label{sect:BadCorrelations}

The W43-MM2\&MM3 ridge presents a large variety of environments, with the MM2 subregion being the most extreme and the Outskirts the least extreme. In detail, the MM2 subregion, compared to the Outskirts, has $\sim$11 times more total gas mass, $\sim$7000 times higher densities, $\sim$15 times greater core density per surface units, $\sim$9 times more gas mass into cores, and displays a very high core mass segregation (see Tables~\ref{tab:subregions chara} and \ref{tab:mass segregation}).

We first searched for correlations of the power-law index of the CMF high-mass end, given in \cref{tab:cmf and s}, with all the parameters listed in \cref{tab:subregions chara}. As a general trend, the more extreme the subregion in terms of mass or associated volume density or in terms of mass into core, the shallower its CMF power-law index. This result is reminiscent of the correlation found between IMF indices and cloud densities of extragalactic starburst environments \citep{marks2012}. While being an extreme cloud of the Milky Way, the W43-MM2\&MM3 ridge has a density, $n_{\rm H_2} \sim 3\times 10^4$~cm$^{-3}$ in $\sim$11~pc$^{-3}$ \citep[see][]{nguyen2013}, just at the level of the lowest densities of the regions studied by \cite{marks2012}. The W43-MM2 subregion, which is about 17 times denser than the global ridge (see \cref{tab:subregions chara}) just reaches the minimum density where an effect on the IMF slope is suspected \citep{marks2012}. 

We then searched for some relationship between the core mass segregation and the CMF shape. Both core mass segregation and top-heavy CMF are indeed, by definition, favored by the presence of high-mass, or at least intermediate-mass, cores. This high-density gas, however, needs to be centrally concentrated to define a single major star-forming site that displays mass segregation, as it is the case of MM2 and to a lesser extent MM3 and MM10 (see \cref{fig:nh2 and cores}). This condition is not a necessary one to build a flat CMF, as proven by the MM12 subregion where no mass segregation is observed (see Tables~\ref{tab:cmf and s}--\ref{tab:mass segregation}). Conversely the MM10 subregion, consisting of a major filament plus a lower density medium, shows core mass segregation while its CMF high-mass end remains close to the Salpeter slope (see \cref{fig:subregion mass vs core mass} and Tables~\ref{tab:cmf and s}--\ref{tab:mass segregation}). In that respect, the link between the level of core mass segregation and the slope of the CMF high-mass end may exist but not as a one-to-one correspondence. 

These general trends could physically constrain models, but we recall that they are currently based on only six data points that correspond to the six subregions of the W43-MM2\&MM3 ridge. We should be able to confirm these trends into definite correlations or to refute them using a much larger sample of subregions that will be defined within the 15 protoclusters imaged by the ALMA-IMF Large Program.

{\renewcommand{\arraystretch}{1.2}%
\begin{table*}[ht]
\centering
\resizebox{\textwidth}{!}{
\begin{threeparttable}[c]
\caption{First and second tails measured for the $\eta$-PDF studied in various star-forming regions and comparison with some numerical simulation models.}
\label{tab:eta-pdf characteristics}
\begin{tabular}{lcccccl}
    \hline\hline
    Region & Reference paper & Fitted range ($s_1$) & $s_1$ & Fitted range ($s_2$) & $s_2$ & Consistent with studies by \\
     & & [$\times10^{21}$ cm$^{-2}$] & & [$\times10^{21}$ cm$^{-2}$] & & \\
    (1) & (2) & (3) & (4) & (5) & (6) & (7)  \\
    \hline 
    \multicolumn{6}{l}{{\bf Low-mass star-forming regions}} \\
    Taurus$^{\dagger}$     & \cite{schneider2022}        &         3--18 &    -2.3 &         18--50 &    -4.4 & \cite{kainulainen2009} \\
    $\rho$-Oph$^{\dagger}$ & \cite{ladjelate2020}        &         7--15 &    -1.3 &         15--65 &    -2.8 & \cite{schneider2022}   \\
    Aquila$^{\dagger}$     & \cite{schneider2022}        &         5--18 &    -2.1 &        18--250 &    -2.4 & \cite{schneider2013,konyves2015}  \\
    \hline
    \multicolumn{6}{l}{{\bf Intermediate-mass star-forming regions}} \\
    Orion B$^{\dagger}$    & \cite{jaupartChabrier2020}  &         2--23 &    -2.0 &   23\st--50\st & -3.0\st & \cite{kainulainen2009,konyves2020}\\
                           &                             &               &         &                &         & \cite{schneider2013,schneider2022}\\
    Orion~A$^{\dagger}$    & \cite{stutzKainulainen2015} &        17--80 &    -0.9 &  80\st--160\st & -3.0\st & \cite{kainulainen2009} \\
    Mon~R2                 & \cite{schneider2015}        &         8--36 &    -2.1 &        36--200 &    -1.0 & \cite{schneider2022}   \\
    \hline
    \multicolumn{6}{l}{{\bf High-mass star-forming regions}} \\
    W3~Main/(OH)           & \cite{riveraIngraham2015}   &        12--36 & -2.1\st &     36--190\st & -1.6\st &                        \\
    NGC~6334               & \cite{russeil2013}          &        11--90 &    -2.0 &  90\st--280\st & -1.3\st & \cite{schneider2022}   \\
    Cygnus-X North         & \cite{schneider2016}        &     12--85\st & -2.3\st &  85\st--280\st & -1.5\st & \cite{schneider2022}   \\
    W43                    & \cite{carlhoff2013}         & 37\st--190\st & -2.5\st & 190\st--660\st & -1.3\st & This article           \\
    \hline
    \multicolumn{6}{l}{{\bf W43-MM2\&MM3 mini-starburst ridge and its subregions}} \\
    W43-Main (without MM1) &    This article             &        40--60 &    -2.9 &        60--400 &    -1.3 & \cite{carlhoff2013}    \\
    MM10                   &    This article             &            -- &      -- &       100--300 &    -2.4 &                        \\
    MM2                    &    This article             &            -- &      -- &      150--1000 &    -0.6 &                        \\
    \hline
    \multicolumn{7}{l}{{\bf Numerical simulations}} \\
    Model \#24             & \cite{kainulainen2013}      &   3\st--40\st &  -1.6\st &       --      &     --  &                        \\
    Model \#C1t03          & \cite{leeHennebelle2018a}   & $10^6-10^{12}$~cm$^{-3}$ & $\equiv$-2 & - &   --   &                        \\
    Model M=3, SFE=20\%    & \cite{jaupartChabrier2020}  &         1--20 &     -2.0 & 20\st--400\st & -1.0\st &                        \\
    \hline
\end{tabular}
\begin{tablenotes}
\item $\dagger$: Low- to intermediate-mass star-forming regions, which exhibit typical CMFs with a high-mass end fitted by a power-law index close to that of the Salpeter slope of the canonical IMF \citep{polychroni2013, konyves2015, marsh2016, konyves2020, ladjelate2020}.
\item \st Values derived graphically from the reference paper.
\end{tablenotes}
\end{threeparttable}
}
\end{table*}}

\begin{figure*}[htbp!]
    \centering
    \includegraphics[width=0.48\textwidth]{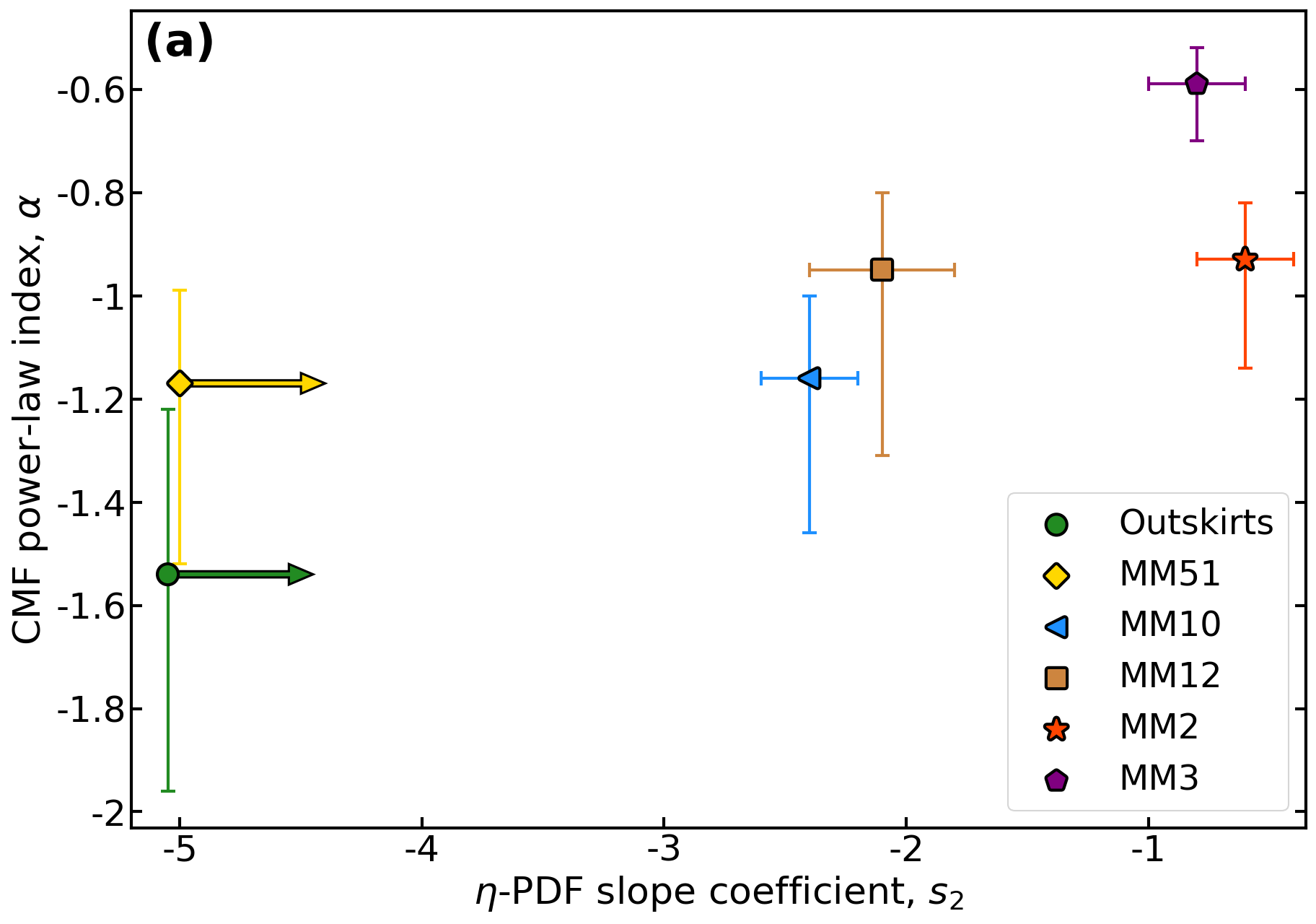} \hskip 0.5cm
    \includegraphics[width=0.48\textwidth]{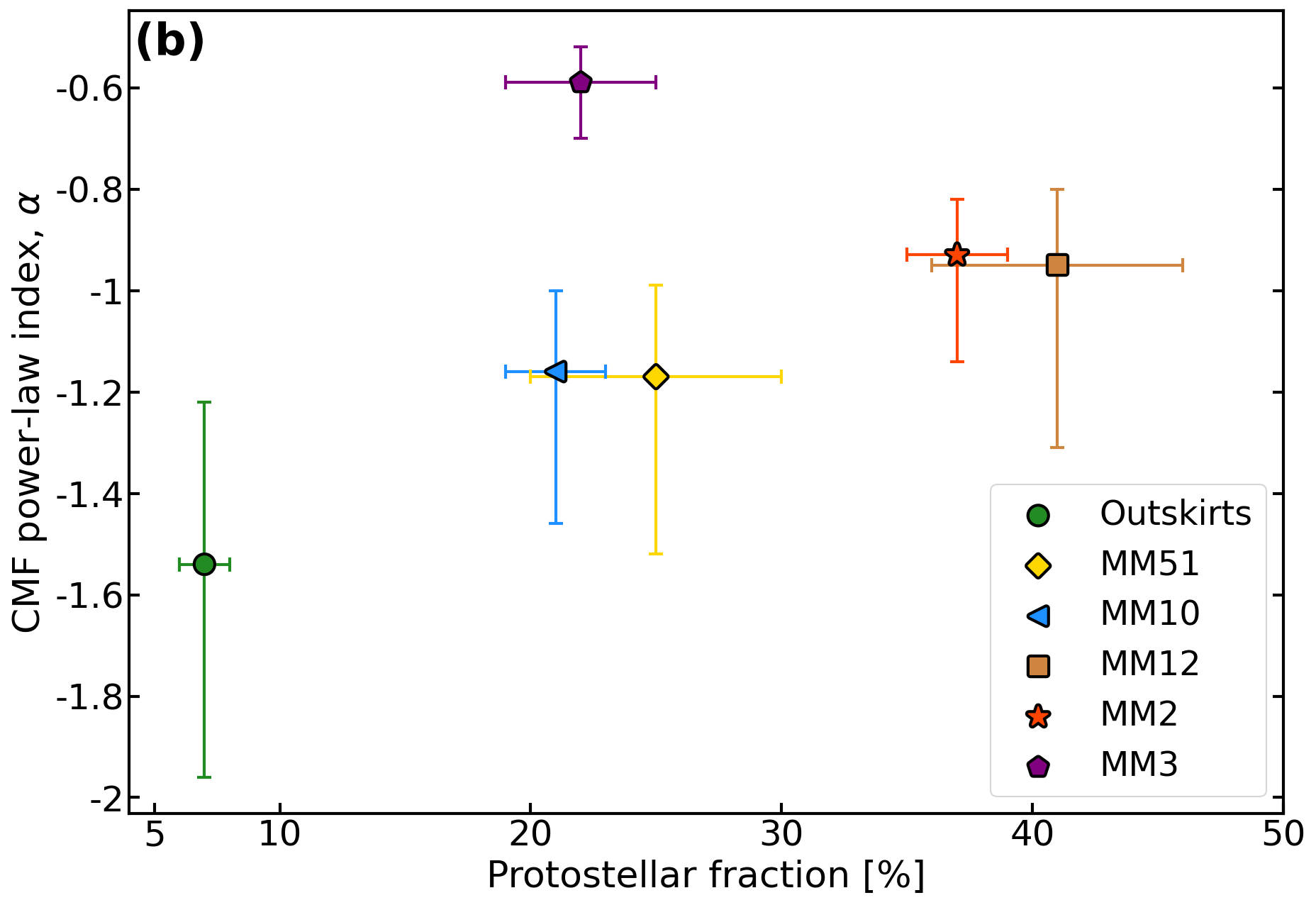}
    \caption{For each W43-MM2\&MM3 subregion, the power-law index of the CMF high-mass end plotted against the slope coefficient of the second $\eta$-PDF tail (\textit{panel a}) and the protostellar fraction (\textit{panel b}). Horizontal arrows represent lower limits for $s_2$ in \textit{panel a}. We avoid fitting the correlation trends observed in \textit{panels a--b} because they rely on too few subregions.}
    \label{fig:correlation alpha vs s and proto}
\end{figure*}

\begin{figure*}[htbp!]
    \centering
    \begin{minipage}{0.485\textwidth}
        \centering
        \includegraphics[width=1.\textwidth]{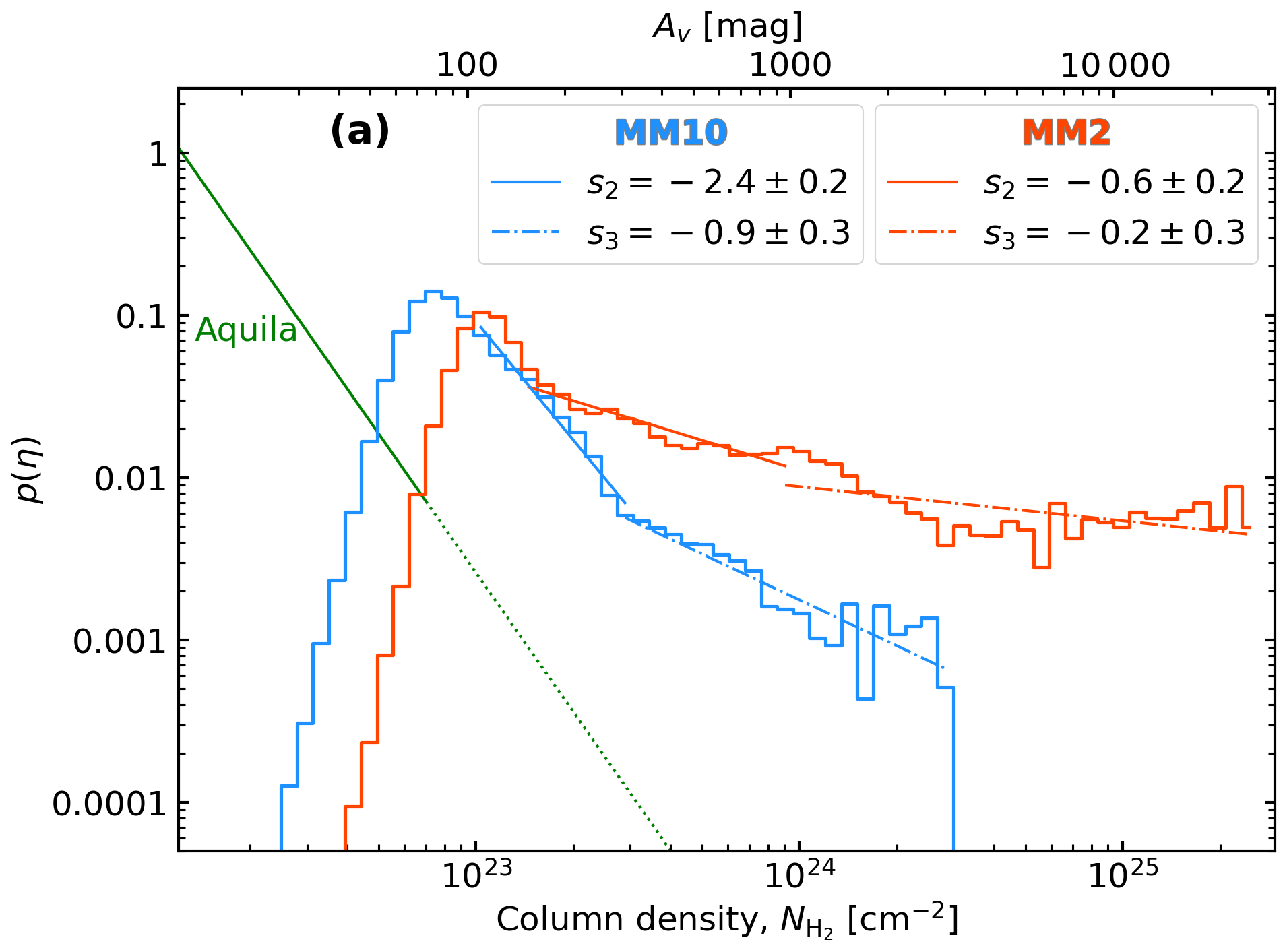}
    \end{minipage}%
    \hskip 0.02\textwidth
    \begin{minipage}{0.475\textwidth}
        \centering
        \vskip 0.5cm 
        \includegraphics[width=1.\textwidth]{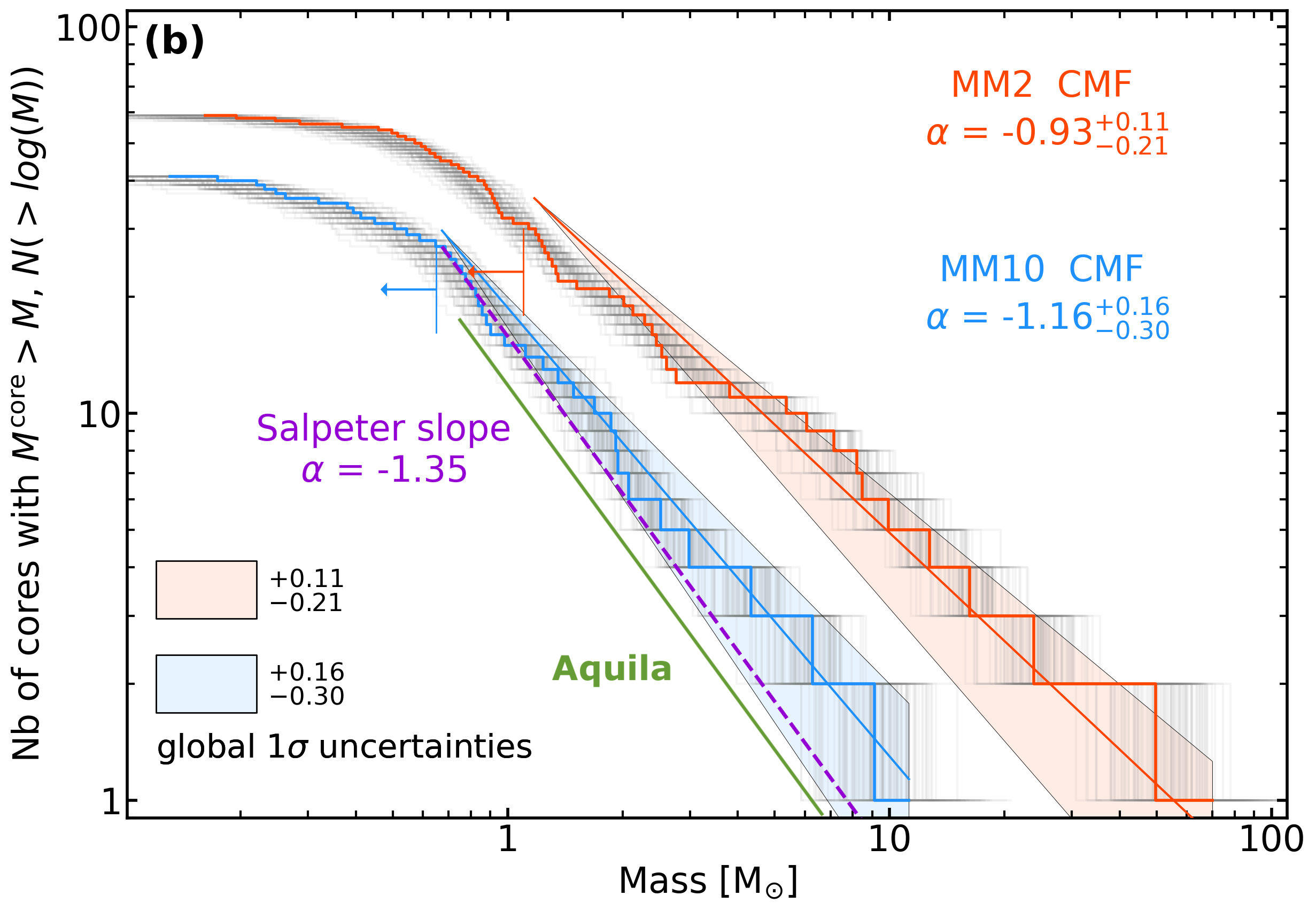}
    \end{minipage}
    \vskip 0.1cm
    \caption{Two prototypical subregions of W43-MM2\&MM3, MM2 (red line), and MM10 (blue line), compared with Aquila (green line). \textit{Panel (a)}: $\eta$-PDF histograms, whose tails (colored segments) are fitted with power laws. Solid and dotted green lines correspond to the observed and extrapolated column density ranges from \cite{konyves2015}. While the $\eta$-PDF of the MM2 subregion displays a flat power-law tail, those of MM10 and Aquila are much steeper. \textit{Panel (b)}: Cumulative CMF histograms, whose high-mass ends are represented by power laws (colored segments) and their $1\sigma$ global uncertainties (colored areas; see the definition in Sect.~\ref{sect:CMFs}). While the MM2 subregion exhibits a top-heavy CMF, MM10 and Aquila have steeper CMF high-mass ends, with power-law indices close to the Salpeter slope (dashed magenta line).}
    \label{fig:PDF CMF mm2 mm10}
\end{figure*}

\subsection{Linking the CMF high-mass end to the \texorpdfstring{$\eta$}{n}-PDF tail}
\label{sect:link CMF-PDF}

\subsubsection{Comparison with observational studies in the literature} \label{sect:comparison with literature}

We here compare the $\eta$-PDF tail of the W43-MM2\&MM3 subregions that are characterized in Sect.~\ref{sect:PDFs} with those obtained from the literature for ten observed star-forming clouds and three simulated clouds \citep[e.g.,][see our \cref{tab:eta-pdf characteristics}]{schneider2022, jaupartChabrier2020}. 
These $\eta$-PDF tails cover different column density ranges depending on the type (low-, intermediate-, and high-mass) of the star-forming regions considered and are quantified by fitting one or two consecutive power laws \citep[as proposed by, e.g.,][]{kainulainen2009, schneider2015}. They are as follows:

\textit{First $\eta$-PDF tail}: Above the transition from a lognormal function to a power-law tail, the first slope of the tails measured in the literature is $s_1 \in [-2.5; -0.9]$, with most values around $s_1 \simeq -2$, (see \cref{tab:eta-pdf characteristics}). 
The first tail of W43-MM2\&MM3 is observed in the $\eta$-PDF of the \textit{Herschel} column density image (see \cref{appendixfig:pdf W43 herschel}b and \cref{tab:eta-pdf characteristics}), but not in the $\eta$-PDFs derived from the ALMA 12~m array alone (see Sect.~\ref{sect:PDFs}).

\textit{Second $\eta$-PDF tail}: At higher column densities, above $40-100\times 10^{21}$~cm$^{-2}$, a flatter second tail is observed for high-mass star-forming regions: $s_2 \in [-1.6; -1.3]$ (see \cref{tab:eta-pdf characteristics}). 
In contrast, the second tails of the low- and intermediate-mass star-forming regions (with the exception of Mon~R2) are steep, indicating a lack of cloud structures at high column densities in maps with spatial resolution about the core size. 
The ALMA $\eta$-PDF built for the W43-MM2\&MM3 subregions have $s_2$ slopes that range from $s_2>-5$ for MM51 and the Outskirts to $s_2=-0.6\pm 0.2$ for MM2. While the $s_2$ tails of the Outskirts and MM51 resemble those of low-mass star-forming regions, those of MM10 and MM12 follow the typical slope of the first tail, and those of MM2 and MM3 are as flat as, or even flatter than, the second slope measured in high-mass star-forming regions. Figure~\ref{fig:correlation alpha vs s and proto} displays, for each W43-MM2\&MM3 subregion, the relation between the power-law index of the CMF high-mass end and the slope coefficient of the second $\eta$-PDF tail, $\alpha$ and $s_2$, and the protostellar fraction listed in \cref{tab:cmf and s}. We observed a correlation trend between these power-law indices but the small number of W43-MM2\&MM3 subregions and the large uncertainties on $\alpha$ and $s_2$ prevent us from making a reliable estimate of the significance and strength of this correlation.

\textit{Third $\eta$-PDF tail}: The $\eta$-PDFs of the W43-MM2\&MM3 subregions reach extremely high column densities, up  to $900-26\,000\times 10^{21}$~cm$^{-2}$), never before reported in other published $\eta$-PDF studies (see \cref{fig:various eta-PDF}). At these column densities corresponding to those of cores, from $250-900\times 10^{21}$~cm$^{-2}$ to $1000-26\,000\times 10^{21}$~cm$^{-2}$, the $\eta$-PDF of the W43-MM2\&MM3 subregions all have $s_3$ slopes that are very flat, $s_3\in [-0.9;-0.1]$ (see \cref{tab:cmf and s}). As suggested by \cite{myers2017}, we could expect the third $\eta$-PDF tail, which is associated with cores, to more strongly correlate with the CMF core distribution than the second tail, which traces the high-column-density gas surrounding cores. 
For the MM2 and MM3 subregions, the second and third tails have their power-law exponents $s_2$ and $s_3$ very close to each other. This suggests that, in these subregions, the distribution of column density within cores originate from the gas distribution in their cloud hubs, in agreement with the dynamical picture of cloud formation \citep[e.g.,][]{smith2009}.

Figure~\ref{fig:PDF CMF mm2 mm10} displays the $\eta$-PDFs and CMFs of two representative subregions of W43-MM2\&MM3, MM2 and MM10, along with those of the reference low-mass star-forming region Aquila \citep{konyves2015,schneider2022}. 
While MM10 has a steep $\eta$-PDF tail and a CMF high-mass end close to the Salpeter slope of the canonical IMF, the MM2 subregion displays a much flatter second $\eta$-PDF tail and a top-heavy CMF (see \cref{fig:PDF CMF mm2 mm10} and \cref{tab:eta-pdf characteristics}). Interestingly, the MM10 subregion has $\eta$-PDF and CMF power-law indices, $s_2=-2.4\pm0.2$ and $\alpha=-1.16_{-0.30}^{+0.16}$, that are similar to those of the low- to intermediate-mass star-forming regions. Taurus, $\rho$~Oph, Aquila, Orion~B, and Orion~A are indeed known to have tails dominated by a steep power law ($s_2 \in [-4.4; -2.4]$; see \cref{tab:eta-pdf characteristics}) and to exhibit typical CMFs with a high-mass end close to the Salpeter slope \citep[$\alpha\simeq -1.35$][]{polychroni2013, konyves2015, marsh2016, konyves2020, ladjelate2020}.

Besides, subregions with the most top-heavy CMFs (MM2 and MM3) have second $\eta$-PDF tail at high column density described by a power law with a slope, $s_2\in[-0.8; -0.6]$, close to but flatter than the one measured for the second power-law tails observed in high-mass star-forming regions (e.g., \citealt{schneider2015, schneider2022}, see also our \cref{tab:eta-pdf characteristics}). The first elements of a physical interpretation for the observed variations of the $\eta$-PDF and CMF shapes are given in Sect.~\ref{sect:interpretation link pdf-cmf}.

\subsubsection{Interpretation of \texorpdfstring{$\eta$}{n}-PDF tails and their link with the CMF} \label{sect:interpretation link pdf-cmf}

It is tempting to make a direct link between the slope of the cloud column density $\eta$-PDF and the power-law index of the CMF high-mass end. 
We here compare the two representative subregions in W43-MM2\&MM3, illustrated in \cref{fig:PDF CMF mm2 mm10}, with published models and discuss various interpretations proposed in observational and theoretical studies.
Analytical studies and numerical simulations often analyze the PDF of the natural logarithm of the normalized volume density, $\phi \equiv \ln{(\rho / \overline{\rho})}$ with $\overline{\rho}$ the mean density and fit the PDF with power laws of exponent $\psi$:
\begin{equation}
    p_\phi = \rho\times p_\rho \propto \rho^\psi .
\end{equation}
In the idealized case of a spherical cloud with a purely radial density distribution, $\rho \propto r^{-a}$, \cite{federrathKlessen2013} have analytically demonstrated the relation that exists between the slope coefficients of $p_\eta$ and $p_\phi$ and that of the volume density, $s$ and $\psi$ versus $a$, respectively: $s = \frac{2}{1-a}$ and $\psi = \frac{-3}{a}$ \citep[see Eqs.~(9) and (11) of][]{federrathKlessen2013}. It leads to the following relation:
\begin{equation}\label{eq:s}
    s=\frac{2}{1+3/\psi}.
\end{equation}

We used \cref{eq:s} to predict, from the power-law exponent, $\psi$, of the models, the power-law exponent $s$ that an $\eta$-PDF tail computed from the simulated volume density cube projected into a column density image would have \citep[e.g.,][see also our \cref{tab:eta-pdf characteristics}]{kainulainen2013,leeHennebelle2018b, jaupartChabrier2020}.
The models of \cref{tab:eta-pdf characteristics} exhibit power-law exponents with $s_1\sim -2$, close to those found for the first $\eta$-PDF tail of low-mass star-forming regions and that of W43-MM2\&MM3 (see \cref{tab:eta-pdf characteristics}). It has been interpreted by many studies as evidence of volume density with a power-law index of $a\simeq 2$ that could be associated with the collapse of a spherically symmetric isothermal cloud \citep[see][and references therein]{schneider2022}.
We however caution that relation of \cref{eq:s} is only valid for the idealized and unrealistic case of a spherical cloud with a radial density distribution. While \cref{eq:s} could also apply, and $s_1= -2$ is expected, for a collection of well-resolved collapsing cores, the complex structure of the cloud at high column density should not be neglected when interpreting the slope of $\eta$-PDF tails.

As for the flat second tails of the $\eta$-PDF obtained for Orion~A, Mon~R2, all high-mass star-forming regions of \cref{tab:eta-pdf characteristics} and the MM2 and MM3 subregions, they have been associated with their hosted hubs. The density profiles of ridges and hubs, when measured, are steeper than the classical $\rho(r)\propto r^{-2}$ profile \citep{hill2011, didelon2015, motte2018a} and interpreted by \cite{motte2018a} as a consequence of adiabatic heating, rotation, or magnetic field, in relation with the observed slow-down of the ridge collapse \citep{wyrowski2016}.
Ridges and hubs are expected to form by dynamical processes such as cloud collision during the initial cloud formation phase or feedback effects associated with the expansion of \hii regions in more evolved clouds \citep[e.g.,][]{motte2018a}. In fact, these second tails were already associated with feedback effects of \hii regions \citep{tremblin2014,riveraIngraham2015} or the young evolutionary stage of a cloud \citep{sadavoy2014, stutzKainulainen2015}. 
In agreement with the interpretations by, for example, \cite{schneider2015} and \cite{motte2018a}, models by \cite{khullar2021} and \cite{donkov2021} showed that the onset of rotation and a change in the equation of state for a hard polytrope both lead to second flatter PDF tails.

 According to \cite{leeHennebelle2018a} and \cite{hennebelle2022}, a cloud with a $\rho$-PDF exhibiting a tail instead of a simple lognormal shape, develops a CMF high-mass end that is shallower than the Salpeter slope. While these simulation studies qualitatively agree with present observations, they associate a top-heavy CMF with a tail slope of $\psi= -1.5$, or potentially in an equivalent manner $s=-2$ (see above). This power-law exponent of the $\eta$-PDF remains steeper than what is observationally found for the exponents of the second power-law tail of high-mass star-forming clouds and second and third tails of the MM2 and MM3 subregions.
A handful of other models predicted a flatter tail \citep[see \cref{tab:eta-pdf characteristics} and, e.g.,][]{kainulainen2013} but their impact on the CMF shape still needs to be investigated.

\subsection{Core population, a witness of the history of cloud and star formation} \label{sect:link with sf history}

Beyond a simple study of the correlation between the distribution in space and mass of the cloud and cores, it is necessary to take into account the history of cloud and star formation to understand the variety of the subregions properties and core populations in W43-MM2\&MM3.
\cref{sect:link CMF-PDF} revealed a correlation pattern between the CMF and $\eta$-PDF. In Sect.~\ref{sect:evolutionary stage}, we define the evolutionary status of the subregions and, in Sect.~\ref{sect:impact on cloud and cores}, make the link between the CMF and $\eta$-PDF quantities to finally propose a cloud and star formation scenario (see \cref{fig:summary}). 

\subsubsection{Evolutionary stage of subregions} \label{sect:evolutionary stage}

In W43-MM2\&MM3 as in the W43-MM1 mini-starburst ridge, clouds are expected to form via a global collapse, which is initiated by colliding flows of H\mbox{\sc i} gas \citep{nguyen2011b, nguyen2013, motte2014, louvet2016}. In dynamical star formation theories, a dynamical cloud assembly is followed by a burst of star formation \citep[e.g.,][]{smith2009,leeHennebelle2018a,vazquez2019, pelkonen2021}. 
This in fact is the favored scenario for the W43 Main cloud and in particular the W43-MM1 and W43-MM2\&MM3 mini-starburst ridges, which are observed to efficiently form stars \citep{motte2003, nguyen2013, louvet2014}. During star formation bursts, clouds actively form cores that should immediately collapse and host protostars in their main accretion phase \citep[e.g.,][]{pelkonen2021,hennebelle2022}. We therefore expect the protostellar core fraction to be higher in the burst phase than in quiescent regions or in the pre-burst phase.
While the burst is a phase when massive cores have not yet had time to develop UC\hii regions, the post-burst phase should be characterized by the presence of \hii regions. Because the W43-MM2\&MM3 subregions are at different evolutionary stages (see below), separating the ridge into subregions allows us to focus on cloud structures that will, do, or did simultaneously form stellar clusters.

Defining the evolutionary stage of the W43-MM2\&MM3 subregions is however not a straightforward task. We focus on durations several free-fall times of the subregions. With volume densities of $n_{\rm H_{2}}\simeq 0.2-5\times10^5$~cm$^{-3}$ (excluding the Outskirts; see \cref{tab:subregions chara}), this would correspond to time spans of $\sim$10$^5-10^6$~yr, with the shortest and longest potentially being for the MM2 and MM51 subregions, respectively. To define the evolutionary stage of these $\sim$0.5--1~pc cloud structures, we quantify the development of the star formation process through three criteria. These are the protostellar core fraction of the subregion, its surface density of cores, both quantified in Sect.~\ref{sect:analysis} (see Tables~\ref{tab:subregions chara} and \ref{tab:mass segregation}), and the potential UC\hii regions. 
Our classification allows us to state whether these subregions are a handful of $10^5$~years before, during, or after their main star formation event, or whether they are and will remain quiescent.
\cref{tab:cmf and s} gives the evolutionary stage of each W43-MM2\&MM3 subregion, as defined below, using the following criteria:

\textit{MM51 and Outskirts}: They qualify as quiescent because their surface number density of cores and their protostellar fraction are lower or equal to the average values measured in the W43-MM2\&MM3 subregions ($\sim$35 cores/pc$^2$ and $\sim$25\%; see Tables~\ref{tab:subregions chara}--\ref{tab:cmf and s}). These two subregions could well remain quiescent, aside from the most active sites of the W43-MM2\&MM3 mini-starburst, never entering a burst mode.

\textit{MM}\textit{10}: Potentially in a pre-burst regime, it has an enhanced surface number density of cores, $\sim$1.3 higher, but a protostellar fraction that remains slightly lower than the average values (see Tables~\ref{tab:subregions chara}--\ref{tab:cmf and s}). Its large number of cores is consistent with MM10 being a site of intense cloud formation by cloud-cloud collision, as evidenced by its strong SiO emission tracing low-velocity shocks \citep[][and Turner et al. in prep. using our ALMA-IMF data]{nguyen2013}.
    We therefore expect MM10 to enter, in the near future, in its main phase of star formation and to form the high-mass cores that it surprisingly lacks at this stage (see Sect.~\ref{sect:spatial variations of core pop}).

\textit{MM12}: It most probably is at the beginning of its burst because it displays an increase of the protostellar core fraction by a factor of $\sim$1.6 but has a surface number density of cores close to the average value measured for the W43-MM2\&MM3 subregions (see Tables~\ref{tab:subregions chara}--\ref{tab:cmf and s}).

    \textit{MM}2: It qualifies as being in its main burst because its surface number density of cores and protostellar core fraction are increased, by factors of $\sim$2.1 and $\sim$1.5 compared to the average values, respectively (see Tables~\ref{tab:subregions chara}--\ref{tab:cmf and s}). Among the 19 protostellar cores discovered in this subregion by \cite{nony2023}, three cores are hot core candidates (see \citealt{pouteau2022} and Bonfand et al. in prep.) powered by $10^3-10^4~L_{\odot}$ protostars \citep{motte2003, bally2010}. These luminosities and the absence of UC\hii region provide evidence that MM2 is at an early stage of the high-mass star formation process.

    \textit{MM3}: It is post-burst because it hosts an UC\hii region and its stellar activity decreased, as proven by their surface number density of cores and protostellar core fraction back to the average values (see \cref{tab:cmf and s}). This UC\hii region developed over the past $\sim$3$\times 10^5$ years \citep{lumsden2013} and originates from the formation of a $\sim$22~$\Msol$\footnote{
        As described in \cite{suarez2023}, we used the equations in \cite{rivera-soto2020} and \cite{martins2005} and estimated a mass of $\sim$22~$\Msol$ for the star ionizing the UC\hii.}
    star (see \cref{fig:subregion mass vs core mass}). We expect MM3 to have already formed a first-generation of stars, which could be searched for with high-sensitivity and high-resolution mid-infrared images of the James Webb Space Telescope (JWST) satellite.    %

\subsubsection{Impact of this evolutionary stage on cloud and cores} \label{sect:impact on cloud and cores}

In the framework of dynamical cloud and star formation models \citep[e.g.,][]{vazquez2019}, the density structure of clouds and distribution of core populations in clouds, both in mass through the CMF and in space and mass through the core mass segregation, should result from the cloud formation, star formation, and evolution processes. This is confirmed in \cref{fig:correlation alpha vs s and proto}b, where the slope of the CMF high-mass end tends to correlate with the protostellar fraction, which is the main criterion used to define the evolutionary stage of a subregion (see Sect.~\ref{sect:evolutionary stage}). This is particularly true when ignoring the MM3 subregion, whose core population has likely changed from its younger stages of burst or pre-burst regimes.

\begin{figure*}
    \centering
    \includegraphics[width=1.\textwidth]{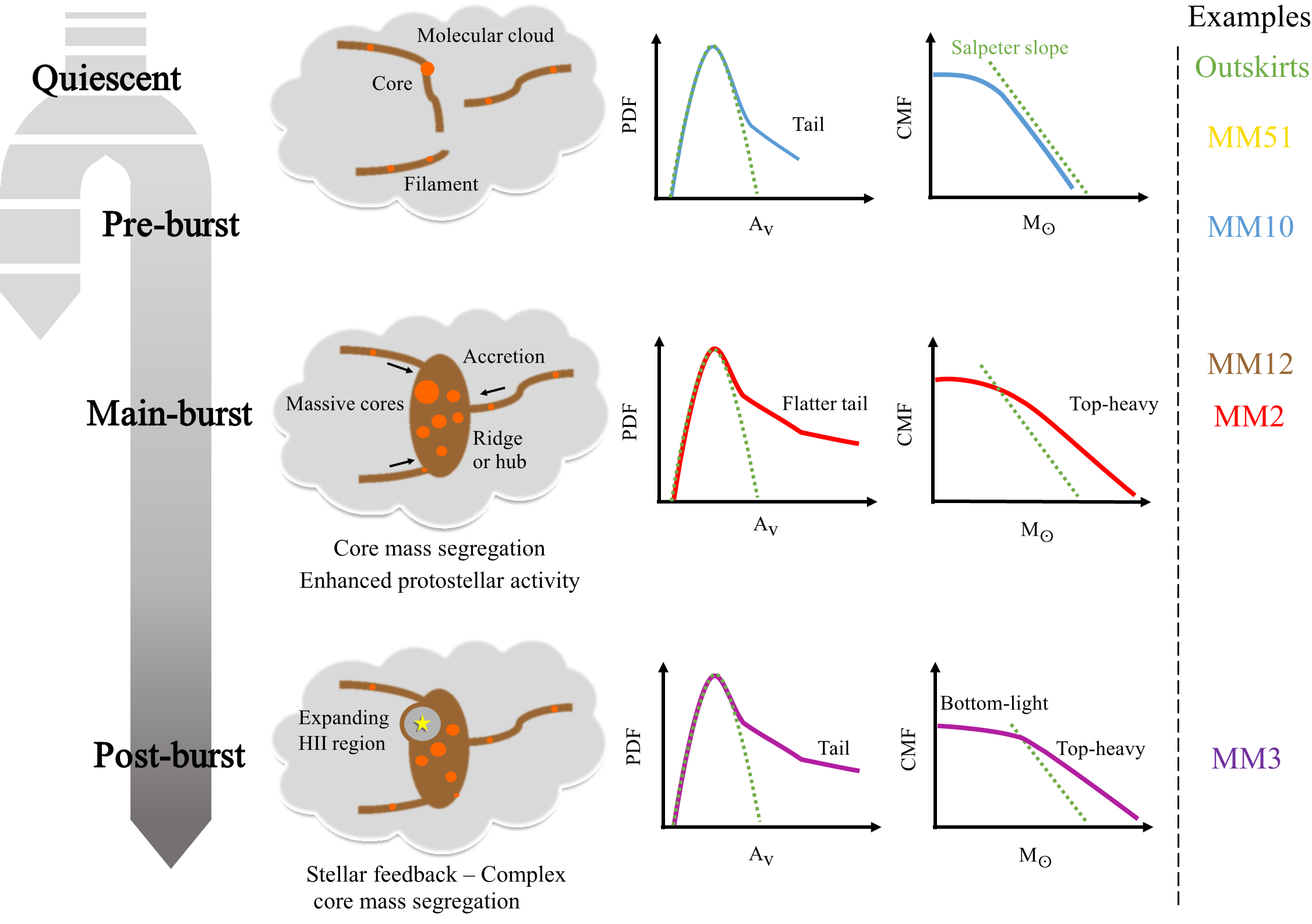}
    \caption{Schematic evolutionary diagram of subregions in dynamical star-forming clouds, which are qualified as either quiescent or as pre-burst, main burst, or post-burst, determined by following the different phases before, during, and after their main star formation event. Their characteristics, including cloud column density, $\eta$-PDF, CMF, and core mass segregation, are illustrated. \textit{Top panels:} Subregions in quiescent or pre-burst regimes present a $\eta$-PDF tail close to that found in low-mass star-forming regions and a CMF high-mass end close to the Salpeter slope. \textit{Central panels:} Subregions in their main burst regime harbor an enhanced star formation activity and a strong core mass segregation. They display a $\eta$-PDF with a much flatter second tail as well as a top-heavy CMF. \textit{Bottom panels:} Post-burst subregions, under the influence of stellar feedback, have less sustained star formation activity and more complex core mass segregation. They also present a $\eta$-PDF with a flat second tail and a CMF that is both top-heavy and bottom-light. The location on the time sequence of the W43-MM2\&MM3 subregions are given to summarize the result of the present article.}
    \label{fig:summary}
\end{figure*}

Figure~\ref{fig:summary} presents a schematic evolutionary diagram of subregions expected in ridges, with typical cloud and core properties, including the shapes of the CMFs and $\eta$-PDFs. 
Ridges are by definition high-density filamentary clouds, which are formed by dynamical processes of cloud formation and where star formation is enhanced \citep{motte2018a}.
We used W43-MM2\&MM3 subregions as examples because their cloud and core properties are well studied (see Sect.~\ref{sect:analysis}) and their evolutionary stages are robustly defined (see Sect.~\ref{sect:evolutionary stage}). The physical characteristics of quiescent subregions and of subregions in their pre-burst, main burst, and post-burst phases, which means before, during, and after their main star formation event, are detailed below.

In quiescent subregions of ridges (see top-panels of \cref{fig:summary})
    the cloud volume densities should be among the lowest and the slopes of their $\eta$-PDF tail among the steepest (like MM51 and the Outskirts in Tables~\ref{tab:subregions chara}--\ref{tab:cmf and s}). Their CMF high-mass end is most probably close to the Salpeter slope of the canonical IMF and the subregions do not present core mass segregation (like MM51 and the Outskirts in \cref{tab:mass segregation}). The quiescent subregions therefore have properties that resemble those of nearby, low- to intermediate-mass star-forming clouds, which are also for the most part quiescent.

    Subregions in pre-burst regime or at the beginning of their main burst (see top and central-panels of \cref{fig:summary}) should have similar cloud properties. Their volume densities are around the average values observed for the ridge subregions but their column density $\eta$-PDFs tail may start flattening (like MM10 and MM12 in Tables~\ref{tab:subregions chara}--\ref{tab:cmf and s}). The case of MM10 is interesting because it contains a filament, with column density enhancement and mass segregation of intermediate-mass cores (see \cref{fig:lambda msr}b). These characteristics combined with the shock evidence found by \cite{nguyen2013} suggest that this filament currently forms through cloud concentration and gas mass inflows. As for MM12, its proximity to the MM2 subregion suggests it formed during the same cloud concentration phase as MM2 and could accrete and concentrate more gas in the near future. Subregions in the transitory phase between pre-burst and burst should have a CMF high-mass end, which slowly gets flatter than the Salpeter slope (like MM12 in \cref{tab:cmf and s}).

The subregions in their main burst (see central panels of \cref{fig:summary}) should have mean cloud volume densities a few times larger and a column density $\eta$-PDF tail much flatter than other ridge subregions (see \cref{fig:various eta-PDF}e in the case of MM2). The high level of core mass segregation observed in MM2 (see \cref{fig:lambda msr}a and \cref{tab:mass segregation}) is in line with that observed after the intense phase of cloud concentration associated with the hierarchical global collapse model \citep{vazquez2019}. Subregions in their main burst should exhibit a CMF high-mass end, which is top-heavy (see, e.g., \cref{fig:PDF CMF mm2 mm10}b). This associated over-numerous population of high-mass cores, with respect to intermediate-mass cores, is a consequence of the high cloud dynamics developing during a burst. Gas streams would indeed preferentially feed high-mass cores, increasing their mass even more \citep[e.g.,][]{smith2009}, while preventing the formation of low-mass cores, whose low density hardly protects them from disruptions by shearing motions \citep{ntormousi2015}.

    Subregions in a post-burst regime (see bottom-panels of \cref{fig:summary}) probably have cloud properties similar to those of subregions in their main burst, both in terms of volume density and slope of the column density $\eta$-PDF tail (see Tables~\ref{tab:subregions chara}--\ref{tab:cmf and s} and \cref{fig:various eta-PDF}f in the case of MM3). However, in the densest parts of the post-burst cloud, young stars should have replaced their first-generation cores and feedback effects due to the expansion of potential \hii regions likely slow further core formation. This first star formation event likely reshaped the CMF and $\Lambda_{\rm MSR}$ functions (see Figs.~\ref{fig:various cmfs}b and \ref{fig:lambda msr}c), with the most massive cores at cloud center being consumed and subsequent formation of low-mass cores being slowed down due to tidal effects of shearing motions \citep{ntormousi2015}.
    %

To confirm the cloud formation scenario of \cref{fig:summary}, it is necessary to trace gas inflow through for instance, C$^{18}$O, DCN, and N$_2$H$^+$ lines, and shocks associated with cloud formation and gas feeding onto cores, notably with SiO and methanol lines \citep[e.g.,][]{peretto2013, louvet2016, csengeri2018}.
The ALMA-IMF survey provides the appropriate lines and associated dynamical studies are planned \citep[see, e.g.,][]{cunningham2023}. The present article  studied two protoclusters that are too young to constrain the latest phases of the evolutionary scenario of \cref{fig:summary}. The ALMA-IMF Large Program imaged four evolved, massive protoclusters, which are associated with developed \hii regions: G010.62, G012.80, G333.60, and W51-IRS2 \citep{motte2022}. They are the focus of ongoing studies for their core population \citep[e.g.,][Armante et al. in prep.]{louvet2023} that will soon give us insights on the latest phases of the CMF evolution.

\section{Summary and conclusions} \label{sect:conclusions}

We have used the most complete and most robust sample of cores obtained to date as part of the ALMA-IMF Large Program to study the dependence of the CMF on the cloud and protocluster properties. Our main results and conclusions can be summarized as follows.

\begin{itemize}
    \item We used a database consisting of the 1.3~mm continuum image  and core catalog produced by Paper~III \citep{pouteau2022} for the W43-MM2\&MM3 mini-starburst ridge. The ALMA image covers the $\sim$10~pc$^2$ cloud and hosts a cluster of 205 cores with sizes of $\sim$3400~au and masses ranging from 0.1~$\Msol$ to 70~$\Msol$ (see Sect.~\ref{sect:obs and catalog} and \cref{fig:nh2 and cores}). The pre-stellar versus protostellar nature of cores is taken from Paper~V \citep{nony2023} to estimate the protostellar fraction. \item Using the {MnGSeg} technique, we separated the W43-MM2\&MM3 ridge into six subregions that are expected to  simultaneously form stellar clusters. They have typical sizes of $0.5-1$~pc  and are called MM2, MM3, MM10, MM12, MM51, and Outskirts (see Sect.~\ref{sect:subregions def} and \cref{fig:mngseg scales}). We then built the column density image of W43-MM2\&MM3 from its continuum image at 1.3~mm and an offset measured on the \textit{Herschel} column density image (see Sect.~\ref{sect:subregions coldens and mass} and Figs.~\ref{fig:nh2 and cores} and \ref{appendixfig:pdf W43 herschel}). These subregions span a large range of masses, column densities, and volume densities (see \cref{tab:subregions chara}).
    \item The measured high-mass end of subregion CMFs in their cumulative form varies from being top-heavy, $\alpha\in [-0.95;-0.59]$, to close to the Salpeter slope, $\alpha\in [-1.54;-1.16]$, (see Sect.~\ref{sect:spatial variations of core pop}, \cref{fig:various cmfs} and \cref{tab:cmf and s}). They are fitted by single power laws of the form $N(> \log M) \propto M^\alpha$ using a bootstrapping method that uses a MLE method and takes the uncertainties on the core masses and on the sample low-number statistics as well as the incompleteness into account.
    \item We then analyzed the variations, in W43-MM2\&MM3, of the cloud density structure and the core mass segregation (see Sects.~\ref{sect:PDFs}--\ref{sect:mass segregation}). The subregion $\eta$-PDFs exhibit a tail up to extremely high column densities, 
    $\NHtwo\sim 2.6\times 10^{25}$~cm$^{-2}$ (see \cref{fig:various eta-PDF}), which is fitted by two power laws, with slope $s_2$ in the range [-2.4 ; -0.6] and $s_3$ in the range [-0.9 ; -0.1] (see Sect.~\ref{sect:PDFs} and \cref{tab:cmf and s}). We then examined the mass segregation of cores in the subregions using the $\Sigma_{\rm 6\,cores}$ and $\Lambda_{\rm MSR}$ indicators (see Sect.~\ref{sect:mass segregation} and Figs.~\ref{fig:sigma6}--\ref{fig:lambda msr}). We find that the MM2, MM3, and MM10 subregions can be mass-segregated, with a very high level for MM2 (see \cref{tab:mass segregation}).
    \item Subregions with top-heavy CMFs display flat $\eta$-PDF tails, which is consistent with those obtained for the second tail of high-mass star-forming regions (see Sect.~\ref{sect:link CMF-PDF} and \cref{tab:eta-pdf characteristics}). In contrast, subregions with a CMF high-mass end close to the Salpeter slope have much steeper $\eta$-PDF tails, reminiscent of those measured for low-mass star-forming regions. In more detail, we observed a correlation trend between the CMF power-law index and the slope coefficient of the second $\eta$-PDF tail that corresponds to the cores' immediate background (see \cref{fig:correlation alpha vs s and proto}a). 
    \item We used the fraction of protostellar cores and the existence of an UC\hii region in MM3 to define the evolutionary stage of subregions, ranging from quiescent to post-burst via pre-burst and burst (see Sect.~\ref{sect:evolutionary stage} and \cref{tab:cmf and s}). Subregions in their main burst of star formation, just at its beginning or its end, are those with the flattest $\eta$-PDF tails and top-heavy CMFs, both likely resulting from an intense cloud formation and concentration at high column density. Subregions in the quiescent or pre-burst regimes have the steepest $\eta$-PDF tails and CMFs close to the Salpeter slope. We find a correlation trend between the power-law index of the CMF at the high-mass end and the protostellar fraction, which is used as a proxy for the star formation history in the W43-MM2\&MM3 subregions.
\end{itemize}

In the framework of dynamical cloud and star formation scenarios, such as in the competitive accretion, global hierarchical collapse, or inertial inflow models, part of the cloud collapses and most of it disperses \citep{smith2009,vazquez2019, pelkonen2021}. At parsec scales and high column density ($>10^{23}$~cm$^{-2}$), cloud ridges created by this global collapse would in turn contain centrally concentrated, collapsing hubs, which are the focus of large amounts of gas, and more weakly concentrated subregions, which should not be much impacted by the hierarchical collapse \citep{motte2018a,vazquez2019}. 
The W43-MM2\&MM3 ridge indeed consists of the weakly concentrated subregions MM51 and Outskirts, the intermediate subregions MM10 and MM12, and the centrally concentrated hubs MM2 and MM3, with MM2 and MM3 likely in their burst and post-burst regimes, respectively (see \cref{fig:summary}).

In the weakly concentrated subregions of W43-MM2\&MM3, star formation may develop in a continuous, almost quasi-static manner. Their more distributed cloud structure would lead to a core population without mass segregation and exhibiting a typical CMF that resembles the canonical IMF.
In contrast, in the centrally concentrated subregions of W43-MM2\&MM3, the cloud $\eta$-PDF has a flat second tail that could be the signature of the onset of rotation. The extreme column density reached in these subregions favors the formation of massive cores. The resulting CMF high-mass end is top-heavy, leading to a star formation burst and a high level of core mass segregation.
When feedback effects begin to set in, the characteristics of the core population, such as CMF and mass segregation, are impacted well before the cloud density structure, including its $\eta$-PDF function.

Since the largest part of a cloud, even if dynamically forming, would more closely resemble the MM51 or Outskirts subregions, we expect the average CMF of the cloud to be typical of low-mass star-forming regions, with a high-mass end close to the Salpeter slope. Top-heavy CMFs would then develop only locally, at a few specific locations in a globally collapsing cloud where column density is greater than $10^{23}$~cm$^{-2}$. 

Better understanding the link of the cloud structure, and in particular its $\eta$-PDF, to the CMF shape and possibly the IMF shape would require theoretical models to specifically simulate and analyze extreme-density clouds either corresponding to the converging points of clouds undergoing a hierarchical global collapse or to clouds heated and compressed by stellar feedback.


\begin{acknowledgements}
    This paper makes use of the ALMA data ADS/JAO.ALMA\#2017.1.01355.L. ALMA is a partnership of ESO (representing its member states), NSF (USA) and NINS (Japan), together with NRC (Canada), MOST and ASIAA (Taiwan), and KASI (Republic of Korea), in cooperation with the Republic of Chile. The Joint ALMA Observatory is operated by ESO, AUI/NRAO and NAOJ.
    This project has received funding from the European Research Council (ERC) via the ERC Synergy Grant \textsl{ECOGAL} (grant 855130), from the French Agence Nationale de la Recherche (ANR) through the project \textsl{COSMHIC} (ANR-20-CE31-0009), and the French Programme National de Physique Stellaire and Physique et Chimie du Milieu Interstellaire (PNPS and PCMI) of CNRS/INSU (with INC/INP/IN2P3).
    YP acknowledges funding from the IDEX Universit\'e Grenoble Alpes under the Initiatives de Recherche Strat\'egiques (IRS) “Origine de la Masse des \'Etoiles dans notre Galaxie” (OMEGa).
    YP, BL and GB acknowledge funding from the European Research Council (ERC) under the European Union’s Horizon 2020 research and innovation programme, for the Project “The Dawn of Organic Chemistry” (DOC), grant agreement No 741002. GB also acknowledges funding from the State Agency for Research (AEI) of the Spanish MCIU through the AYA2017-84390-C2-2-R grant and from the PID2020-117710GB-I00 grant funded by MCIN/ AEI  /10.13039/501100011033 . 
    RGM, TN, and DDG acknowledge support from UNAM-PAPIIT project IN108822. RGM is also supported by CONACyT Ciencia de Frontera project ID 86372. TN acknowledges support from the postdoctoral fellowship program of the UNAM.
    AGi acknowledges support from the National Science Foundation under grant No. 2008101.
    TCs and MB have received financial support from the French State in the framework of the IdEx Universit\'e de Bordeaux Investments for the future Program.
    PS was supported by a Grant-in-Aid for Scientific Research (KAKENHI Number 18H01259) of the Japan Society for the Promotion of Science (JSPS). P.S. and H.-L.L. gratefully acknowledge the support from the NAOJ Visiting Fellow Program to visit the National Astronomical Observatory of Japan in 2019, February.
    AS gratefully acknowledges funding support through Fondecyt Regular (project code 1180350), from the ANID BASAL project FB210003, and from the Chilean Centro de Excelencia en Astrof\'isica y Tecnolog\'ias Afines (CATA) BASAL grant AFB-170002. 
    FL acknowledges the support of the Marie Curie Action of the European Union (project \textsl{MagiKStar}, Grant agreement number 841276).
    SB acknowledges support from the French Agence Nationale de la Recherche (ANR) through the project \textsl{GENESIS} (ANR-16-CE92-0035-01).
    TB acknowledges the support from S. N. Bose National Centre for Basic Sciences under the Department of Science and Technology, Govt. of India.
    LB gratefully acknowledges support by the ANID BASAL projects ACE210002 and FB210003. 
    K.T. was supported by JSPS KAKENHI (Grant Number 20H05645).
\end{acknowledgements}

\bibliographystyle{aa}
\bibliography{ALMA-IMF-PaperVI}

\begin{appendix}

\section{Complementary figures} \label{appendixsect:complementary figures}

Appendix~\ref{appendixsect:complementary figures} shows in \cref{appendixfig:PPMAP with NH2} the temperature map with the $\NHtwo$ contours of Figs.~\ref{fig:nh2 and cores}b--c, and in \cref{appendixfig:pdf W43 herschel} the comparison of the column density $\eta$-PDFs derived from \cref{fig:nh2 and cores} and \textit{Herschel} data \citep{nguyen2013}. 
It also displays the fitted functions to the bootstrapping distributions of power-law indices of the CMF high-mass end measured in the MM2, MM3, MM10, and Outskirts subregions (see \cref{appendixfig:EMG subreg}).

\begin{figure*}[hbp!]
    \centering
    \includegraphics[width=0.9\textwidth]{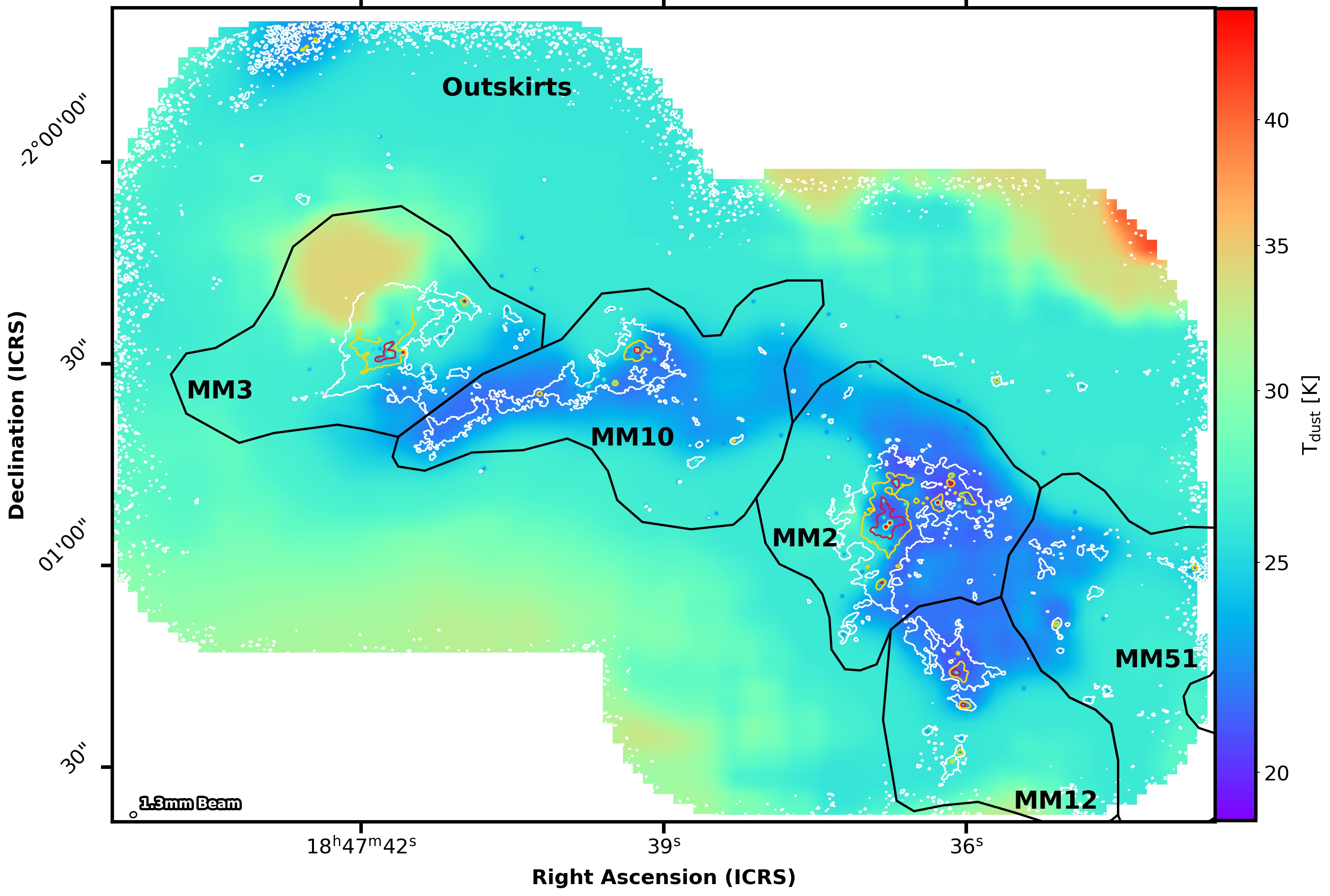}
    \caption{Dust temperature image of the W43-MM2\&MM3 protocluster cloud, taken from \cite{pouteau2022}. It combines a $2.5\arcsec$ resolution dust temperature image computed from a Bayesian spectral energy distribution fit for fluxes ranging from 70~$\mu$m to 3~mm (see Dell'Ova et al. in prep.) with the central heating and self-shielding of protostellar and pre-stellar cores, respectively, at $0.46\arcsec$ resolution \citep[see Sect.~4.2 of][for more details]{pouteau2022}. White, orange, and red contours correspond to the $\NHtwo$ levels defined in Figs.~\ref{fig:nh2 and cores}b--c. Black polygons indicate the subregions defined in \cref{sect:subregions def}, and the ellipse in the bottom-left corner corresponds to the 1.3~mm beam.}
\label{appendixfig:PPMAP with NH2}
\end{figure*}

\begin{figure*}[htbp!]
  \includegraphics[width=0.48\textwidth]{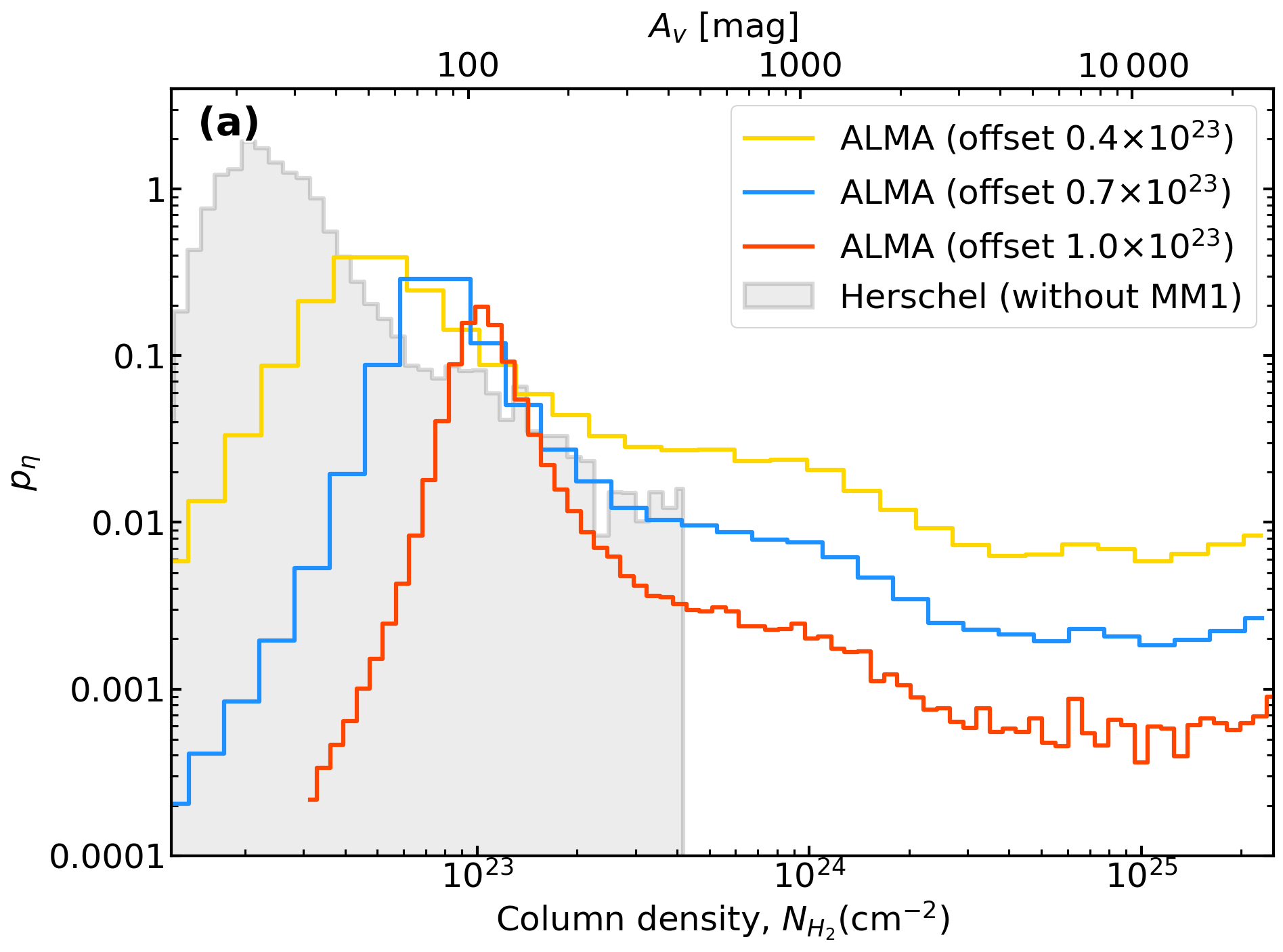} \hskip 0.5cm
  \includegraphics[width=0.48\textwidth]{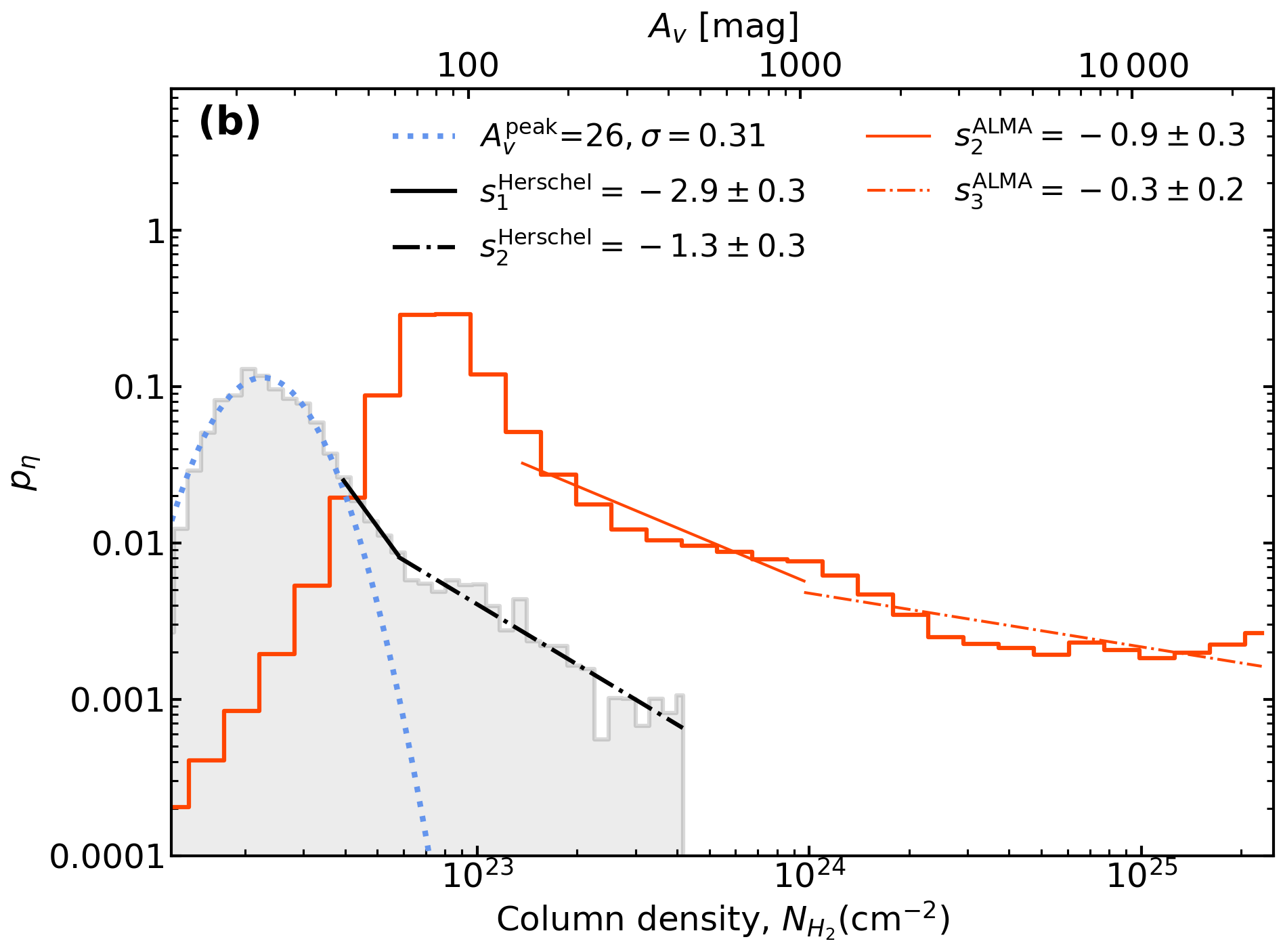}
  \caption{Comparison of the $\eta$-PDFs produced from the column density images of \textit{Herschel}/HOBYS covering W43-Main ($25\arcsec$ beam, without W43-MM1, gray-shaded histogram) and of ALMA-IMF toward W43-MM2\&MM3 ($0.51\arcsec\times0.40\arcsec$ beam, colored histograms). \textit{Panel (a):} Determination of the background column density of W43-MM2\&MM3 that best reconciles the ALMA and \textit{Herschel} $\eta$-PDFs: $0.7\times 10^{23}$~cm$^{-2}$. The $\eta$-PDFs are built with this column density offset (blue histogram) and the lower and upper limits of the offset uncertainty (yellow and red histograms, respectively). The \textit{Herschel} $\eta$-PDF has been multiplied by 15 to emphasize its relative agreement with the ALMA $\eta$-PDF. \textit{Panel (b):} $\eta$-PDF tails of \textit{Herschel}/HOBYS and ALMA-IMF with consistent power-law indices: $s_2^{\rm Herschel}=-1.3\pm 0.3$ (dotted black segment) and $s_2^{\rm ALMA}=-0.9\pm 0.3$ (continuous red segment). The first $\eta$-PDF tail with power-law index $s_1$ (continuous black segment) is only traced by the \textit{Herschel} image; the third tail with power-law index $s_3$ (dotted red segment) is only traced by ALMA.}
\label{appendixfig:pdf W43 herschel}
\end{figure*}

\begin{figure}[htbp!]
    \centering
    \includegraphics[width=0.49\textwidth]{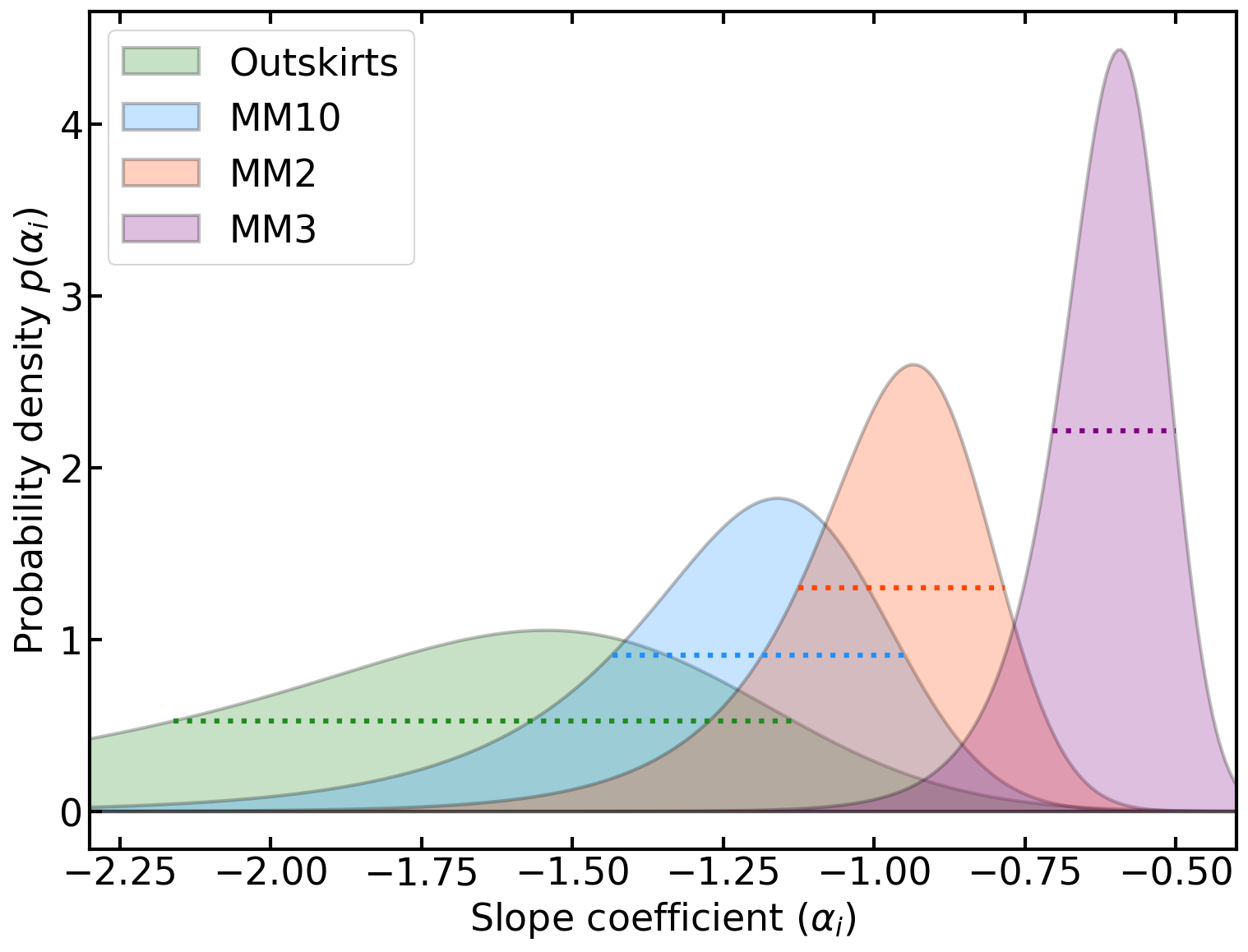}
    \caption{Bootstrapping probability density functions of the power-law indices measured for the CMF high-mass end of the MM2, MM3, MM10, and Outskirts subregions. Each distribution is fitted by an EMG with negative skewness. According to their $1\sigma$ width (dotted segments), subregions appear distinct, except for MM10, whose highest probability part of the bootstrapping distribution overlaps with both the MM2 and Outskirts subregions.}
    \label{appendixfig:EMG subreg}
\end{figure}

\end{appendix}

\end{document}